\documentclass[traditabstract]{aa}  
\usepackage{natbib}
\usepackage{graphicx}
\usepackage{txfonts}
\usepackage{color}
\usepackage{supertabular,longtable}

\begin{document} 

   \title{On the frequency dependence of p-mode frequency shifts induced by magnetic activity in {\it Kepler} solar-like stars}
   \titlerunning{Frequency shifts induced by magnetic activity in  \emph{Kepler} solar-like stars}
  
  \author{D. Salabert\inst{1,2}
         \and C. R\'egulo\inst{3,4}
         \and F. P\'erez Hern\'andez \inst{3,4} 
         \and R. A. Garc\'ia\inst{1,2}
          }

     \institute{IRFU, CEA, Universit\'e Paris-Saclay, F-91191 Gif-sur-Yvette, France\\
            \email{david.salabert@cea.fr}
            \and
        Universit\'e Paris Diderot, AIM, Sorbonne Paris Cit\'e, CEA, CNRS, F-91191 Gif-sur-Yvette, France
             \and Instituto de Astrof\'{\i}sica de Canarias, E-38200, La Laguna, Tenerife, Spain
              \and Universidad de La Laguna, Dpto. de Astrof\'{\i}sica, E-38205, La Laguna, Tenerife, Spain
               } 

   \date{Received XX XX XX; accepted XX XX XX}
   
     \abstract
     {The variations of the frequencies of the low-degree acoustic oscillations in the Sun induced by magnetic activity show a dependence with radial order. The frequency shifts are observed to increase towards higher-order modes to reach a maximum of about 0.8~$\mu$Hz over the 11-yr solar cycle. A comparable frequency dependence is also measured in two other main-sequence solar-like stars, the F-star HD\,49933, and the young 1-Gyr-old solar analog KIC\,10644253, although with different amplitudes of the shifts of about 2~$\mu$Hz and 0.5~$\mu$Hz respectively. Our objective here is to extend this analysis to stars with different masses, metallicities, and evolutionary stages. From an initial set of 87 {\it Kepler} solar-like oscillating stars with already known individual p-mode frequencies, we identify five stars showing frequency shifts that can be considered reliable using selection criteria based on Monte Carlo simulations and on the photospheric magnetic activity proxy $S_\text{ph}$. The frequency dependence of the frequency shifts of four of these stars could be measured for the $l=0$ and $l=1$ modes individually. Given the quality of the data, the results could indicate that a different physical source of perturbation than in the Sun is dominating in this sample of solar-like stars.}
     
      \keywords{stars: oscillations - stars: solar type - stars: activity - methods: data analysis}
 
  \maketitle

%
%
\section{Introduction}
The first detection of the signature in the acoustic (p) oscillation parameters of a magnetic activity cycle was observed with the Convection, Rotation, and planetary Transits \citep[CoRoT;][]{baglin06} space telescope in the F-type solar-like star HD\,49933 \citep{garcia10}. It demonstrated that asteroseismology can provide new and original constraints for the study of stellar magnetic activity and dynamo models under conditions different from those of the Sun. Since then, additional studies were performed using the long and continuous observations collected by the NASA {\it Kepler} telescope \citep{borucki10}. Changes in the characteristics of the p-mode frequencies as a consequence of their surface magnetism have been studied now in a handful of solar-like stars \citep[][Karoff et al. submitted; Santos et al., in preparation]{salabert16,regulo16,kiefer17}, suggesting that the frequency shifts decrease with stellar age and rotation period \citep{kiefer17}. 

The frequency dependence of the p-mode frequency shifts can provide insights on the structural changes of the stellar interior with magnetic activity. In the case of the Sun, the frequency shifts are known to increase with frequency as first shown by \citet{libbrecht90} and \citet{anguera92} for the intermediate- and low-degree modes respectively. Such dependence is explained as a consequence of changes in the outer layers of the Sun just beneath the solar photosphere \citep[e.g.,][]{libbrecht90,chaplin98,salabert04}. Moreover, similar frequency dependence of the shifts was also found in HD\,49933 \citep{salabert11} and in the G0-type KIC\,10644253 \citep{salabert16}. This latter star is particularly interesting as this young solar analog (1 Gyr-old) has an amplitude of the frequency shifts comparable to the Sun \citep{salabert16}. \citet{metcalfe07} and \citet{chaplin07a} formulated two different predictions of the expected frequency shifts based on scaling relations with the strength of the \ion{Ca}{ii} H and K emissions. For a given stellar age, the predicted shifts proposed by \citet{metcalfe07} get larger for hotter stars, while the ones by \citet{chaplin07a} show almost no variation with temperature. The results for the F-star HD\,49933 corroborate the formulation from \citet{metcalfe07}, while given the approximations considered and the observational uncertainties, the results for KIC\,10644253 cannot favor any of the two predictions. Nonetheless, the analysis from \citet{kiefer17} of a sample composed of 24 solar-like stars would support the scaling relation proposed by \citet{metcalfe07}.

The light-curve variability associated with the presence of spots rotating on the stellar surface provides another manner to study the global magnetic activity of stars. To cancel out the other sources of stellar variability at different timescales, such as the granulation or the oscillations, one needs to take into account the rotation period of the star when calculating such global photometric activity proxy, referred as the $S_\mathrm{ph}$ index \citep{mathur14a}. Several studies of the magnetic activity of main-sequence stars \citep{mathur14b,garcia14a,ferreira15,salabert16} observed with CoRoT and {\it Kepler} were performed based on the $S_\mathrm{ph}$. However, the $S_\text{ph}$ index is dependent on the inclination angle of the rotation axis of the star in respect to the line of sight, and thus represents a lower limit of stellar activity. This is of course supposing that starspots are formed over comparable latitunal bands as for the Sun. Nevertheless, the distribution of the spin orientation of field stars is observed to be consistent with an isotropic distribution \citep[see e.g.,][]{corsaro17,kuszlewicz17}, resulting in a most observed inclination angle close to 90$^{\circ}$, hence perpendicular to the line of sight. On the other hand, in exo-planet systems the distribution could differ from isotropy \citep[see e.g.,][]{winn06,hirano14}. The observed isotropy for field stars implies that the $S_\text{ph}$ index thus measured represents the actual level of activity for most of the stars and not a lower limit. Furthermore, comparisons with several standard solar activity proxies sensitive to different layers of the solar photosphere and chromosphere show that both the long- and short-term magnetic variabilities are well monitored with the $S_\text{ph}$ proxy \citep{salabert17}.

In this work, we want to extend this analysis to a larger sample of solar-like stars with different fundamental parameters. The objective here is to make progress in the understanding of the physical mechanisms responsible of the perturbations inducing the variability of the p-mode oscillations with magnetic activity. We started with a sample of 87 solar-like pulsating stars observed by {\it Kepler} and for which the peak-fitting analysis of the individual modes can be found in the literature \citep{app12,lund17}. The reason for selecting this sample is threefold: (1) it allowed us to study stars from the three following categories depending to their seismic and spectroscopic properties: the simple stars, the F-type stars, and the mixed-mode stars \citep[for more details, see][]{app12}; (2) the characterization of the individual modes of these stars is already available and
their fundamental stellar properties determined \citep{mathur12,metcalfe14,creevey17,silva17}; (3) these stars were observed in short cadence \citep[SC; for more details, see][]{gilliland10} for at least 1~year. The SC {\it Kepler} observations actually cover more than 3.5~years for most of these 87 stars. 

The paper is organized as follows. In Section\,\ref{sec:obs}, we describe the {\it Kepler} observations used here. The methods applied to derive the frequency shifts of the acoustic oscillations and the contemporaneous photometric magnetic activity proxy $S_\text{ph}$ are described in Section\,\ref{sec:activity}. In Section\,\ref{sec:criteria}, we define selection criteria based on Monte Carlo simulations to determine if the extracted variations of the frequency shifts are likely to be genuine signatures of magnetic variability or most likely due to noise. This is complemented by a selection criterion measuring the correlation between the frequency shifts and the $S_\text{ph}$ proxy. Based on these selection criteria, only few stars are found to show frequency shifts related to magnetic activity as presented in Section\,\ref{sec:results}. A discussion on the frequency dependence of the frequency shifts is proposed in Section\,\ref{sec:model}. Conclusions are presented in Section\,\ref{sec:conclusion}.

\begin{table}[t]
\caption{List of the 87 solar-like oscillating stars observed by  {\it Kepler} analyzed in this work. The full table will be available in the on-line version.}
\label{table:1}
\centering 
\begin{tabular}{l c c l} 
\hline\hline
KIC     &  \multicolumn{1}{c}{Years} &   \multicolumn{1}{c}{Duty cycle} &   \multicolumn{1}{l}{Category}\\
& &  \multicolumn{1}{c}{(\%) }& \\
\hline
1435467   &    3.73     &    82.7	 &      F-like\\
2837475   &    3.81     &    81.1   &   	F-like\\
3424541   &    3.73     &    82.7   &  	F-like\\
3427720   &    4.00     &    77.5   &      simple\\
3456181   &    2.54     &    51.3   &      F-like\\
3632418   &    3.73     &    82.7   &      simple\\
3656476   &    4.00     &    58.2   &      simple\\
3733735   &    3.81     &    81.1   & 	    F-like\\
3735871   &    3.89     &    79.4   & 	    F-like\\
4914923   &    4.00     &    71.2   &      simple\\
5184732   &    3.81     &    67.6   &      simple\\
5607242   &    3.89     &    79.4   &   mixed-modes\\
5773345   &    2.65     &    48.8   &      F-like\\
5950854   &    2.11     &    57.1   &      simple\\
5955122   &    4.00     &    77.5   &   mixed-modes \\ 
6106415   &    3.72     &    59.7   &      simple\\
6116048   &    3.81     &    81.1   &      simple\\
6225718   &    3.65     &    77.3   &      simple\\
6508366   &    4.00     &    77.5	 &      F-like\\
6603624   &    4.00     &    77.5   &      simple\\
6679371   &    3.73     &    82.7	 &      F-like\\
6933899   &    3.89     &    79.4   &      simple\\
...		  &				  &				 &		      \\
 \hline
\label{table:table_kic}
\end{tabular}
\tablefoot{The total length of the {\it Kepler} observations in fractional years and the associated duty cycles are given as well, along the category of each star.} 
\end{table}

%
%
\section{Observations}
\label{sec:obs}
Our initial sample consists in the 61 solar-like oscillating stars observed by {\it Kepler} analyzed in \citet{app12} complemented by 26 dwarfs from the LEGACY sample of 66 stars in \citet{lund17} and not included in \citet{app12}. These 87 stars containing main-sequence and slightly more evolved stars are listed in Table~\ref{table:table_kic}, where the total length of the observations and the associated duty cycles are given as well. This set of stars consists of {\it Kepler} targets which were observed for at least 12 months in SC \citep[$\Delta t \simeq 58.9$\,sec,][]{gilliland10}. The details of the observed and missing quarters can be found in Table~1 of \citet{lund17}. These stars are categorized within three groups: (1) the simple stars with a clear mode identification; (2) the F-like stars for which the mode identification is ambiguous; and (3) the more evolved mixed-mode stars for which the acoustic modes do not follow the asymptotic relation.

In this work, we used both the SC\footnote{The Data Release 25 (DR25) SC data were used in this analysis.} to perform the asteroseismic analysis of the frequency shifts of the acoustic oscillations, and the long-cadence \citep[LC, $\Delta t \simeq 29.4$\,min,][]{jenkins10} observations to study the temporal evolution of the photometric activity proxy $S_\text{ph}$ derived from the light-curve fluctuations. Both SC and LC observations were corrected for instrumental problems with the {\it Kepler} Asteroseismic Data Analysis and Calibration Software \citep[KADACS,][]{garcia11}. All the small gaps in the SC data were finally interpolated using an in-painting technique \citep{garcia14b,pires15}.

%
%
\section{Measurement of the magnetic activity variability}
\label{sec:activity}

\subsection{Extraction of the oscillation frequency shifts}
\subsubsection{Cross-correlation analysis (Method \#1)}
\label{sec:cross}  
The cross-correlation method returns a global value of the frequency shift for all the visible modes in the analyzed frequency range of the power spectrum (Method \#1). It was introduced by \citet{palle89} to measure the variations of the solar oscillation frequencies of the $l=0,1$, and 2 acoustic modes during cycles 21 and 22 from one single ground-based instrument located in Tenerife. The method has since then been widely used to study the frequency shifts from helioseismic \citep[see][and references therein]{chaplin07b} and asteroseismic \citep{garcia10,regulo16,salabert16,kiefer17} observations.

For each of the 87 stars in our sample, the {\it Kepler} SC time series were split into contiguous 90-day-long subseries, shifted by 30 days. The power spectrum of each 90-day subset was then cross correlated with a reference spectrum taken as the averaged spectrum of the independent spectra. The cross correlations were computed over frequency ranges, $\Delta\nu_\text{range}$, centered around the frequency of the maximum of the p-mode power excess, $\nu_\text{max}$, taken from \citet{app12} and \citet{lund17}. As the large frequency separation between consecutive radial orders decreases as $\nu_\text{max}$ goes towards lower frequencies, three frequency ranges to compute the cross correlation were defined as follows: (1) if $\nu_\text{max} >1100\,\mu$Hz , $\Delta\nu_\text{range} = \pm\,400\,\mu$Hz around $\nu_\text{max}$; (2) if $800\,\mu$Hz $\leq \nu_\text{max} \leq 1100\,\mu$Hz,  $\Delta\nu_\text{range} = \pm\,300\,\mu$Hz; and (3) if $\nu_\text{max} < 800\,\mu$Hz , $\Delta\nu_\text{range} = \pm\,200\,\mu$Hz. Doing so, we ensure to select all the visible modes for each star, although for the cross-correlation method, the p modes with lower heights --located at each end of the frequency range-- have very low weight and the obtained results are thus mostly sensitive to the higher modes around $\nu_\text{max}$.

To estimate the frequency shift $\delta\nu(t)$, the cross-correlation function was fitted with a Gaussian profile, whose centroid provides a measurement of the mean $l=0,1,$ and 2 mode frequency shift over the considered frequency range. The associated 1$\sigma$ uncertainties were obtained through Monte Carlo simulations as follows:  the reference spectrum was shifted in frequency by the mean frequency shift obtained for each subseries; then 500 simulated power spectra were computed by multiplying the shifted reference spectrum by a random noise with a distribution following a $\chi^2$ with 2 degrees of freedom. For each sub spectrum, the standard deviation of all the shifts obtained from the 500 simulated spectra with respect to the corresponding reference spectrum was adopted as the 1$\sigma$ error bar of the fit. A detailed description of the method can be found in \citet{regulo16}.

\subsubsection{Individual mode peak-fitting analysis (Method \#2)}
\label{sec:pkf}
As explained in Section\,\ref{sec:cross}, the cross-correlation method returns a global value of the frequency shift for all the visible $l=0,1$ and 2 modes in the power spectrum averaged over a given frequency range. The peak-fitting analysis provides the possibility to study separately the modes with different angular degrees and to have access to the frequency dependence of the frequency shifts (Method\,\#2). It gives access as well to the other acoustic parameters, such as the mode heights and widths for instance. 

For the peak-fitting analysis, 180-day-long subseries of the SC data were used with an overlap of 30 days. This length provides enough frequency resolution to retrieve the individual modes with accuracy. It offers also the possibility to extract the frequency shifts over independent subseries compared the cross-correlation analysis (Method\,\#1). The power spectrum of each 180-day subseries was then obtained in order to extract estimates of the oscillation parameters. The individual frequencies of the angular degrees $l=0$, 1, and 2 provided by \citet{app12} and by \citet{lund17} were taken as initial guesses. The analysis was performed by fitting sequences of successive series of $l=0$, 1, and 2 modes using a maximum-likelihood estimator as in \citet{salabert11} for the CoRoT target HD\,49933 and in \citet{salabert16} for the young solar-analog {\it Kepler} target KIC\,10644253. The individual $l=0$, 1, and 2 modes were modeled using a Lorentzian profile for each of the angular degrees. Therefore, neither rotational splitting nor inclination angle were included in the fitting model.  The height ratios between the $l=0$, 1, and 2 modes were fixed to 1, 1.5, and 0.5 respectively \citep{ballot11}, and only one linewidth was fitted per radial order $n$. The natural logarithms of the mode height, linewidth, and background noise were varied resulting in normal distributions. The formal uncertainties in each parameter were then derived from the inverse Hessian matrix. The mode frequencies thus extracted were checked to be consistent within 1$\sigma$ errors with \citet{app12} and \citet{lund17}. 

The mean temporal variations of the frequencies were defined as the differences between the frequencies observed at different dates and reference values of the corresponding modes. The set of reference frequencies was taken as the average over the entire observations. The frequency shifts thus obtained were then averaged over the same frequency ranges as the ones used for the cross-correlation analysis defined in Section~\ref{sec:cross}. As the cross-correlation method returns mean frequency shifts unweighted over the analyzed frequency range, we decided to use unweighted averages of the frequency shifts extracted from the individual frequencies. We also note that we did not use the frequencies of the $l = 2$ modes because of their lower signal-to-noise ratios. 

Additionally, we analyzed the temporal variations of the power and damping of the p-mode oscillations. Indeed, these physical properties are known to be sensitive to magnetic activity as observed in the Sun \citep[see e.g.,][]{palle90,elsworth93,chaplin00,howe03,salabert06} and stars \citep[][Karoff et al., submitted]{garcia10,kiefer17}. The results for two of the stars analyzed in this work are presented in Appendix~\ref{sec:height}. A more complete analysis will be available in Santos et al. (in preparation).

\subsection{Determination of the photometric activity proxy}
\label{sec:sph} 
The stellar magnetic variability was also measured through the photometric activity proxy $S_\text{ph}(t)$, which is derived by means of the surface rotation, $P_\mathrm{rot}$. The $S_\mathrm{ph}(t)$ proxy is defined as the mean value of the light curve fluctuations estimated as the standard deviations calculated over subseries of length $5 \times P_\mathrm{rot}$. Out of the 87 solar-like oscillating stars in our sample, the temporal evolution of the $S_\mathrm{ph}(t)$ proxy was estimated from the {\it Kepler} LC data for the 53 of them which have estimates of $P_\mathrm{rot}$ available \citep{garcia14a}.

%
%
\section{Definition of the selection criteria}
\label{sec:criteria}
We defined three selection criteria to determine whether the variations observed in the extracted frequency shifts, $\delta\nu(t)$, are likely to be actual signatures of magnetic variability or to be due to noise. The two first criteria are based on Monte-Carlo simulations applied to observations of the Sun. The first selection criterion, $\lambda_1$, evaluates the significance of frequency shifts with a period longer than the observations, while the second criterion, $\lambda_2$, assess the variations with a cycle period shorter than the observations.
The third criterion, $\lambda_3$, compares the frequency shifts with the contemporaneous temporal variability of the photometric activity proxy $S_\textrm{ph}(t)$. Since both the cross-correlation analysis (Method\,\#1) and peak-fitting analysis (Method\,\#2) used to extract the frequency shifts return similar results within their errors, the selection criteria were characterized using the results from Method\,\#1.

\begin{table}
\caption{Detection percentage $p(\lambda_1)$ to have the selection criterion $\lambda_1$ fulfilled, based on simulations of linearly varying frequency shifts.}
\label{table:simu}
\centering 
\begin{tabular}{r l l l l l l } 
\hline\hline
& \multicolumn{6}{c}{$\overline{\sigma_{\delta\nu}}$ ($\mu$Hz) }\\
\hline
               & 0.09  & 0.16 &0.25 & 0.35 &0.61 & 0.70\\
$\delta\nu^\textrm{max}$  ($\mu$Hz)& & & & & & \\
\hline
0.0  \vline   & 23   & 19  &  19  &  19  &  12  &  6  \\
0.1  \vline   & 85   & 51  &  29  &  20  &  11  &  3  \\
0.2  \vline   & 84   & 40  &  32  &  20  &  16  &  5  \\
0.3  \vline   & 97   & 76  &  40  &  31  &  18  &  3  \\   
0.4  \vline   & 100  & 88  &  53  &  43  &  19  &  14  \\
0.5  \vline   & 100  & 95  &  74  &  51  &  24  &  8  \\  
0.6  \vline   & 100  & 100 &  84  &  57  &  32  &  10  \\
0.7  \vline   & 100  & 99  &  87  &  71  &  26  &  11  \\
0.8  \vline   & 100  & 100 &  97  &  75  &  50  &  13  \\  
\hline
\end{tabular}
\tablefoot{The mean error $\overline{\sigma_{\delta\nu}}$ depends on the different simulated levels of noise, $\sigma_0$: $\overline{\sigma_{\delta\nu}}= 0.09\,\mu$Hz corresponds to $5\sigma_0$; $\overline{\sigma_{\delta\nu}}= 0.16\,\mu$Hz to $10\sigma_0$; $\overline{\sigma_{\delta\nu}}= 0.25\,\mu$Hz to $15\sigma_0$; $\overline{\sigma_{\delta\nu}}= 0.35\,\mu$Hz to $20\sigma_0$; $\overline{\sigma_{\delta\nu}}= 0.61\,\mu$Hz  to $25\sigma_0$; and $\overline{\sigma_{\delta\nu}}= 0.70\,\mu$Hz  to $30\sigma_0$. The parameter $\delta\nu^\textrm{max}$ is the maximum linear frequency shift introduced in the associated simulation.}
\end{table}

\begin{figure*}[tbp]
\begin{center} 
\includegraphics[width=1\textwidth,angle=0]{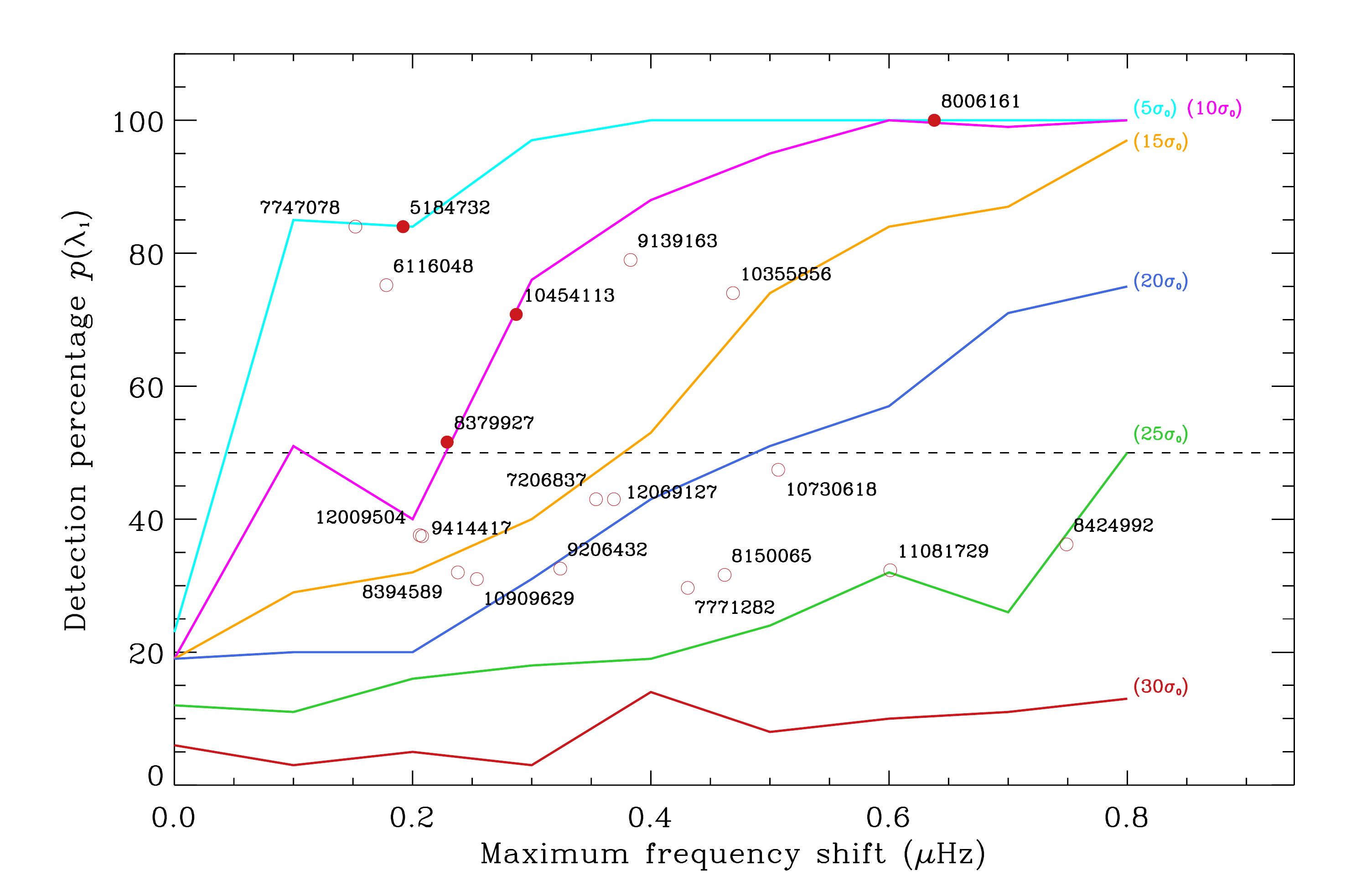}
\end{center}
 \caption{Detection percentage $p(\lambda_1)$ as a function of the maximum linear frequency shift $\lvert\textrm{Max}(\delta\nu)\rvert_\textrm{L}$ obtained from 100 realizations of each simulated spectrum (see Section~\ref{sec:lambda1}). Each solid line corresponds to a given (color-coded) level of noise $\sigma_0$. The opened red circles correspond to the {\it Kepler} stars which successfully passed the selection criterion $\lambda_1$, while the filled red circles indicate the stars which passed the selection criterion $\lambda_2$ as well (see Section~\ref{sec:results}). The $y$ coordinate of each star represents the percentage $p(\lambda_1)$ of being detected according to the simulations. The horizontal dashed line indicates the 50\% detection threshold.}
\label{fig:simu}
\end{figure*} 

\begin{table}
\caption{Detection percentage $p(\lambda_2)$ to have the selection criterion $\lambda_2$ fulfilled, based on simulations of sinusoidal varying frequency shifts. The results for simulated periodicity of 4~years are presented here.}
\label{table:simu2}
\centering 
\begin{tabular}{r l l l l l l } 
\hline\hline
& \multicolumn{6}{c}{$\overline{\sigma_{\delta\nu}}$ ($\mu$Hz) }\\
\hline
& 0.09  & 0.16 &0.25 & 0.35 &0.61 & 0.70\\
$\delta\nu^\textrm{cycle}$  ($\mu$Hz)& & & & & & \\
\hline
0.0  \vline   & 5    & 9    &  9   &  5   &  2  &  0  \\		
0.2  \vline   & 97   & 50   &  19  &  6   &  3  &  0  \\
0.3  \vline   & 100  & 84   &  38  &  21  &  2  &  0  \\   
0.4  \vline   & 100  & 100  &  69  &  29  &  7  &  0  \\
\hline
\end{tabular}
\tablefoot{The mean error $\overline{\sigma_{\delta\nu}}$ depends on the different simulated levels of noise, $\sigma_0$: $\overline{\sigma_{\delta\nu}}= 0.09\,\mu$Hz corresponds to $5\sigma_0$; $\overline{\sigma_{\delta\nu}}= 0.16\,\mu$Hz to $10\sigma_0$; $\overline{\sigma_{\delta\nu}}= 0.25\,\mu$Hz to $15\sigma_0$; $\overline{\sigma_{\delta\nu}}= 0.35\,\mu$Hz to $20\sigma_0$; $\overline{\sigma_{\delta\nu}}= 0.61\,\mu$Hz  to $25\sigma_0$; and $\overline{\sigma_{\delta\nu}}= 0.70\,\mu$Hz  to $30\sigma_0$. The parameter $\delta\nu^\textrm{cycle}$ is the amplitude of the frequency shifts along an activity cycle introduced in the associated simulation.}	
\end{table}


\subsection{Simulations with the photometric VIRGO data}
\label{sec:virgosimu}
The selection criteria $\lambda_1$ and $\lambda_2$ were characterized through Monte-Carlo simulations using 4 years of photometric space-based observations of the Sun collected by the Sun Photometers (SPM) of the Variability of Solar Irradiance and Gravity Oscillations \citep[VIRGO;][]{frohlich95} instrument on board the Solar and Heliospheric Observatory \citep[SoHO;][]{domingo95} satellite. The selected VIRGO/SPM time series, long of 1440~days, starts on February 8, 2007 and is centered around the deep and long minimum of the solar cycle~23 \citep[e.g.,][]{salabert09}. {It was split into 16 subseries of 90 days, and the associated power spectra computed.}

The outline of the Monte-Carlo simulations is the following: (1) a reference power spectrum was defined as the mean spectrum of the 16 VIRGO/SPM spectra; (2) starting from this reference spectrum, 16 power spectra were constructed by shifting their p-mode frequencies from one spectrum to the next in both a linear and sinusoidal ways; (3) finally, additional noise following a $\chi^2$ distribution and proportional to the photon noise, $\sigma_0$, estimated from the high-frequency region of the reference power spectrum was included to the power spectra. 
For validation purposes, we checked that  the selection criteria do not return positive detections for variability of the frequency shifts extracted from the 16 VIRGO/SPM power spectra taken during solar minimum and used in the Monte-Carlo simulations.

\subsection{Selection criterion $\lambda_1$}
\label{sec:lambda1}
The absolute minimum-to-maximum temporal variations of the frequency shifts along the total duration of the observations, $\lvert\textrm{Max}(\delta\nu)\rvert_\textrm{L}$, is estimated from the linear regression of the temporal variations of the extracted frequency shifts, $\delta\nu(t)$. The selection criterion $\lambda_1$ is fulfilled when:
\begin{equation}
\lvert\textrm{Max}(\delta\nu)\rvert_\textrm{L} > \overline{\sigma_{\delta\nu}},
\end{equation}
where $\overline{\sigma_{\delta\nu}}$  corresponds to the mean value of the uncertainties on the extracted frequency shifts.

The criterion $\lambda_1$ was assessed through simulations of linear frequency shifts introduced in the 90-day-long VIRGO/SPM reference spectrum (see Section~\ref{sec:virgosimu}). Maximum frequency shifts $\delta\nu^\textrm{max}$ between 0.0 and 0.8\,$\mu$Hz were modeled along the 16 contiguous spectra. Different frequency shifts were included by steps of 0.1\,$\mu$Hz. All the p-mode frequencies were shifted by the same amount, meaning that no frequency dependence of the frequency shift was introduced.
In addition, noise levels between 5$\sigma_0$ and 30$\sigma_0$ in steps of 5$\sigma_0$ were introduced, where $\sigma_0= 0.4$\,ppm. The frequency shifts of these simulated spectra were then extracted using Method\,\#1 as described in Section~\ref{sec:cross}. The mean errors on the frequency shifts, $\overline{\sigma_{\delta\nu}}$, extracted from the simulated spectra depend on the introduced level of noise and lie between 0.09 and 0.70\,$\mu$Hz, which is comparable to the errors obtained with real data. 

The results obtained for 100 realizations of each simulated spectrum are shown in Fig.~\ref{fig:simu}, while Table~\ref{table:simu} provides the corresponding detection percentage, $p(\lambda_1)$, to have the criterion $\lambda_1$ fulfilled. Each line represents the number of successful detections as a function of a given level of noise. When no frequency shift was introduced, the numbers given in Table~\ref{table:simu} represent the percentage of false positives, which ranges from 23\% to 6\%. These false positives are higher when the noise is lower. Indeed, any small shift created by the noise is likely to be larger than the mean error, $\overline{\sigma_{\delta\nu}}$, when this one is small. On the other hand, when the noise is low and the maximum frequency shift  $\lvert\textrm{Max}(\delta\nu)\rvert_\textrm{L}$ is larger than 0.3\,$\mu$Hz, the temporal variability of the p-mode frequency shifts is detected 100\% of the time. We note that we checked that similar results were obtained if a larger number of realizations was used.

\begin{table*}
\caption{List of stars fulfilling the selection criterion $\lambda_1$ for showing temporal variability of their p-mode frequency shifts.}
\centering 
\begin{tabular}{l l c c c c c } 
\hline\hline
KIC & \multicolumn{1}{c}{$\lvert\textrm{Max}(\delta\nu)\rvert_\textrm{L}$} & \multicolumn{1}{c}{$\overline{\sigma_{\delta\nu}}$}  & \multicolumn{1}{c}{$p(\lambda_1)$}   &  \multicolumn{1}{c}{$\langle S_\text{ph} \rangle$}  & \multicolumn{1}{c}{$P_\text{rot}$} & Category\\

& \multicolumn{1}{l}{($\mu$Hz)}   & \multicolumn{1}{c}{($\mu$Hz)} & \multicolumn{1}{c}{(\%)}& \multicolumn{1}{c}{(ppm)} &  \multicolumn{1}{c}{(days)}\\			
\hline

   5184732$^\ast$      &    0.192   &     0.077     &    84.0      & 222.0 & 19.79\,$\pm$\,2.43 & simple \\
   6116048$^\dagger$     &    0.178   &      0.093    &    75.2     &  87.4 & 17.26\,$\pm$\,1.96 & simple \\ 
   7206837          &    0.354   &      0.280    &    43.0     &  246.6 & 4.04\,$\pm$\,0.28 &simple \\
   7747078         &    0.152   &      0.077    &    84.0      &  176.0 & 17.71\,$\pm$\,1.78 & mixed-modes\\
   7771282          &    0.431   &      0.396    &    29.7     &  117.6 & 11.88\,$\pm$\,0.91 & F-like\\
   8006161$^\ast$       &    0.638   &      0.131    &   100.0      &  763.5 & 29.79\,$\pm$\,3.09 & simple\\
   8150065          &    0.462   &      0.400    &    31.6      & &  & simple\\
   8379927$^{\ast,\flat}$        &    0.229   &      0.138    &    51.6      & &  & simple\\
   8394589          &    0.238   &      0.196    &    32.0      & &  & simple\\
   8424992          &    0.749   &      0.449    &    36.2      & &  & simple\\
   9139163$^\ddagger$         &    0.383   &      0.194    &    78.9     & &  & simple\\
   9206432          &    0.324   &      0.297    &    32.6      &  96.7 & 8.80\,$\pm$\,1.06 & F-like\\
   9414417          &    0.206   &      0.172    &    37.6     &  146.5 & 22.77\,$\pm$\,2.37 & F-like\\
   10355856       &    0.469   &      0.227    &    74.0      &   321.5 & 4.47\,$\pm$\,0.31 & F-like\\
   10454113$^\ast$     &    0.287   &      0.173    &    70.8      &  327.3 & 14.61\,$\pm$\,1.09 & simple\\
   10730618         &    0.507   &      0.421    &    47.4     & & & F-like\\ 
   10909629         &    0.254   &      0.250    &    31.0      & 55.1 & 12.37\,$\pm$\,1.22 & F-like \\ 
   11081729        &    0.601   &      0.521    &    32.3     &   272.2 & 2.74\,$\pm$\,0.31& F-like\\
   12009504         &    0.208   &      0.188    &    37.4      & 155.3 & 9.39\,$\pm$\,0.68 & simple\\
   12069127         &    0.369   &      0.262    &    43.0    &   86.6 & 4.47\,$\pm$\,0.31 & F-like\\
\hline                                                 
\end{tabular}
\label{table:shifts}
\tablefoot{The column $\lvert\textrm{Max}(\delta\nu)\rvert_\textrm{L}$ corresponds to the maximum linear frequency shift measured for the stars. The column $\overline{\sigma_{\delta\nu}}$ gives the mean errors of the extracted frequency shifts. The column $p(\lambda_1)$ gives the percentage for a given star to be detected according to the simulations. The column $\langle S_\textrm{ph} \rangle$ corresponds to the mean value of the $S_\textrm{ph}(t)$ proxy calculated over the duration of the observations. The surface rotation periods, $P_\mathrm{rot}$, were measured by \citet{garcia14a}. When no measurements of $P_\mathrm{rot}$ are available, the corresponding values of $P_\mathrm{rot}$ and $\langle S_\textrm{ph} \rangle$ are left blank. The $\ast$ symbol indicates the stars which fulfilled the selection criterion $\lambda_2$ too. The star KIC\,9139163 flagged by the $\ddagger$ symbol has frequency shifts with a temporal periodicity close to the {\it Kepler} orbital period. The star KIC\,6116048 marked with the $\dagger$ symbol is a seismic solar analog \citep{salabert16}. The star KIC\,8379927 marked with the $\flat$ symbol is part of a spectroscopic binary system \citep{pourbaix04}.  }
\end{table*}

\subsection{Selection criterion $\lambda_2$}
\label{sec:lambda2}
The differences between the 25\% higher and lower percentiles of the extracted temporal variations of the frequency shifts $\delta\nu(t)$ provide a manner to evaluate the existence of variability shorter than the length of the observations, i.e. 4 years in the case of the {\it Kepler} satellite.
The selection criterion $\lambda_2$ is thus defined to be fulfilled when: 
\begin{equation}
  [\delta\nu(t)^{25^\textrm{th\_high}} -\delta\nu(t)^{25^\textrm{th\_low}}] > 3 \times \overline{\sigma_{\delta\nu}},
\end{equation}
where $\overline{\sigma_{\delta\nu}}$  corresponds to the mean errors on the extracted frequency shifts. It thus allows us to select stars showing sinusoidal behavior, i.e. cycle-like, of the frequency shifts, that could not be detected with the first criterion $\lambda_1$ defined to measure the linear deviation from zero of the frequency shifts over the total length of the observations.

To characterize the $\lambda_2$ criterion, sinusoidal frequency shifts were introduced with cycle amplitudes, $\delta\nu^\textrm{cycle}$, of 0.2, 0.3, and 0.4 $\mu$Hz and periods of 1, 2, 3, and 4 years  in the VIRGO/SPM reference power spectrum (see Section~\ref{sec:virgosimu}). The same levels of noise as the ones applied in Section~\ref{sec:lambda1} were introduced.  The frequency shifts of these simulated spectra were obtained using Method\,\#1 as described in Section~\ref{sec:cross}. The results obtained from 100 realizations of each simulated spectrum are shown in Table~\ref{table:simu2} which provides the corresponding detection percentage $p(\lambda_2)$  to have the criterion $\lambda_2$ fulfilled. Table~\ref{table:simu2} was obtained for a simulated period of 4 years, and similar values are obtained for periods of 1, 2, and 3 years. The selection criterion $\lambda_2$ thus defined is quite restrictive. When the introduced noise produces mean errors on the frequency shifts $\overline{\sigma_{\delta\nu}}$ larger than 0.16\,$\mu$Hz and that the introduced cycle amplitudes are smaller than 0.3\,$\mu$Hz, less than 20\% of the stars are recovered. On the other hand, the number of false positives is never higher than 10\% in all the simulated cases. However, if the criterion $\lambda_2$ is relaxed to a threshold of $2 \times \overline{\sigma_{\delta\nu}}$, the number of false positives becomes larger than 70\%.

\begin{figure*}[tbp]
\begin{center} 
\includegraphics[width=0.33\textwidth,angle=0]{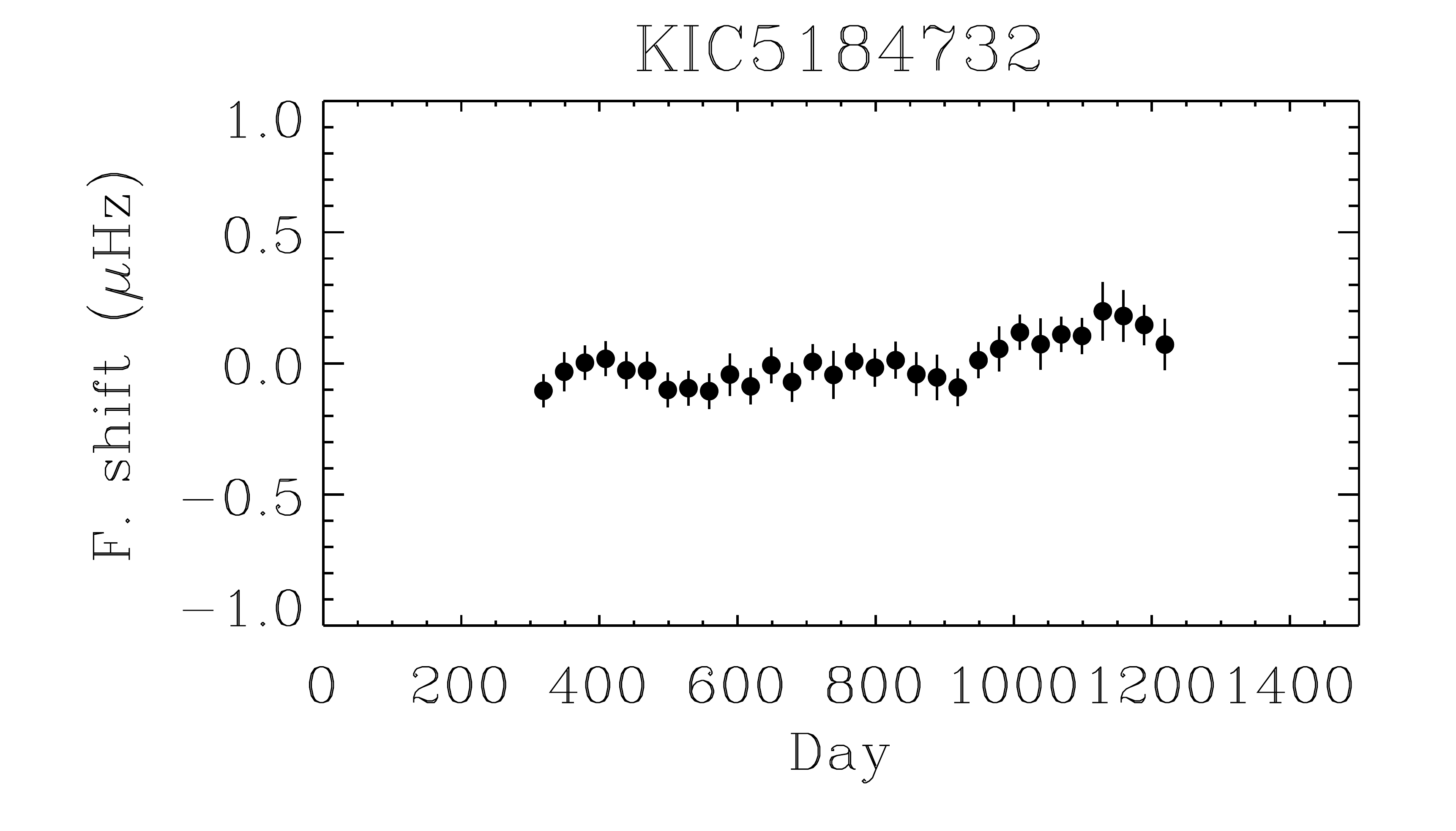}
\includegraphics[width=0.33\textwidth,angle=0]{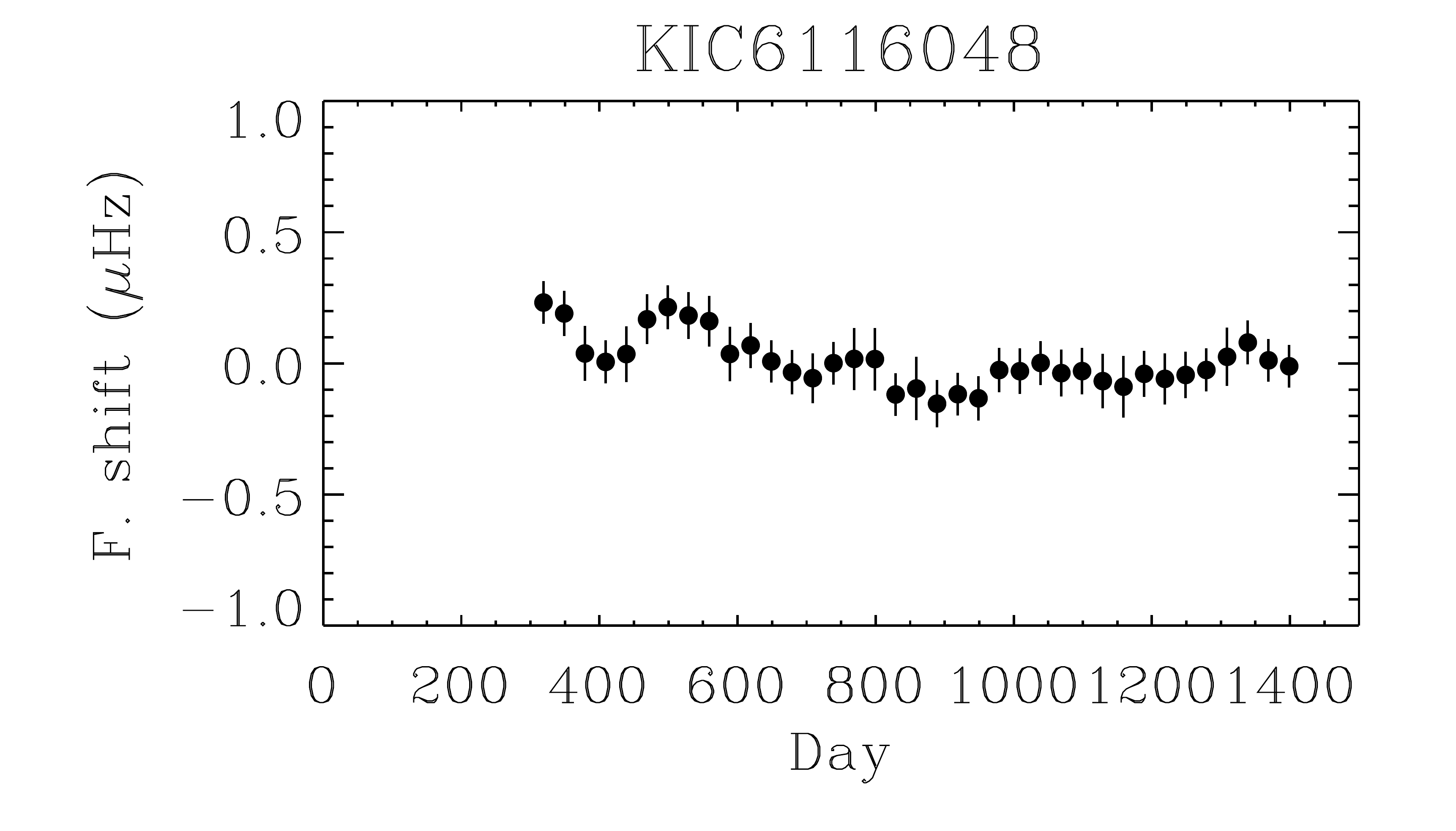}
\includegraphics[width=0.33\textwidth,angle=0]{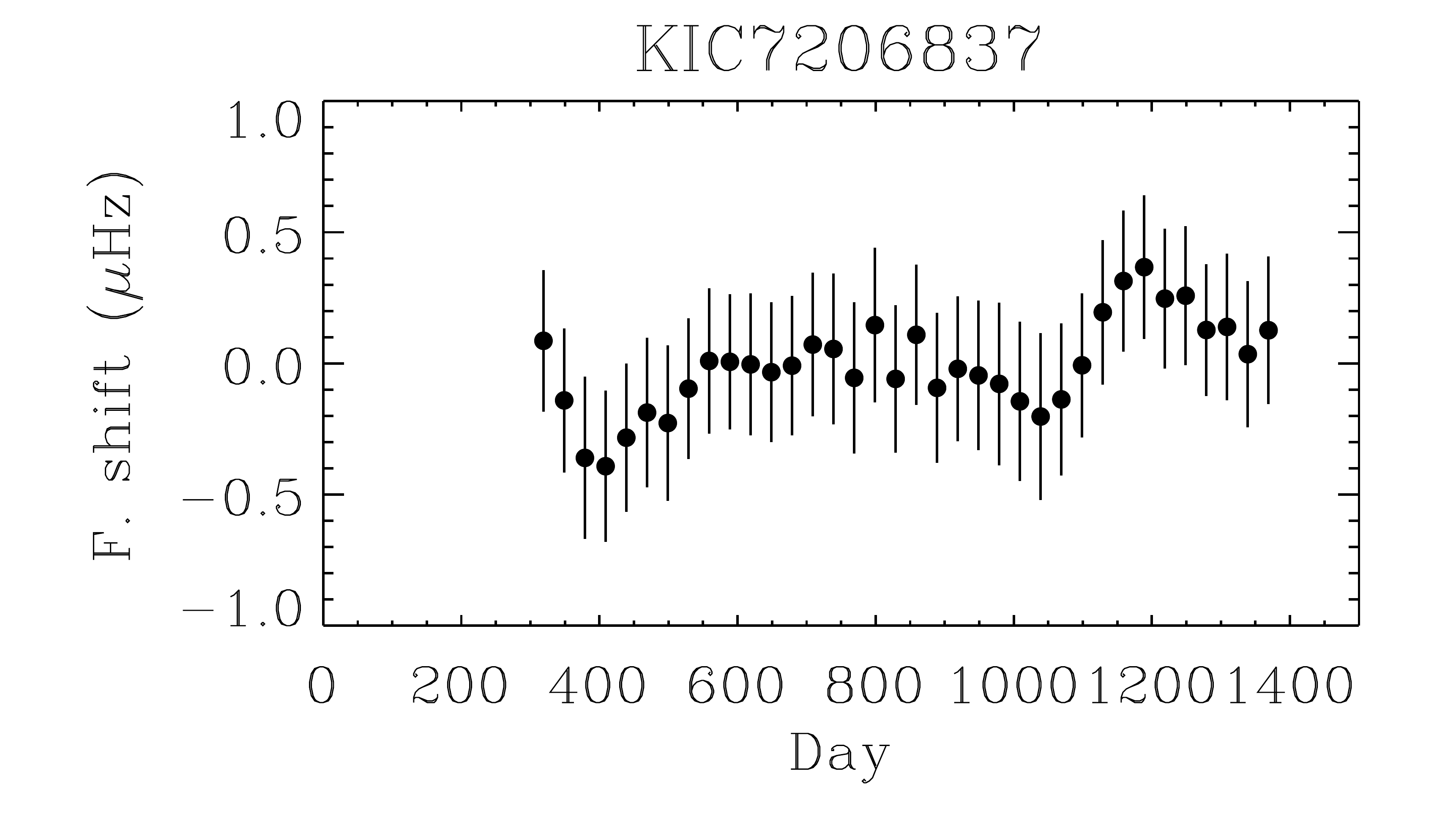}

\includegraphics[width=0.33\textwidth,angle=0]{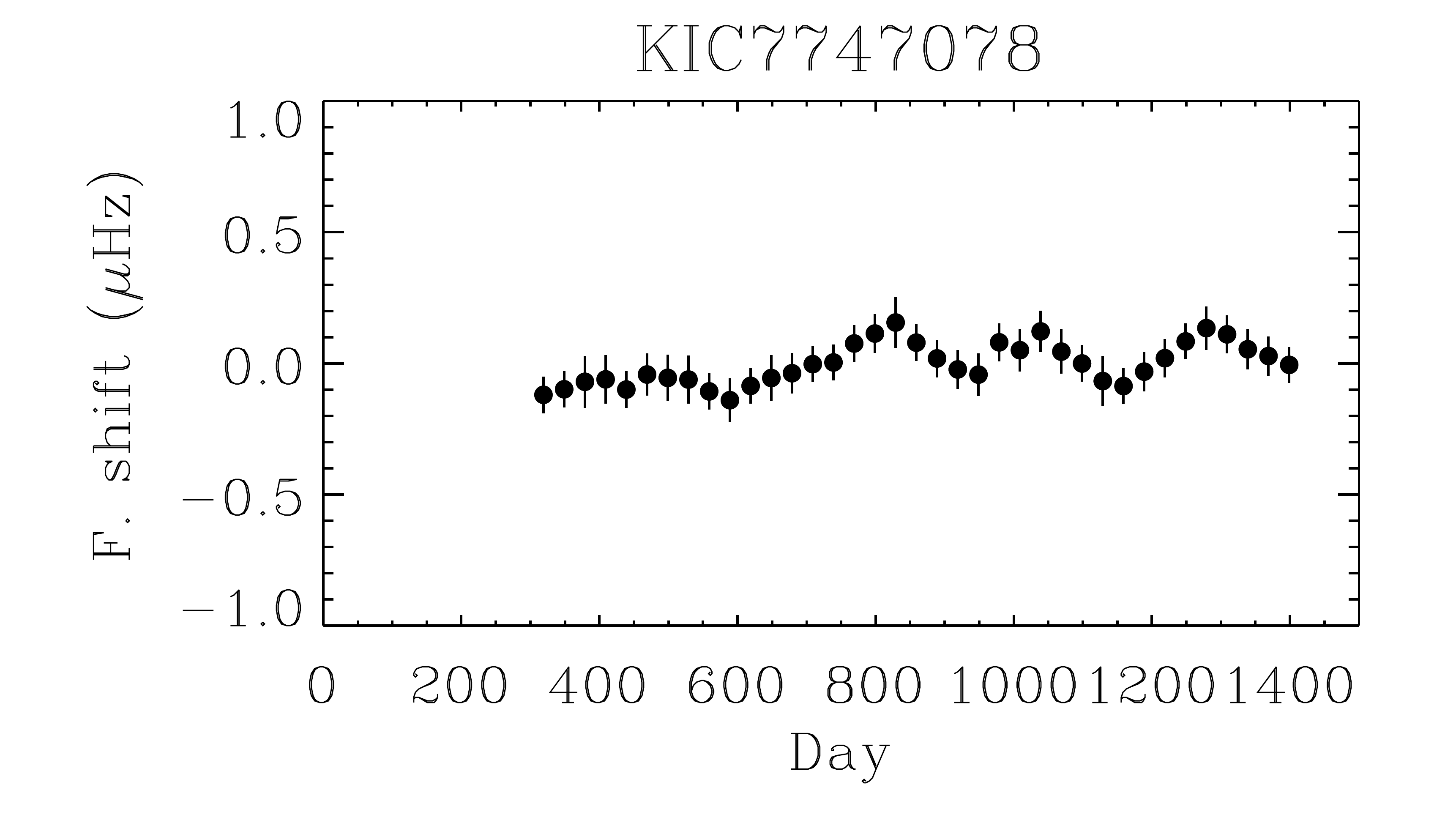}
\includegraphics[width=0.33\textwidth,angle=0]{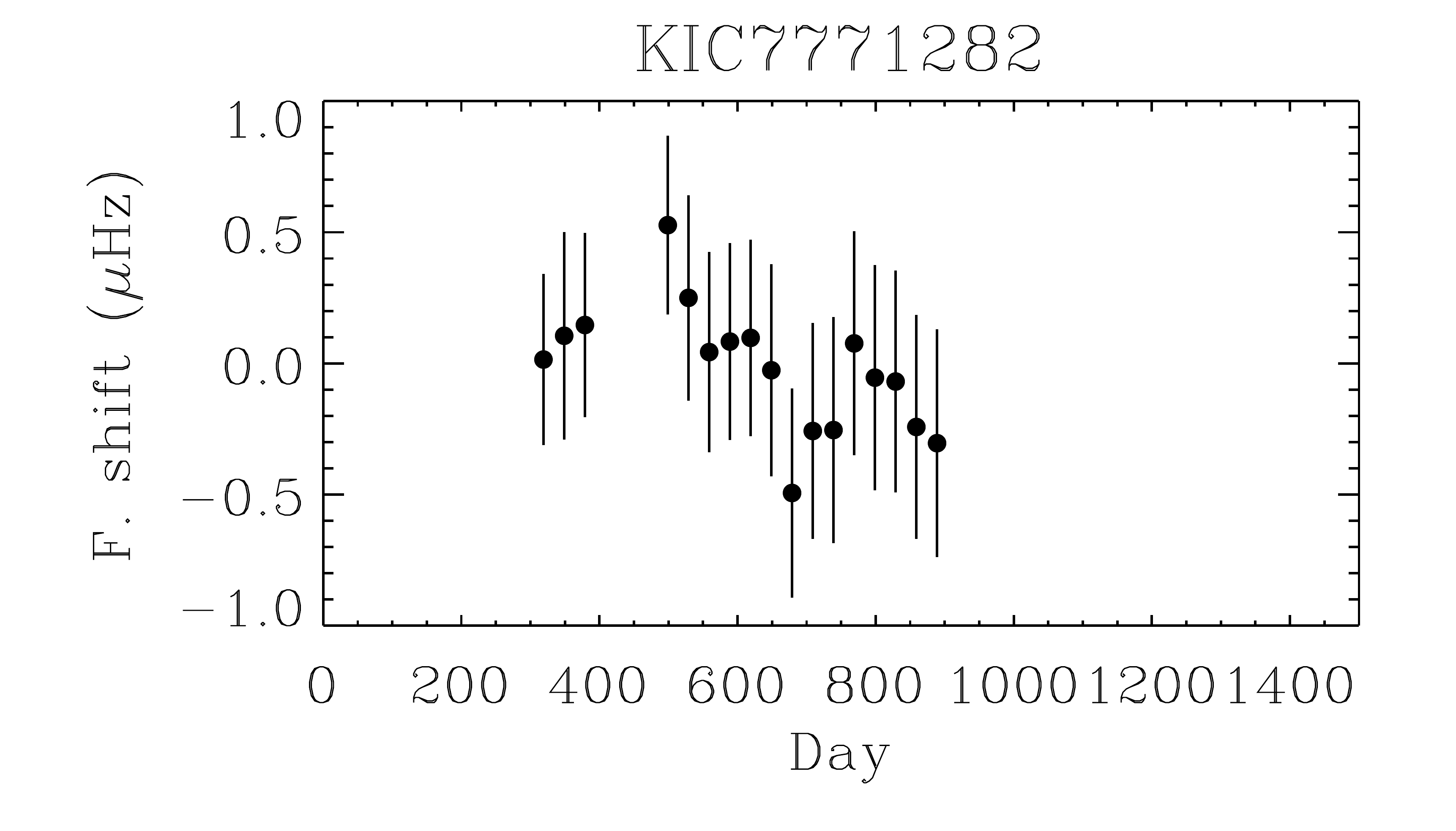}
\includegraphics[width=0.33\textwidth,angle=0]{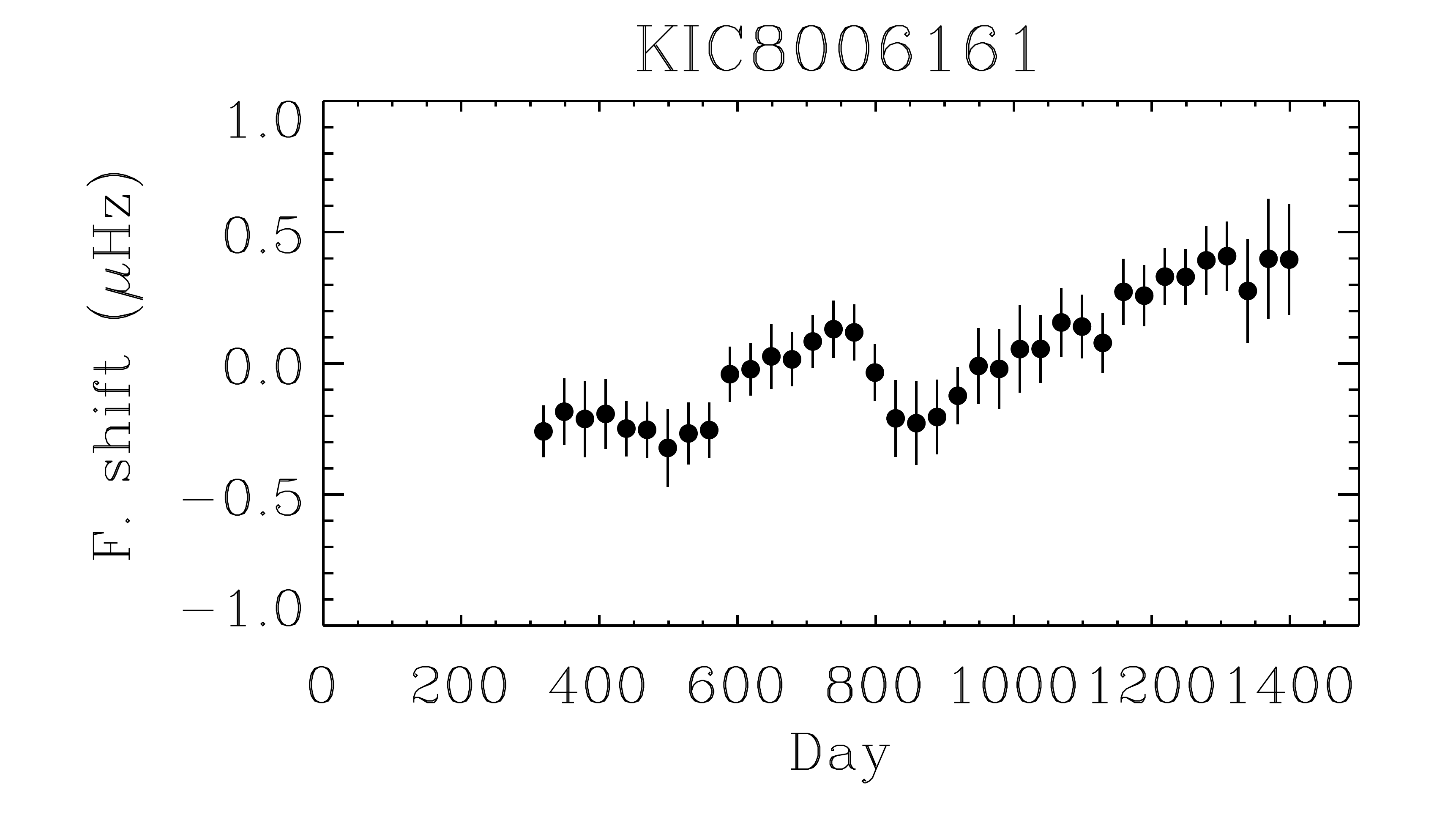}

\includegraphics[width=0.33\textwidth,angle=0]{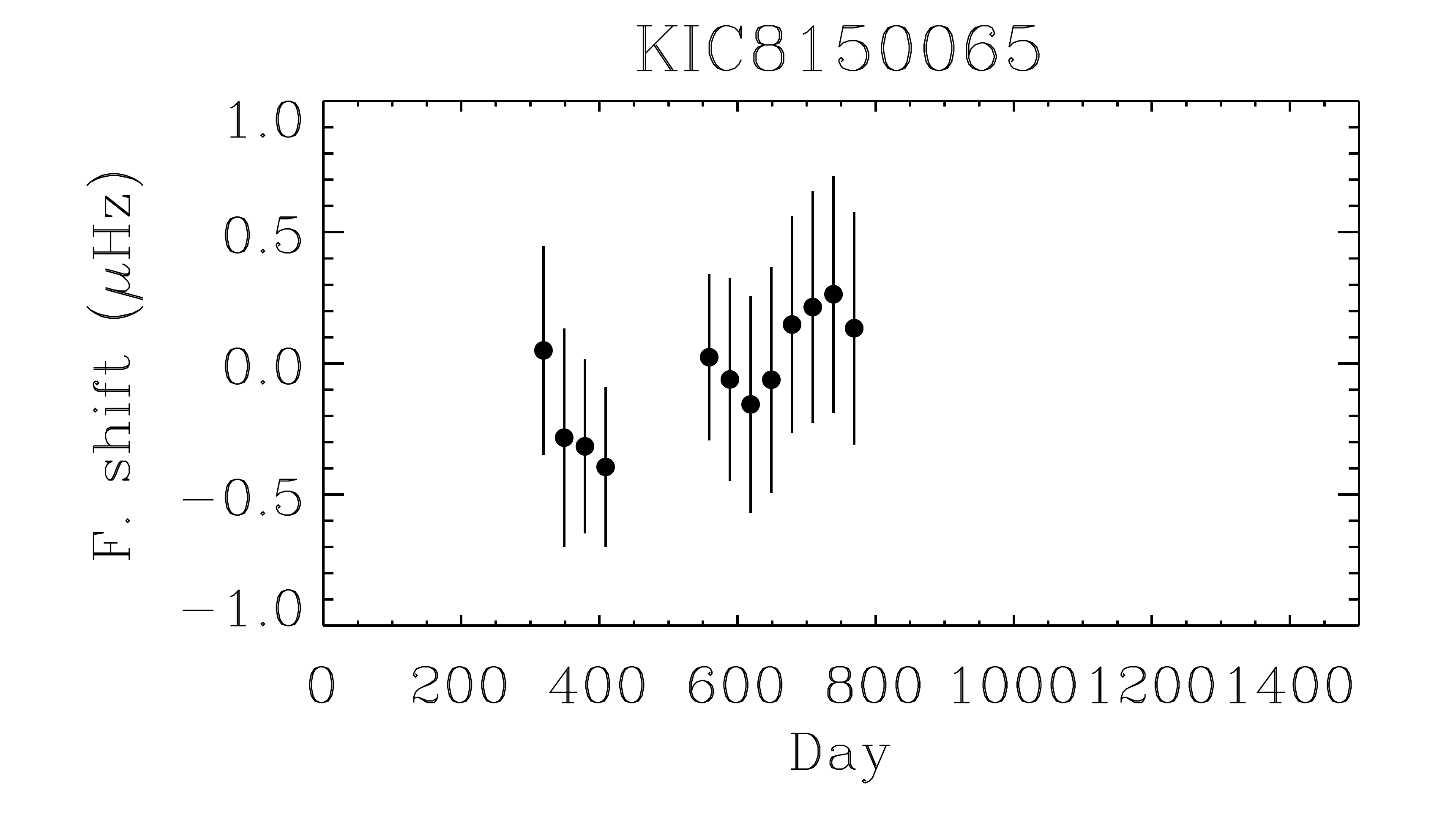}
\includegraphics[width=0.33\textwidth,angle=0]{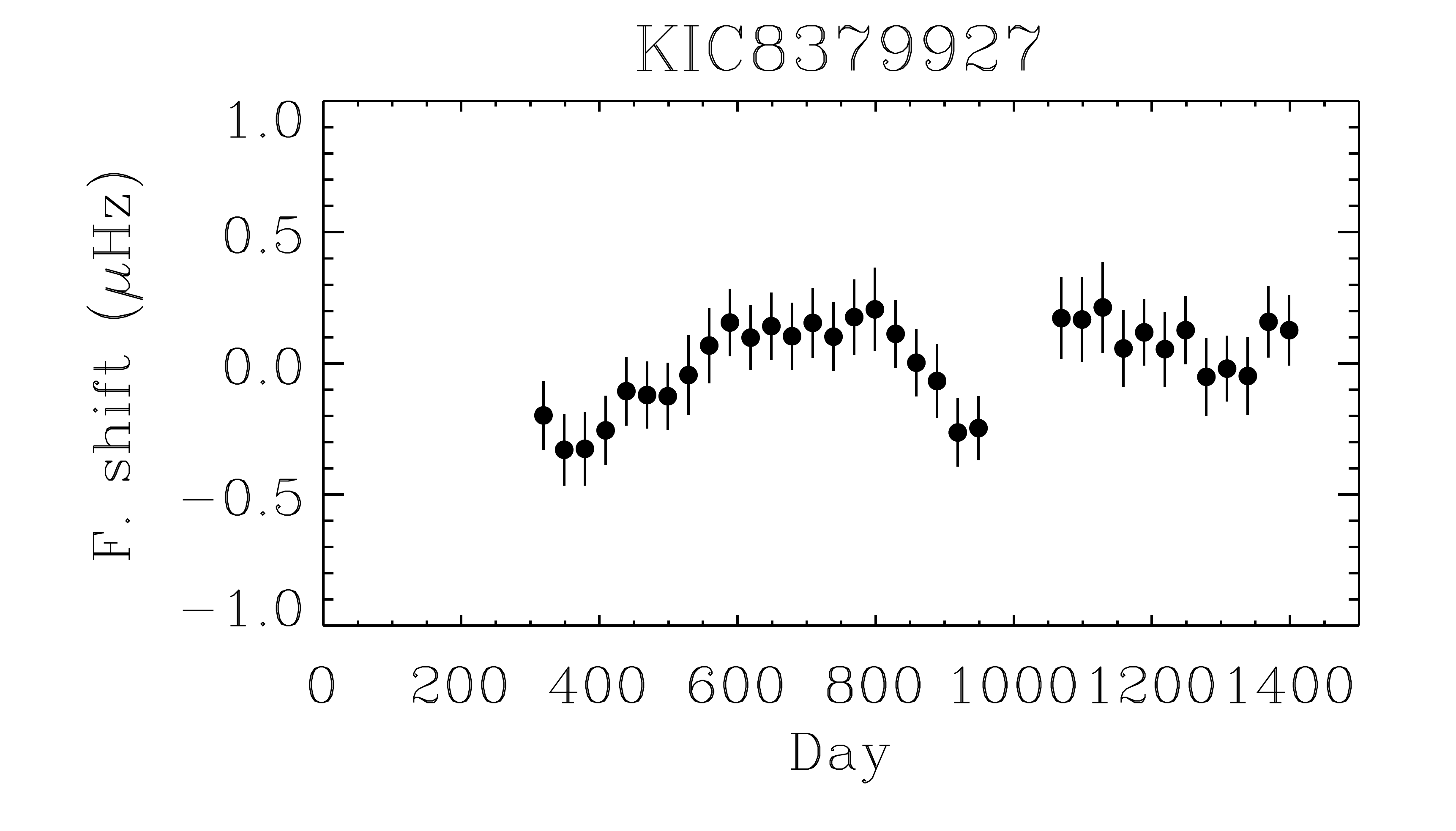}
\includegraphics[width=0.33\textwidth,angle=0]{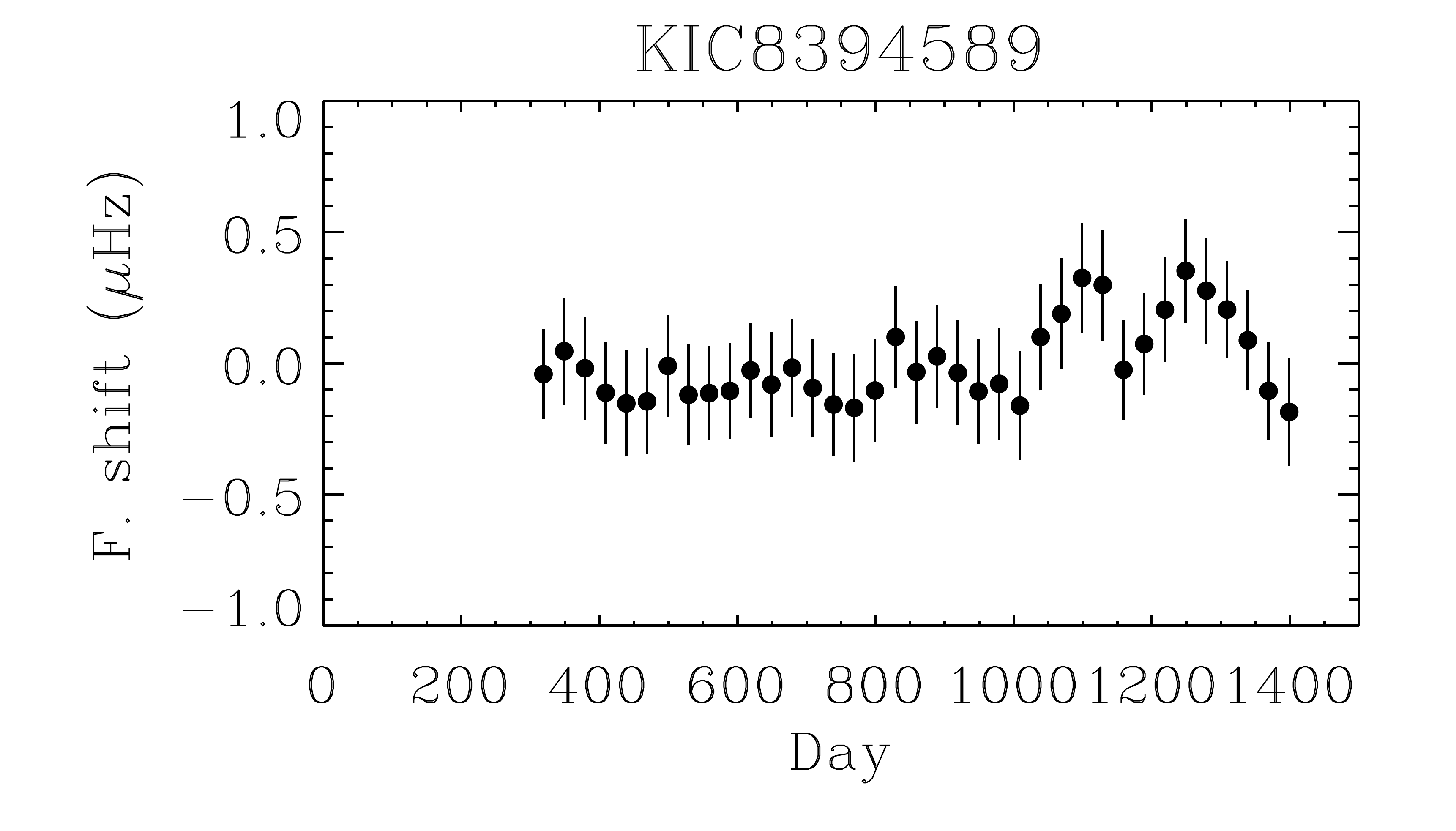}

\includegraphics[width=0.33\textwidth,angle=0]{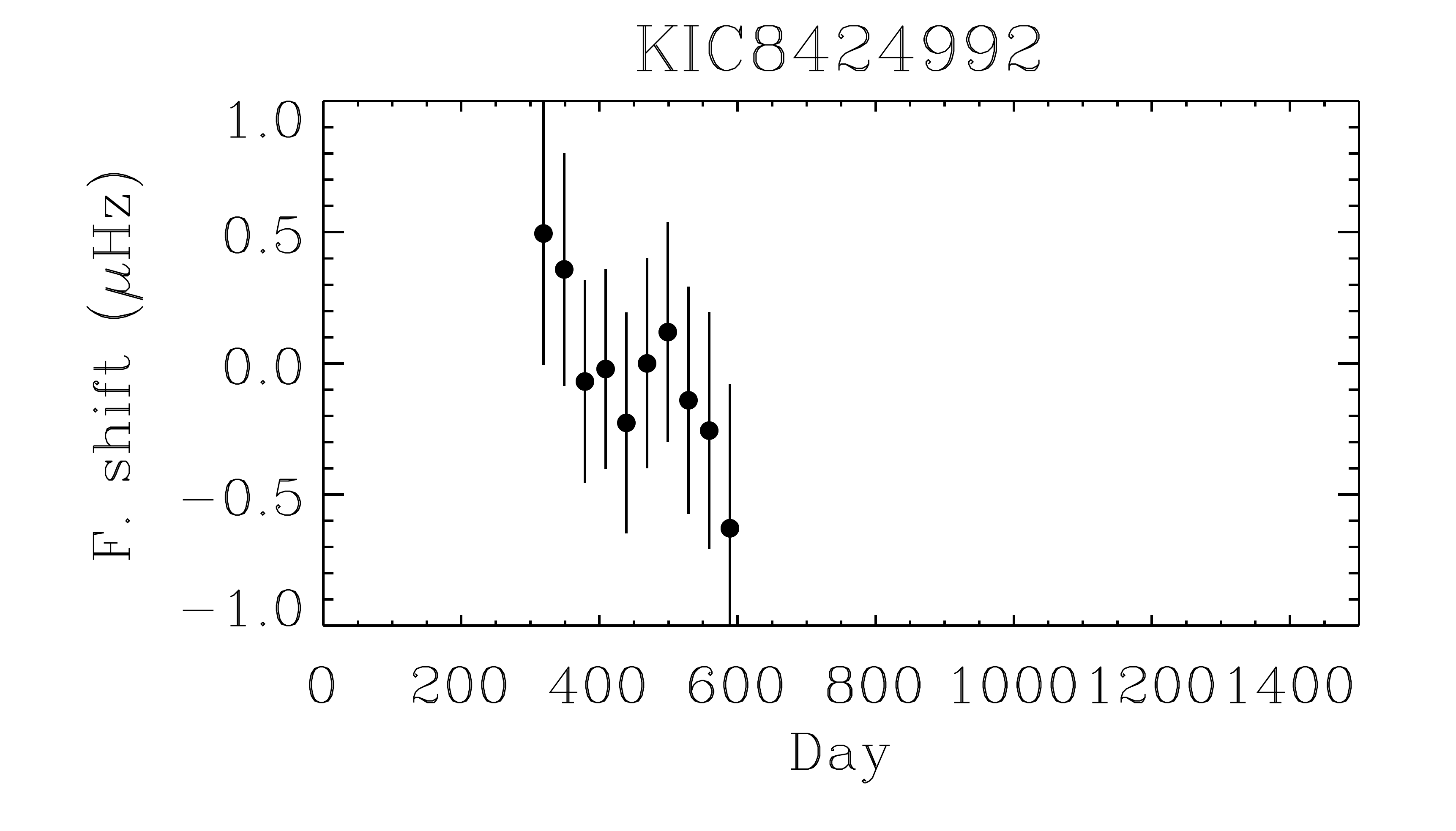}
\includegraphics[width=0.33\textwidth,angle=0]{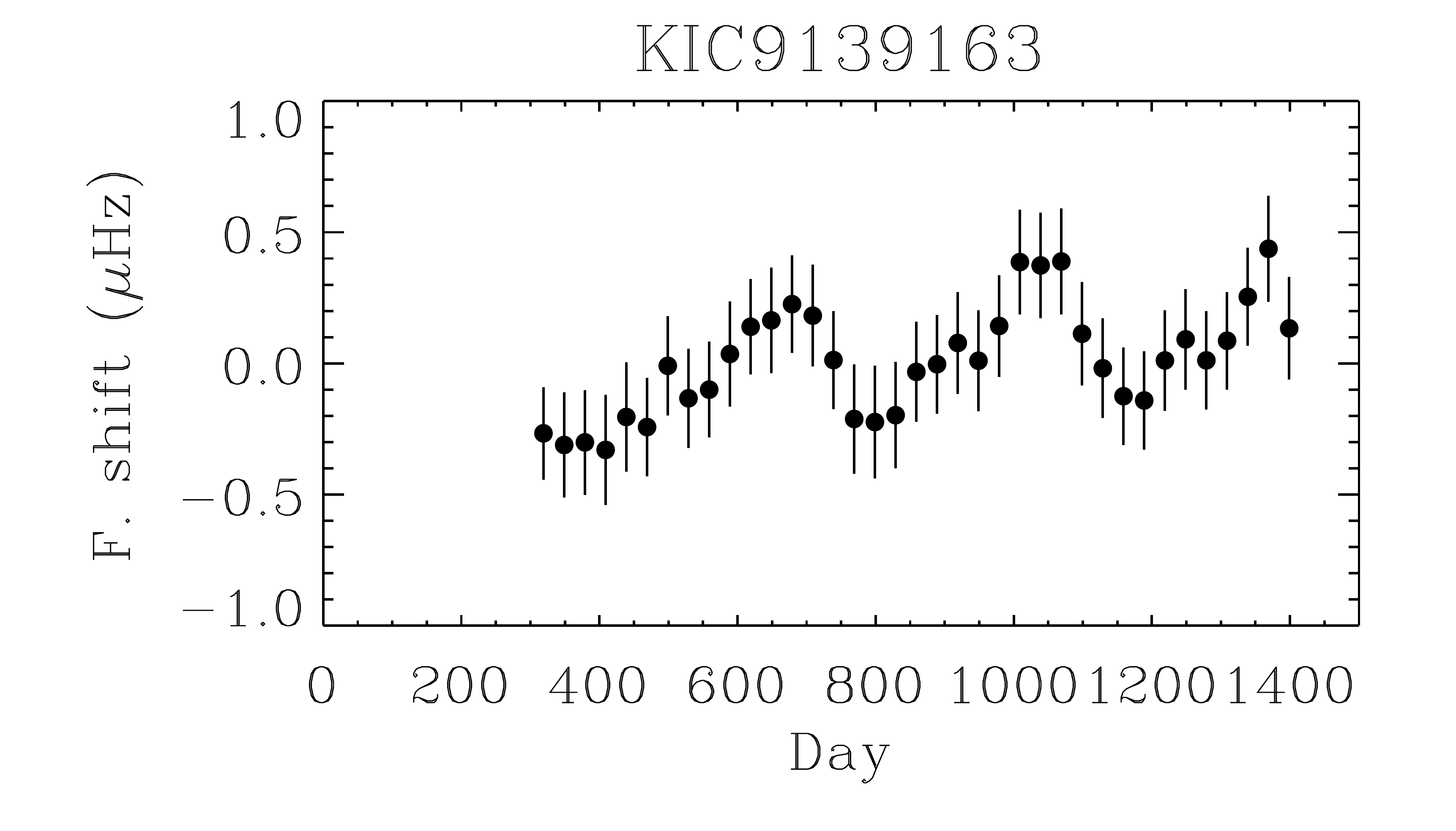}
\includegraphics[width=0.33\textwidth,angle=0]{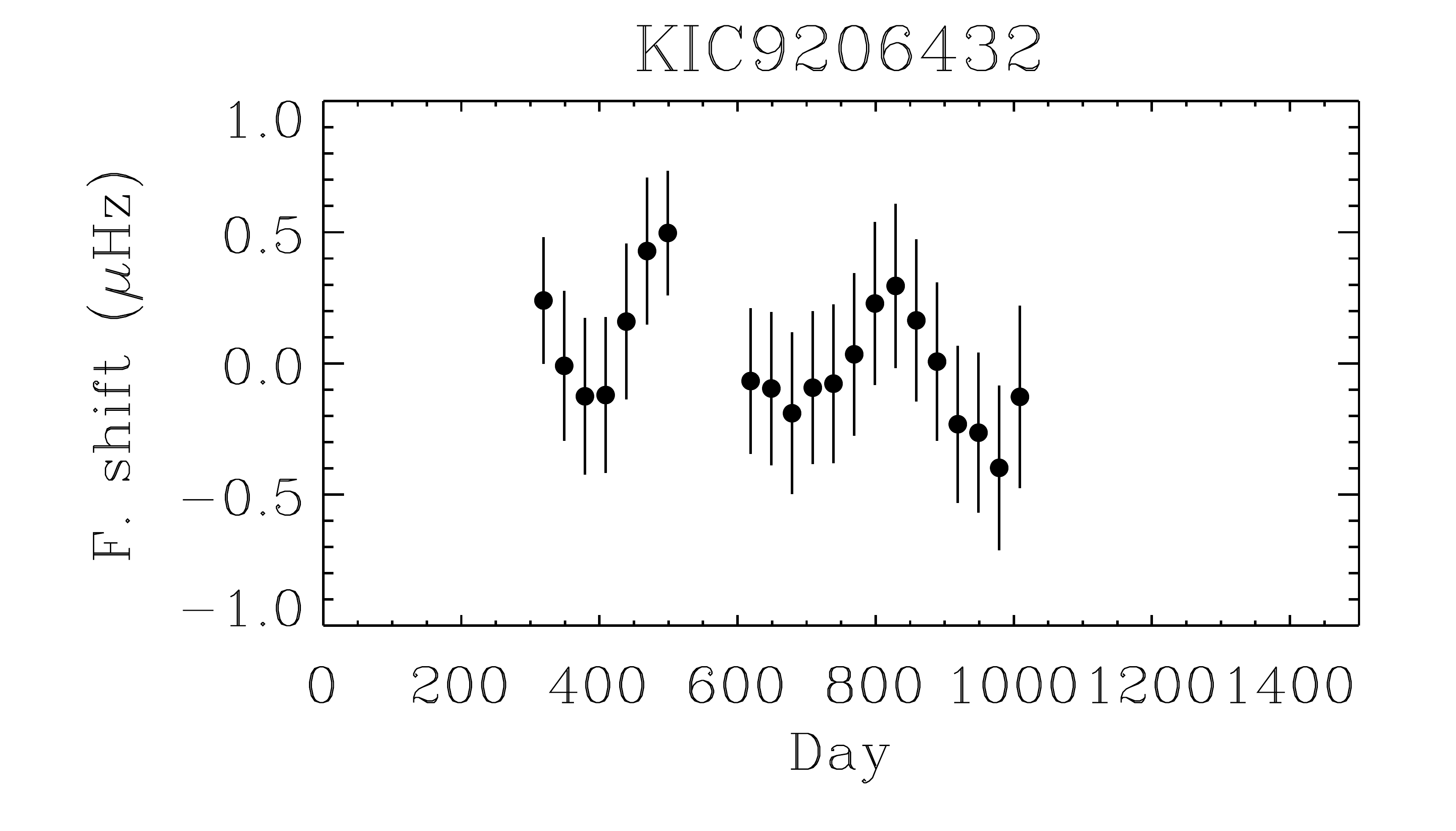}

\includegraphics[width=0.33\textwidth,angle=0]{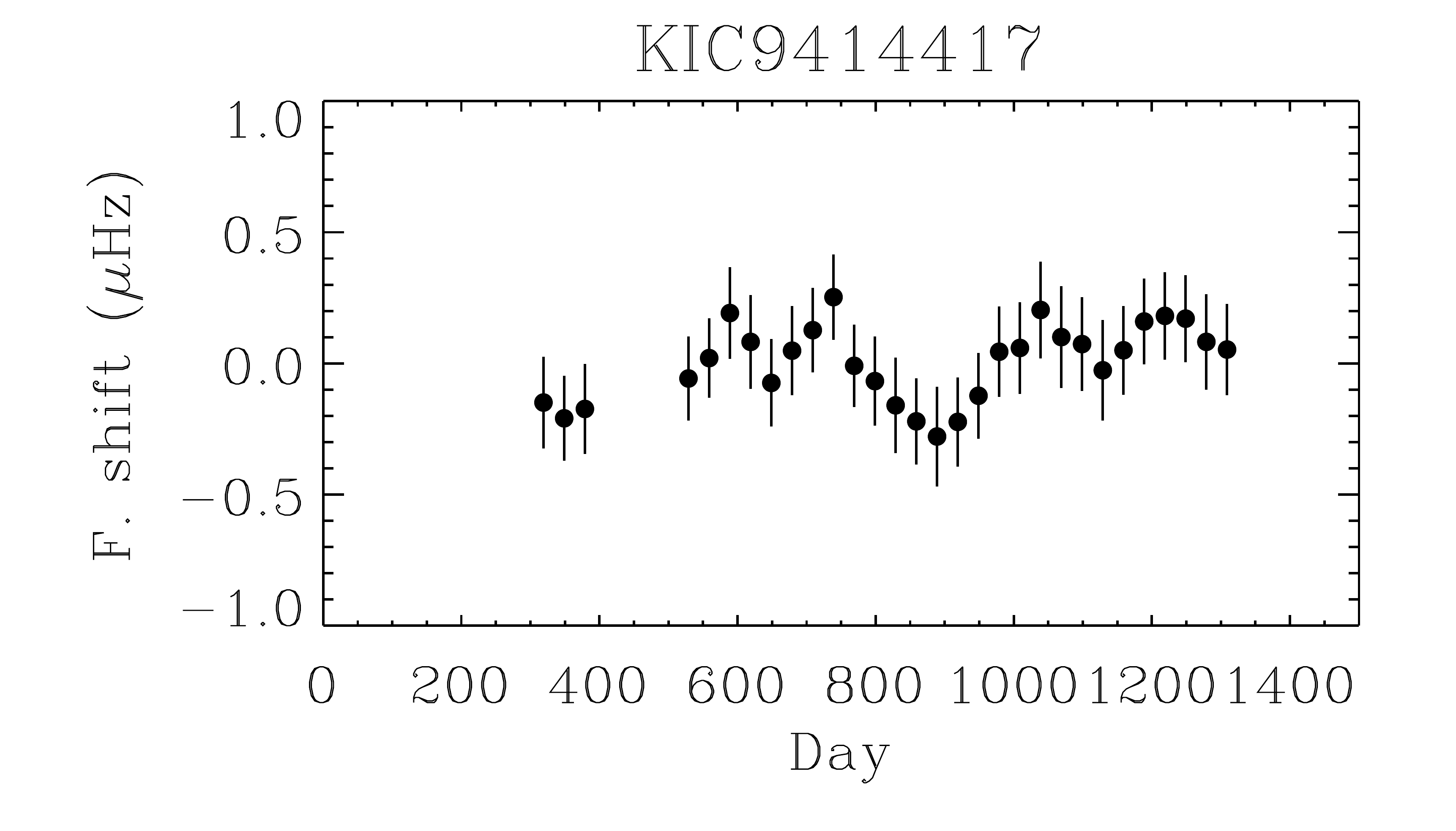}
\includegraphics[width=0.33\textwidth,angle=0]{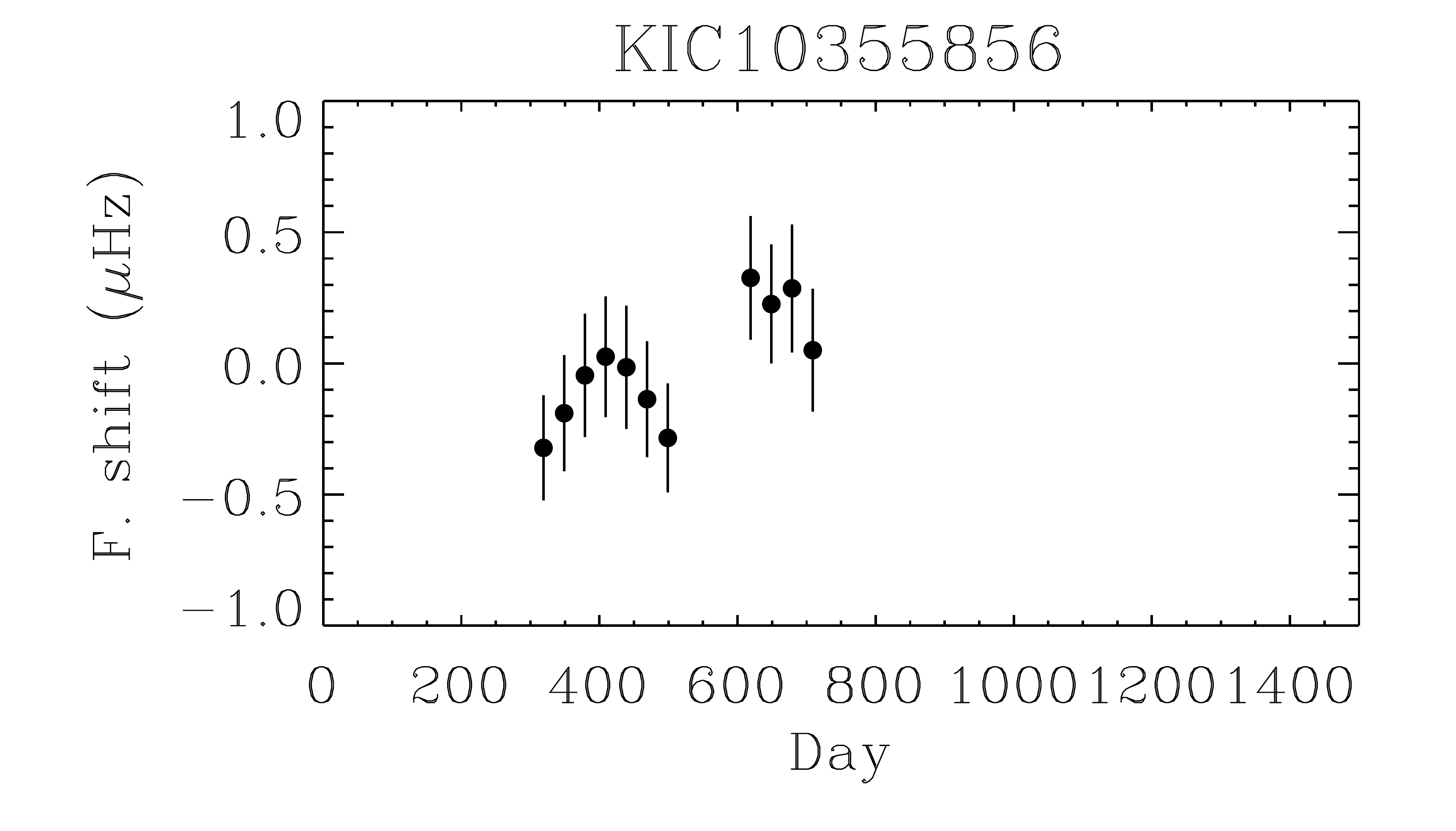}
\includegraphics[width=0.33\textwidth,angle=0]{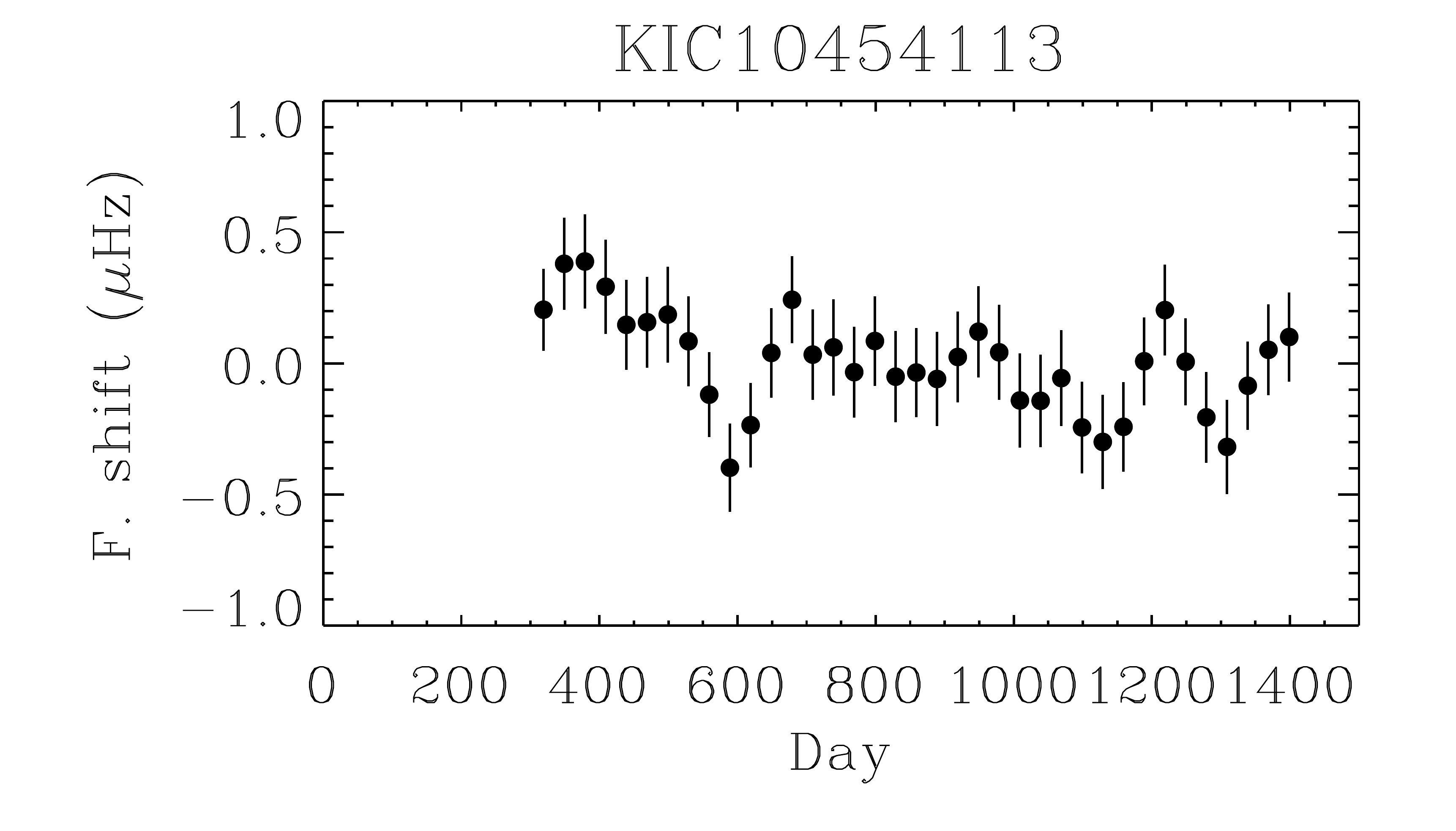}

\includegraphics[width=0.33\textwidth,angle=0]{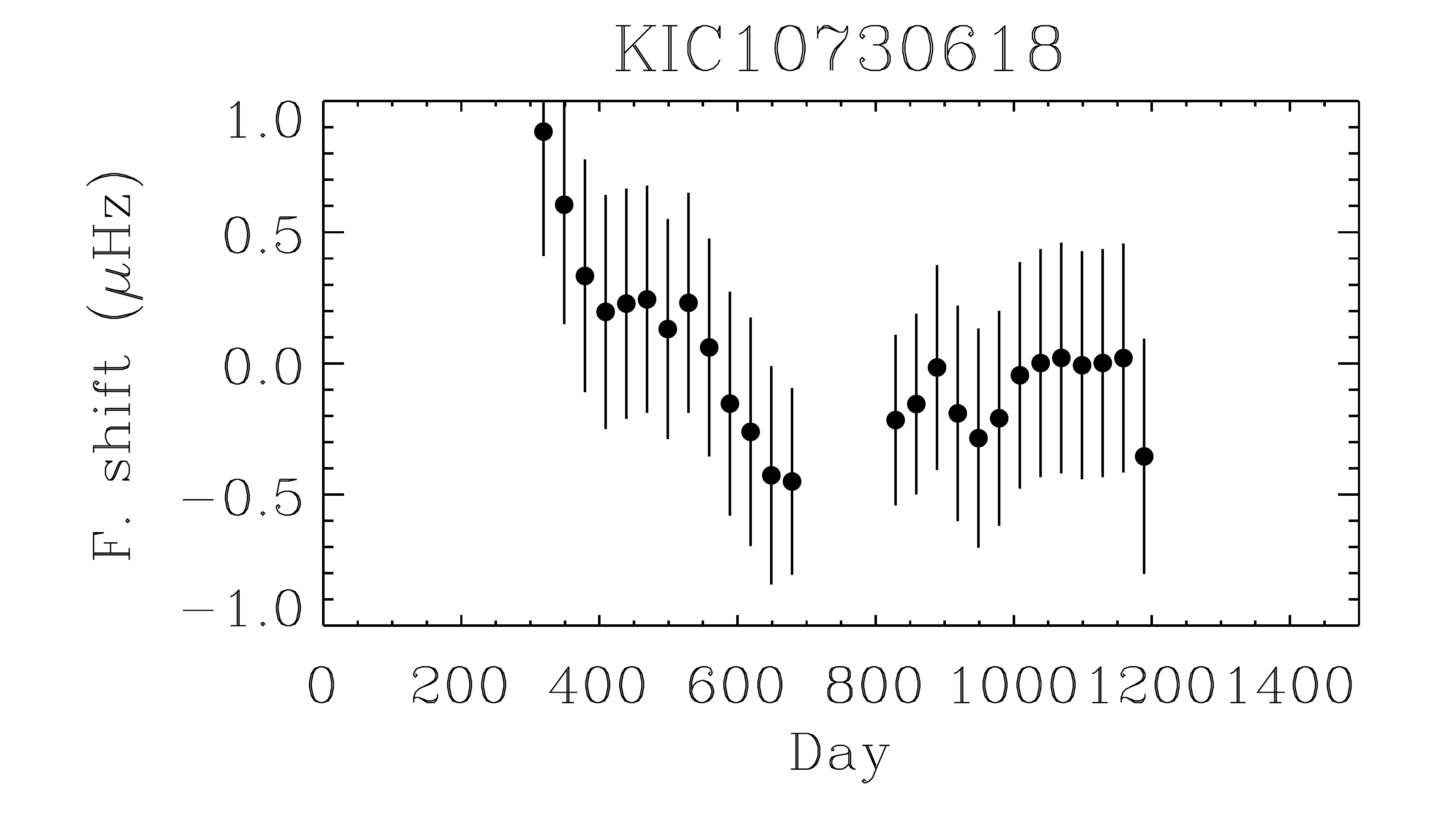}
\includegraphics[width=0.33\textwidth,angle=0]{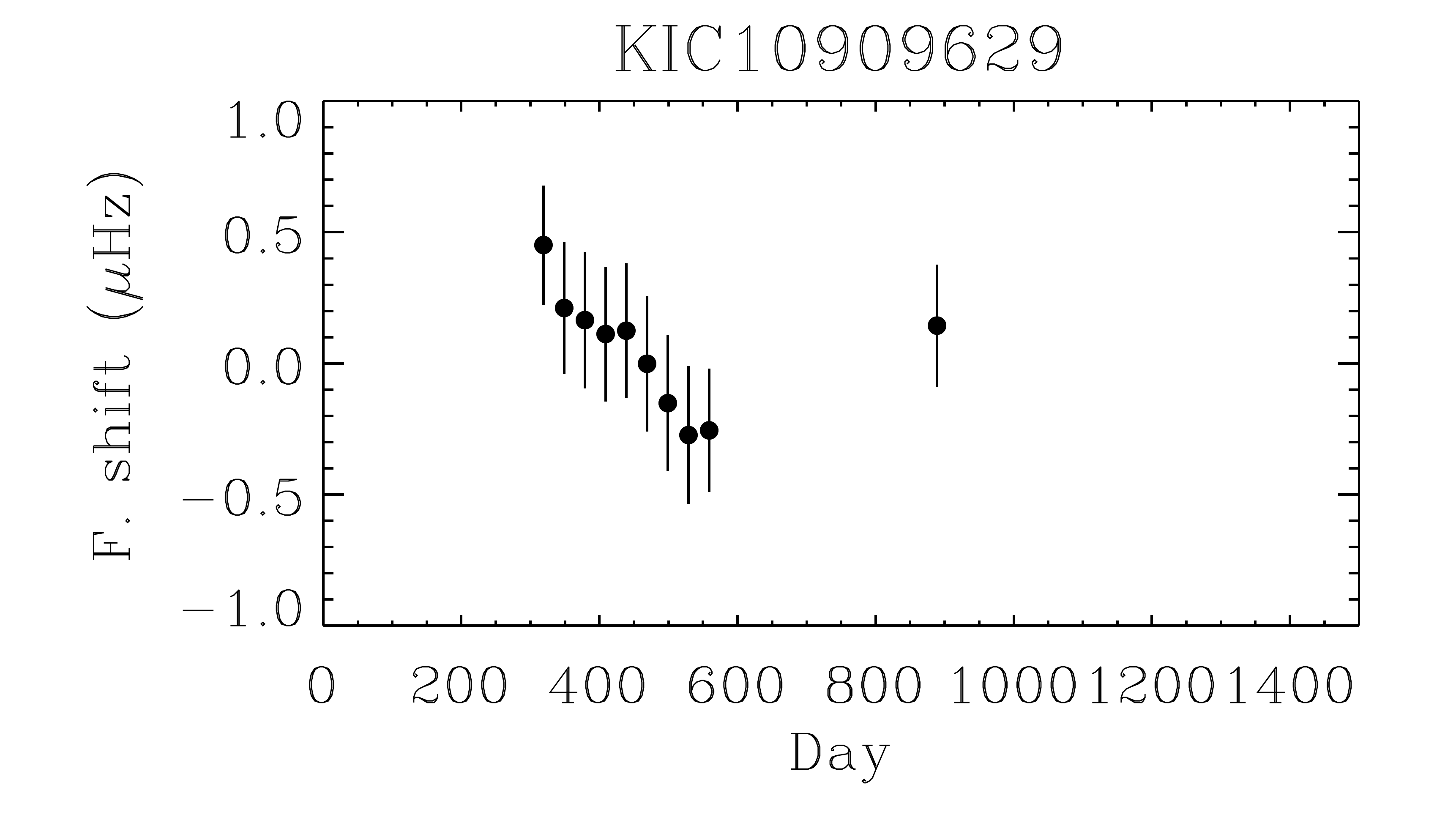}
\includegraphics[width=0.33\textwidth,angle=0]{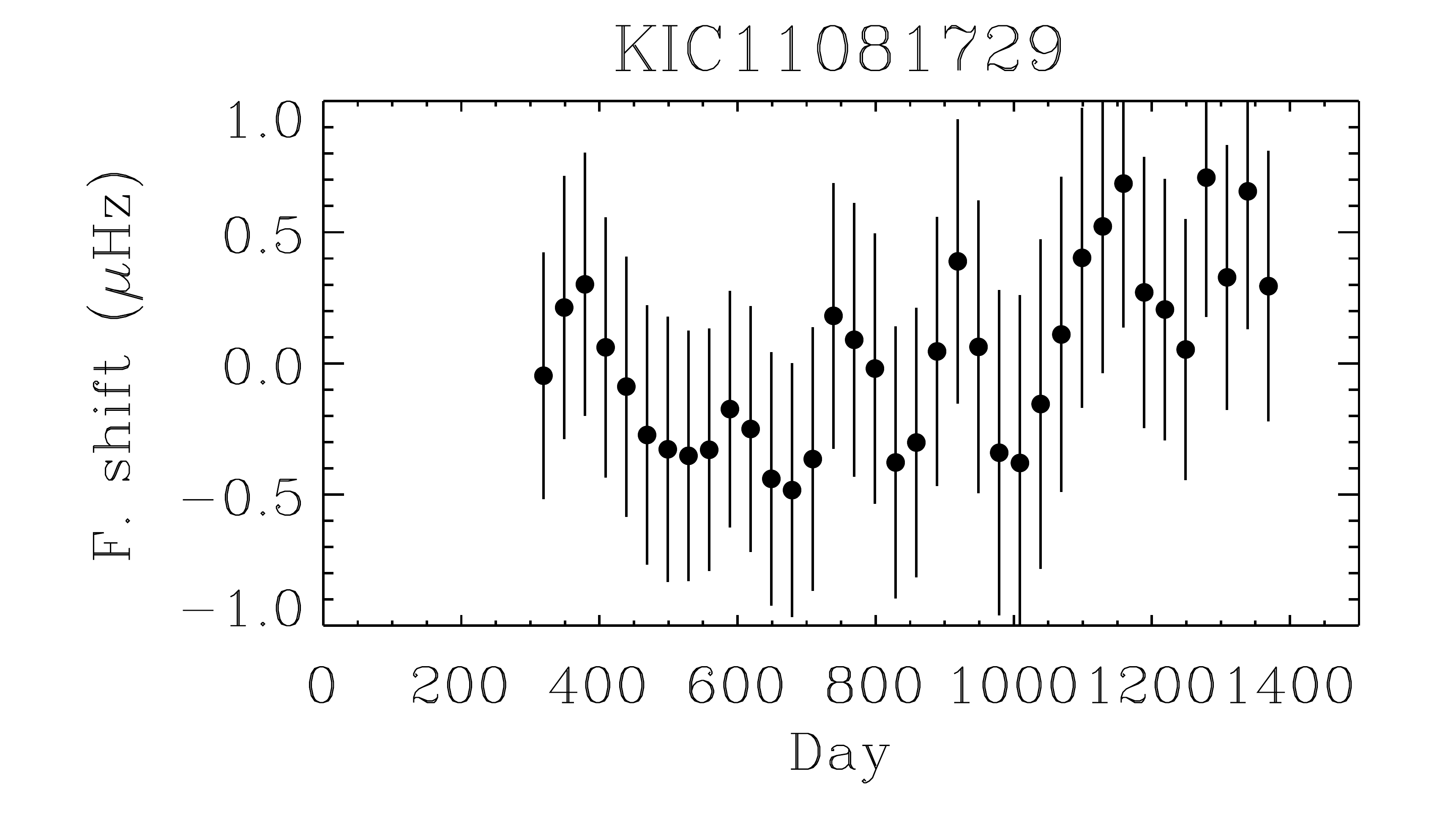}

\includegraphics[width=0.33\textwidth,angle=0]{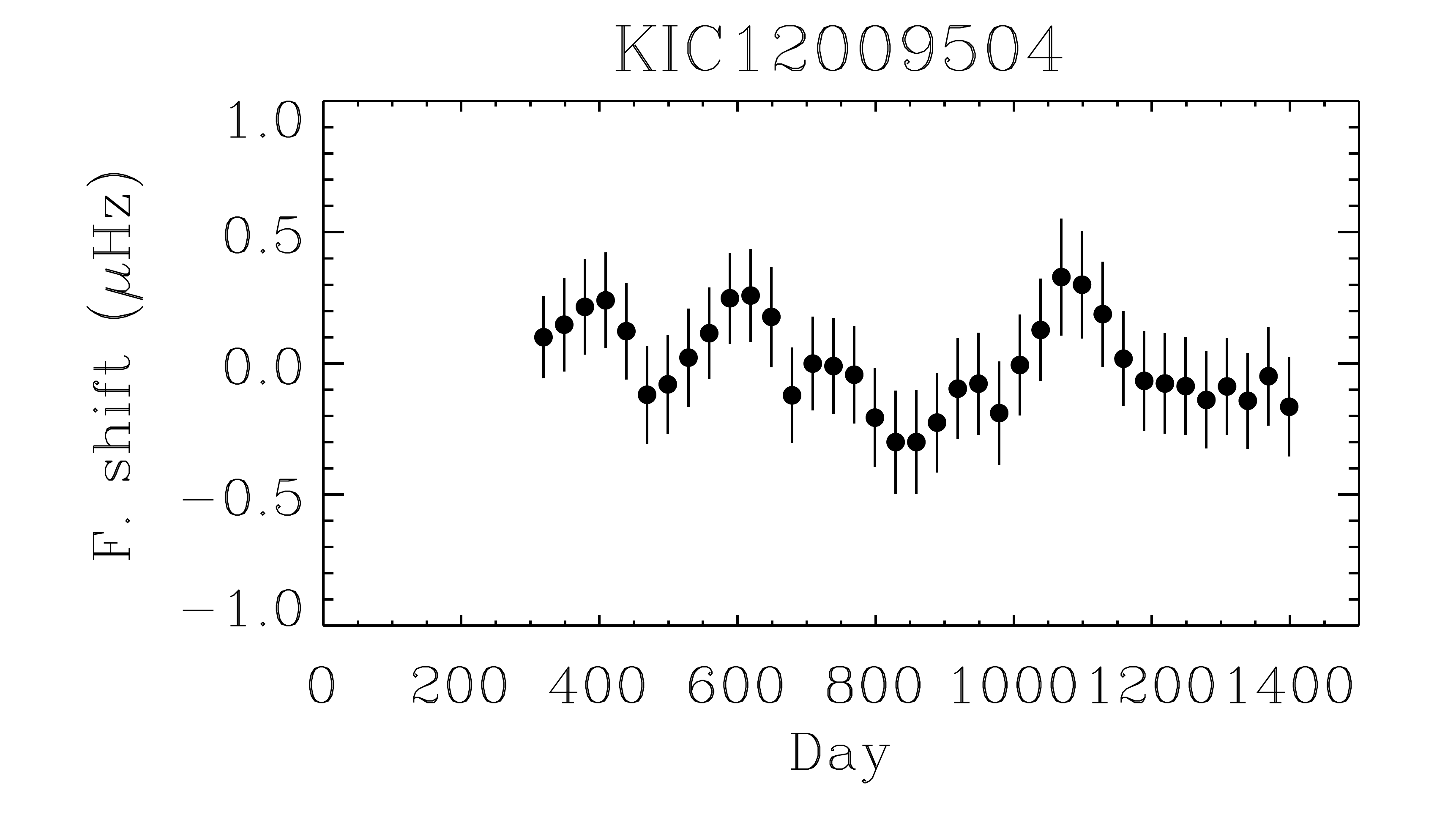}
\includegraphics[width=0.33\textwidth,angle=0]{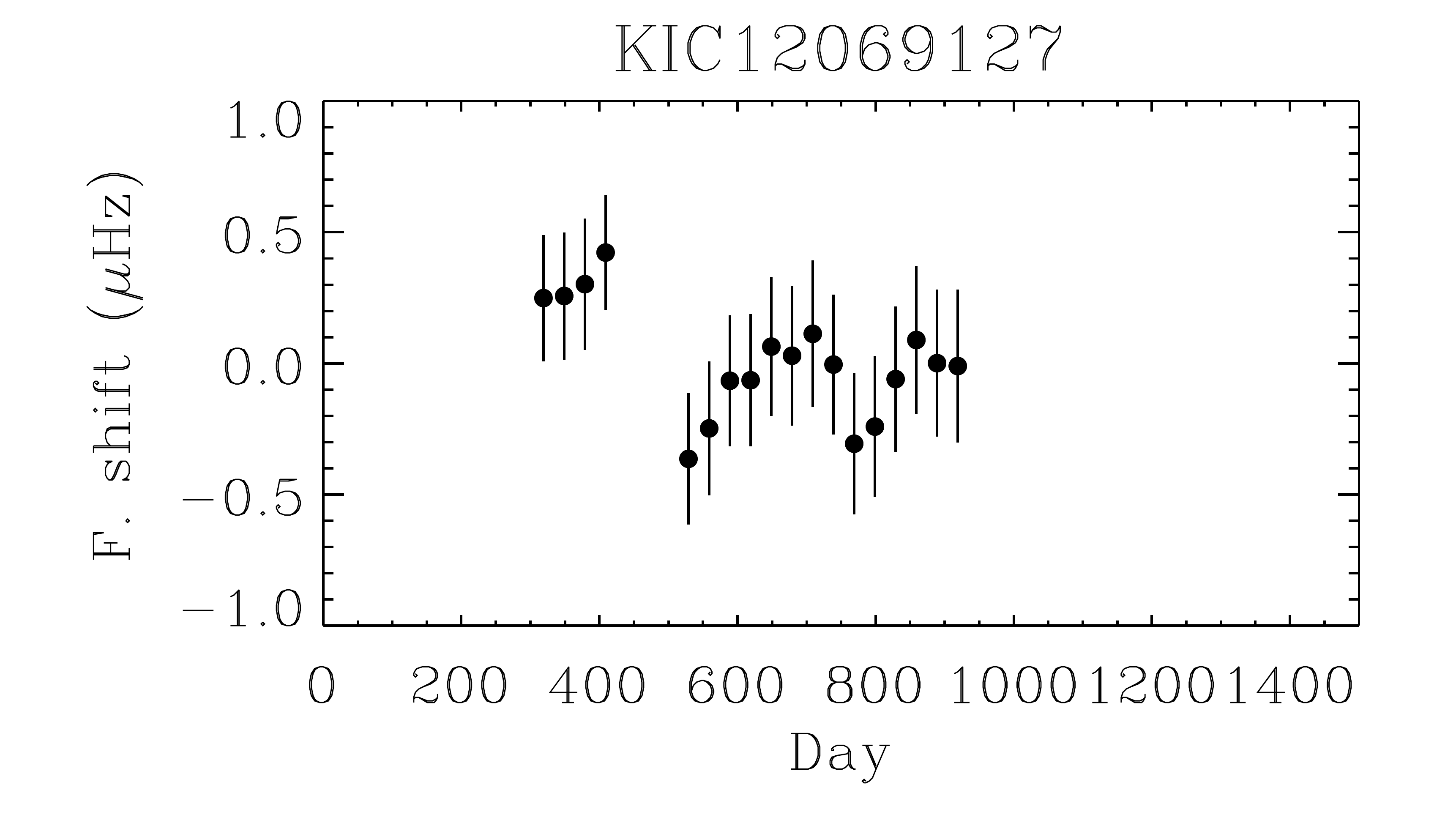}

\end{center}
\caption{\label{fig:cross}  
Temporal variability of the p-mode frequency shifts (in $\mu$Hz) extracted from the cross-correlation analysis (Method\,\#1) of the solar-like pulsating {\it Kepler} stars fulfilling the selection criterion $\lambda_1$ (see Table~\ref{table:shifts}).}
\end{figure*} 

\subsection{Selection criterion $\lambda_3$}
\label{sec:lambda3}
The $\lambda_3$ criterion checks the Spearman's correlation coefficients, $r$, between the temporal variations of the extracted frequency shifts and the photometric activity proxy $S_\textrm{ph}(t)$ estimated with the contemporaneous {\it Kepler} LC data. The $S_\textrm{ph}(t)$ proxy was estimated for the stars with a known rotation period, $P_\mathrm{rot}$ \citep{garcia14a}, over subseries of $5 \times P_\mathrm{rot}$. The criterion $\lambda_3$ helps to recover stars having photospheric magnetic variability but that did not pass the criteria $\lambda_1$  and $\lambda_2$ because their large errors on the frequency shifts due to low signal-to-noise ratio of the p-mode oscillations. The selection criterion $\lambda_3$ was fulfilled when:
\begin{equation}
r[\delta\nu(t),S_\textrm{ph}(t)] > 0.5
\end{equation}

The $S_\textrm{ph}(t)$ proxy was thus interpolated to the same dates as the ones used to measure the frequency shifts (Method\,\#1, subseries of 90 days). We note also that the coefficients $r$ were computed using independent points only.

%
%
\section{Results}
\label{sec:results}
Out of the 87 stars in the initial sample defined in Section~\ref{sec:obs}, 20 candidates (11 simple stars, 8 F-like stars, and 1 mixed-mode star; see Table~\ref{table:shifts}) fulfill the selection criterion $\lambda_1$ for showing temporal variability of their frequency shifts. These stars were included in Fig.~\ref{fig:simu} by interpolating their maximum linear frequency shift $\lvert\textrm{Max}(\delta\nu)\rvert_\textrm{L}$ and their levels of noise. Their positions on the y-axis indicate their associated percentage, $p(\lambda_1)$, of being detected according to the simulations (see Section~\ref{sec:lambda1} and Table~\ref{table:shifts}). The corresponding mean values of the temporal evolution of the $S_\text{ph}(t)$ proxy (see Section~\ref{sec:lambda3}), along the associated rotation periods, $P_\textrm{rot}$, used to compute them, are also given in Table~\ref{table:shifts}. 

Out of these 20 stars, 8 of them have a detection percentage $p(\lambda_1)$ higher than 50$\%$. With the application of the selection criterion $\lambda_2$ (see Section~\ref{sec:lambda2}), 4 stars are selected. These stars have a detection percentage $p(\lambda_2)$ above 50\% and an associated detection percentage of false positives lower than 10\%, given their mean errors on the extracted frequency shifts $\overline{\sigma_{\delta\nu}}$ being smaller than  0.2 $\mu$Hz (Table~\ref{table:simu2} and Table 4). These 4 stars were also selected with the criterion $\lambda_1$.

Figure~\ref{fig:cross} shows the frequency shifts extracted from the cross-correlation analysis (Method\,\#1) of the 20 stars fulfilling the selection criterion $\lambda_1$ (Table~\ref{table:shifts}). Comparable results are obtained with the peak-fitting analysis (Method\,\#2). The temporal variations and associated uncertainties are quite different from one star to the other, but all the shifts are within $\pm 1\mu$Hz over the duration of the {\it Kepler} mission. Some stars have very small errors of the shifts, such as KIC\,5184731, KIC\,6116048, and KIC\,7747078. The stars KIC\,7771282, KIC\,8150065, KIC\,8424992, KIC\,10355856, KIC\,10909629, and KIC\,12069127, which were observed for less than three years show nonetheless signatures of frequency shifts. The star KIC\,9139163, flagged in Table~\ref{table:shifts}, has an interesting and  well defined periodicity of the frequency shifts of about 385\,days, which is surprisingly close to the {\it Kepler} orbital period. To evaluate the origin of such coinciding periodicity, any possible effects of the line-of-sight \citep{davies14} and Earth-trailing orbit in the {\it Kepler} observations on the p-mode frequencies need to be investigated in more details. 

The p-mode oscillations of F-like stars with short lifetimes result in broad linewidths in the spectral domain. The errors on the extracted frequency shifts of these stars are thus much larger than for the simple stars whose oscillations have longer lifetimes. Uncovering significant frequency shifts larger than their errors is then more difficult to achieve for F-like stars, although their shifts are expected to be larger than for the simple stars \citep{metcalfe07}. In consequence, except KIC\,10355856, the 8 F-like stars selected with the criterion $\lambda_1$ have detection percentages lower than 50\%. One evolved star with mixed modes, KIC\,7747078, also shows evidence for frequency shifts. This particular case is discussed in the following Section~\ref{sec:mixed}.

Furthermore, we note that two stars, the F-like star KIC\,3733735 \citep{regulo16} and the simple star KIC\,10644253 \citep{salabert16}, known to show changes in the acoustic frequencies correlated to the photospheric magnetic activity measured with the $S_\textrm{ph}$ proxy were not selected with the criteria $\lambda_1$ and $\lambda_2$. Indeed, the errors on the extracted frequency shifts of these stars are large due to the low signal-to-noise ratio of their p-mode oscillations. 
The combination of the two criteria $\lambda_1$ and $\lambda_2$ is very restrictive allowing us to select only the best candidates among a large sample of stars to perform a detailed study of their frequency shifts. The stars thus selected do not constitute a full set of the active solar-like stars observed by {\it Kepler} but the best candidates which fulfill these objective criteria. An exhaustive analysis of the temporal variations of the frequency shifts of all the {\it Kepler} solar-like pulsating stars will be presented in Santos et al. (in preparation).

\begin{figure}[tbp]
\begin{center} 
\includegraphics[width=0.49\textwidth,angle=0]{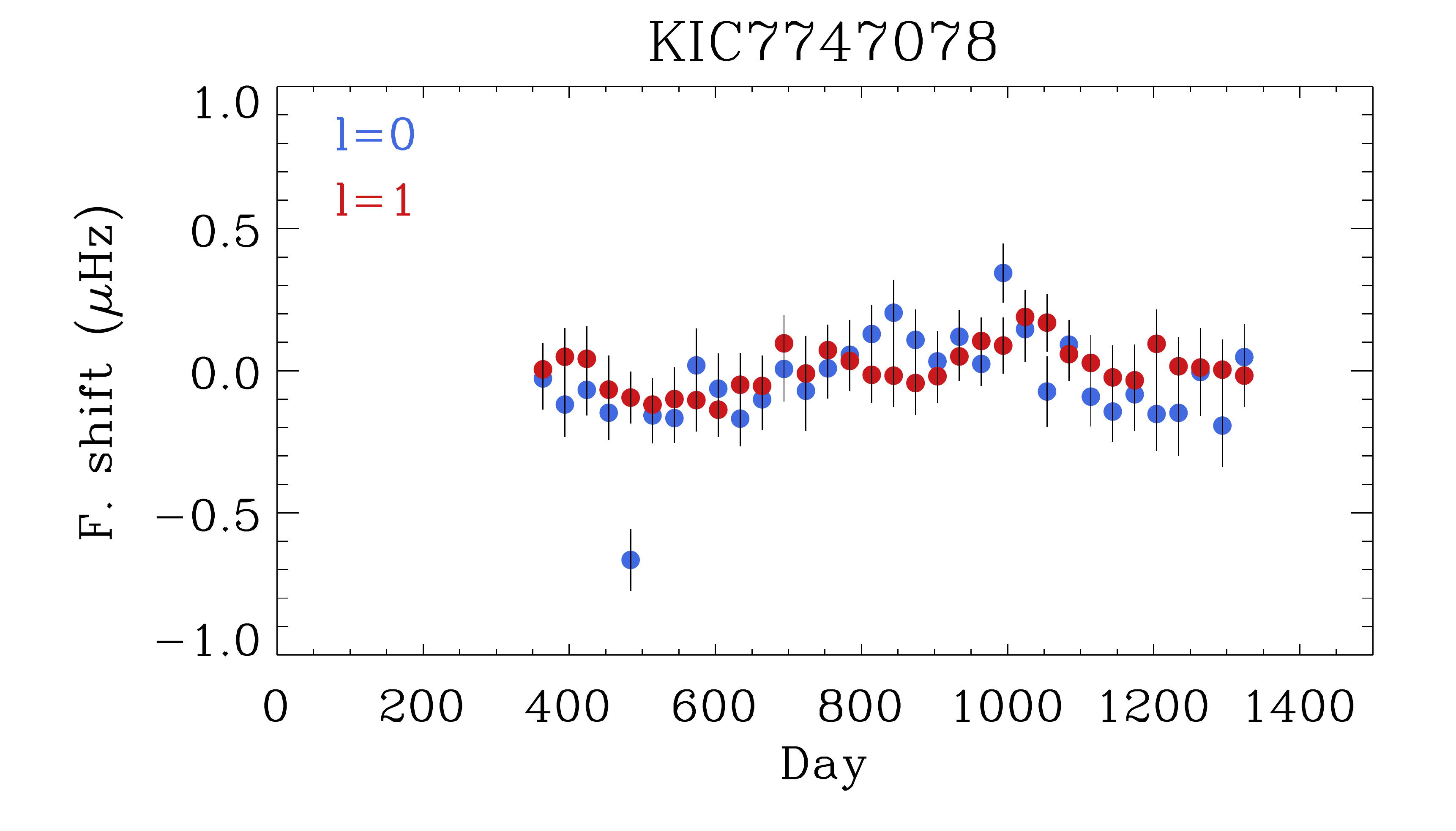}
\end{center}
\caption{Temporal variability of the frequency shifts (in $\mu$Hz) of the individual $l=0$ (blue) and $l=1$ (red) modes measured for the evolved mixed-mode star KIC\,7747078 extracted from the peak-fitting analysis (Method\,\#2).
 \label{fig:mixed}}
\end{figure} 

\begin{figure*}[tbp]
\begin{center} 
\includegraphics[width=0.49\textwidth,angle=0]{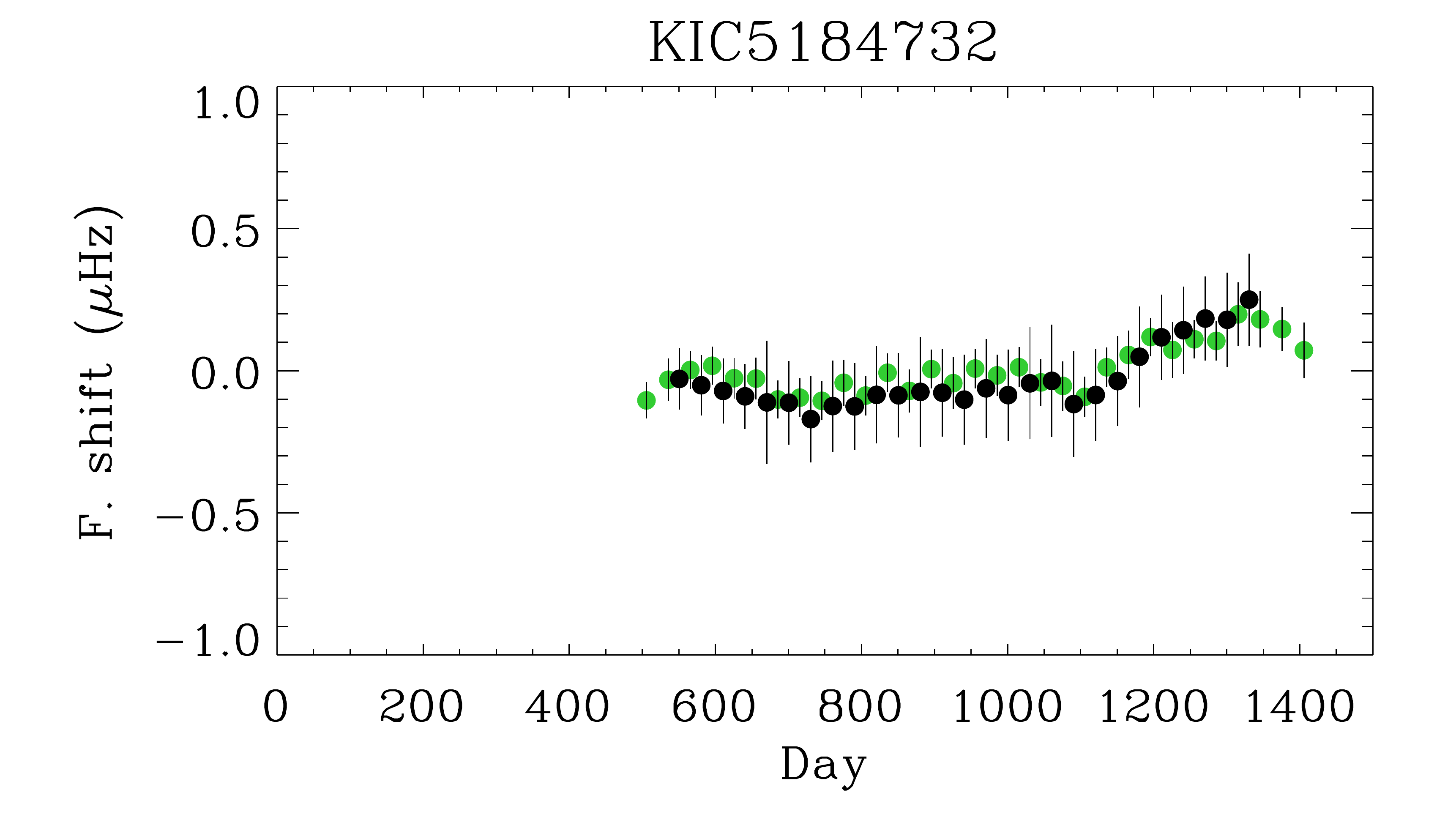}
\includegraphics[width=0.49\textwidth,angle=0]{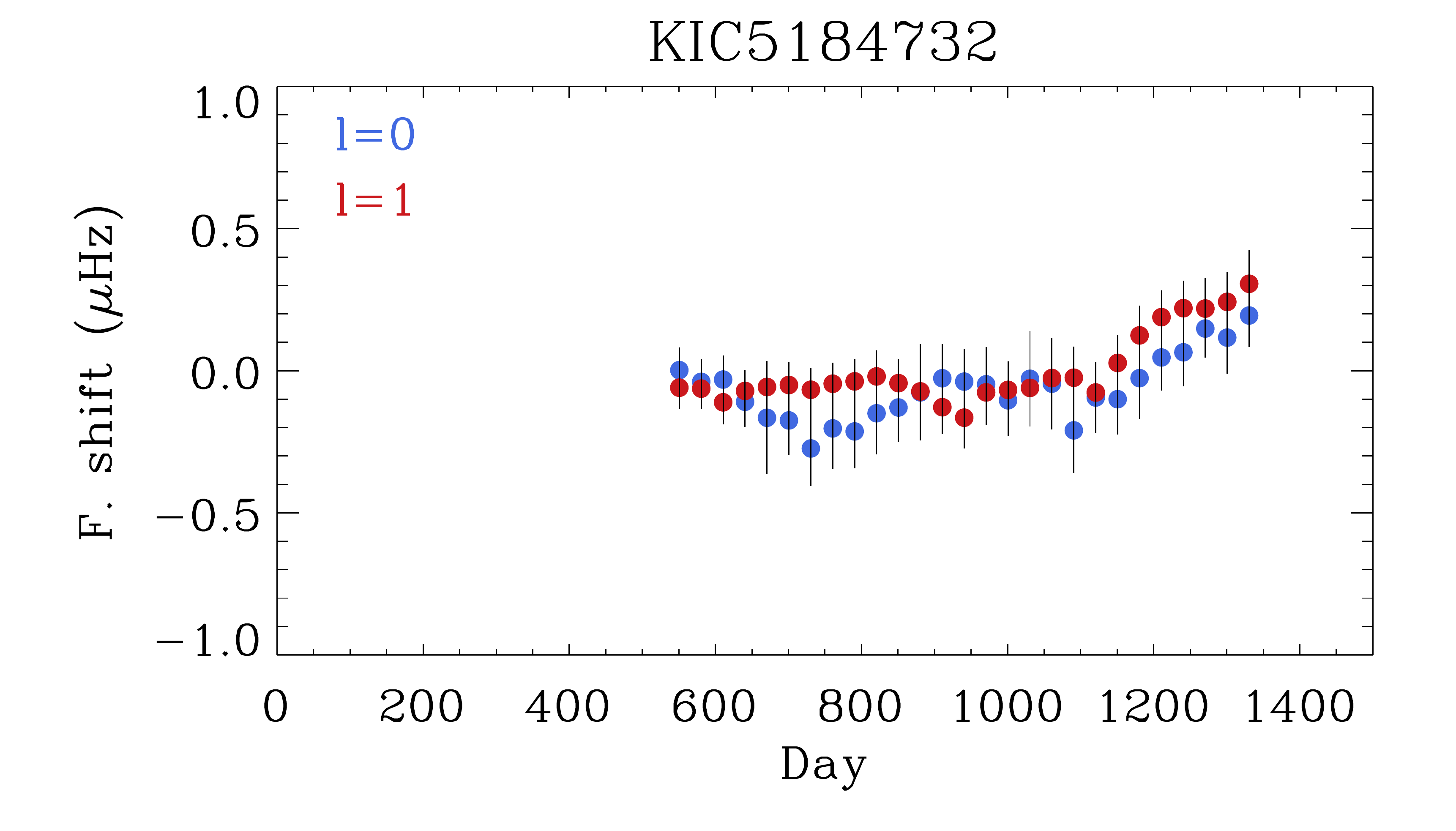}
\includegraphics[width=0.49\textwidth,angle=0]{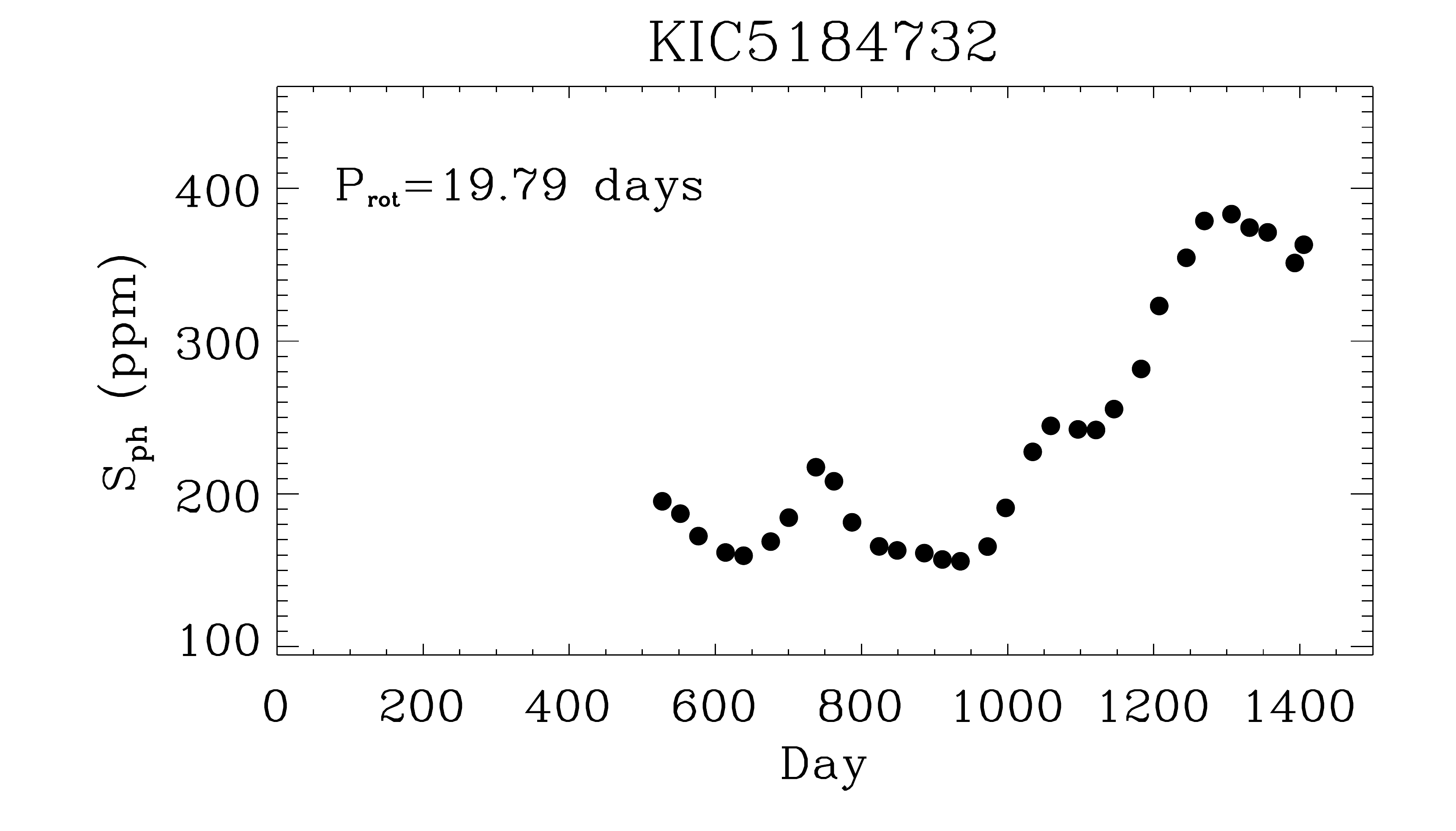} 
\includegraphics[width=0.49\textwidth,angle=0]{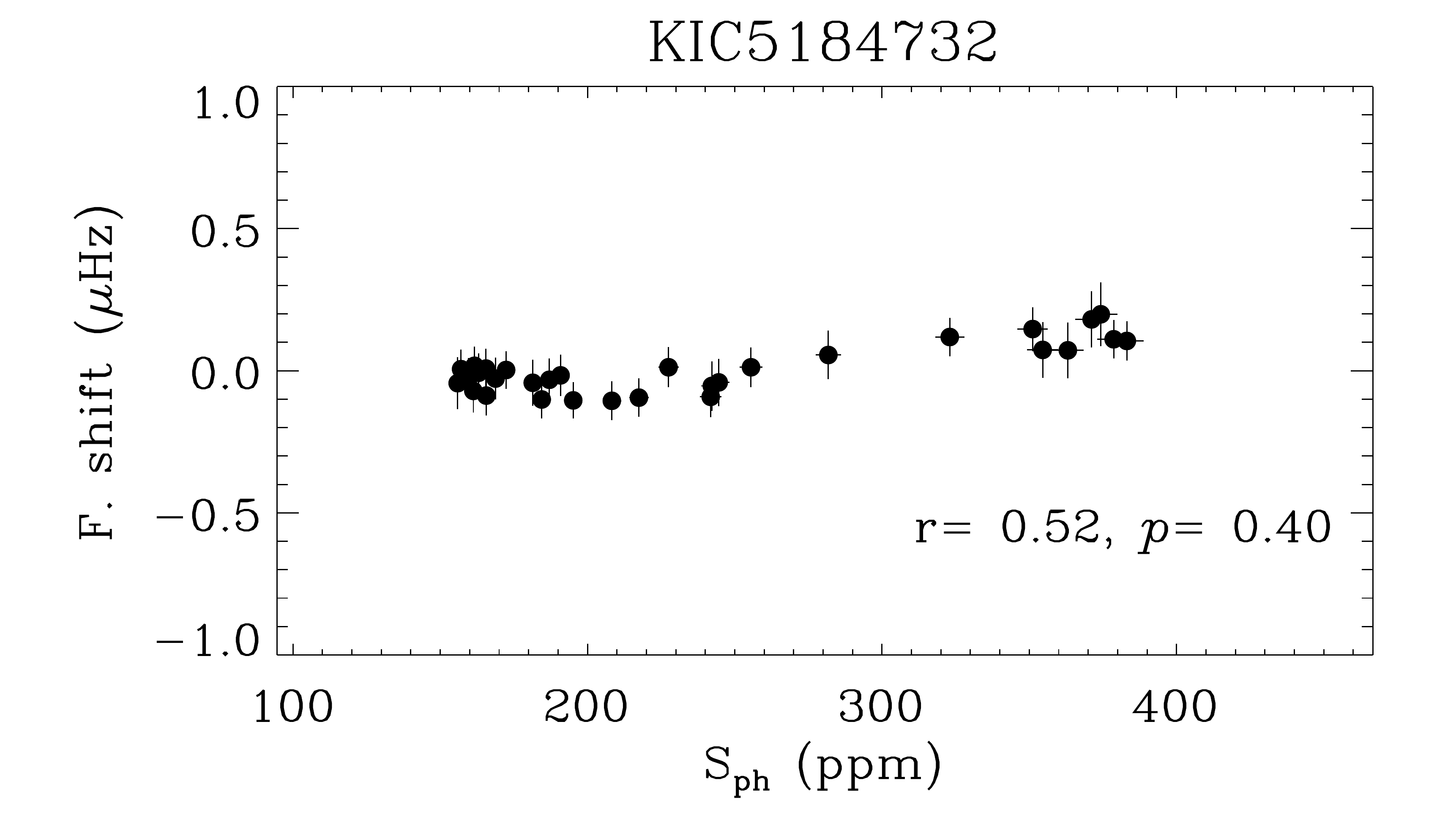}
\end{center}
\caption{\label{fig:kic51} {\it Upper-left panel}: Comparison of the temporal variability of the frequency shifts (in $\mu$Hz) of KIC\,5184732 extracted from the cross-correlation (Method\,\#1, in green) and peak-fitting (Method\,\#2, in black) analyses. {\it Upper-right panel}: Frequency shifts of the individual $l=0$ (blue) and $l=1$ (red) modes from Method\,\#2. {\it Bottom-left panel}: Temporal variability of the photospheric activity proxy $S_\textrm{ph}$ (in ppm). {\it Bottom-right panel}: Frequency shifts extracted from Method\,\#1 as a function of the photospheric magnetic proxy $S_\textrm{ph}$.  The corresponding correlation coefficient $r$ and two-sided significance of the deviation from zero $p$ are also indicated.}
\end{figure*} 

\begin{figure*}[tbp]
\begin{center} 
\includegraphics[width=0.49\textwidth,angle=0]{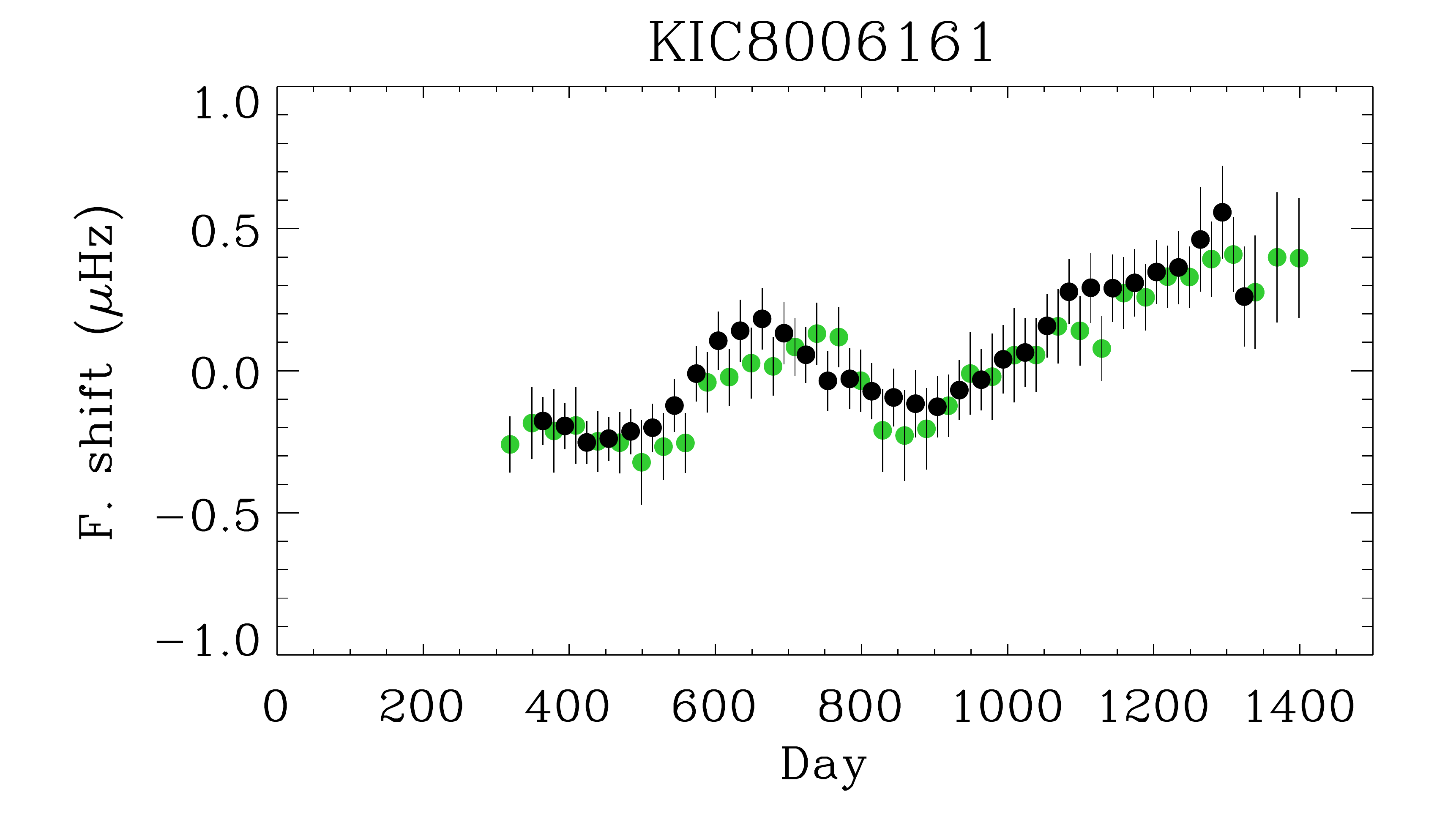}
\includegraphics[width=0.49\textwidth,angle=0]{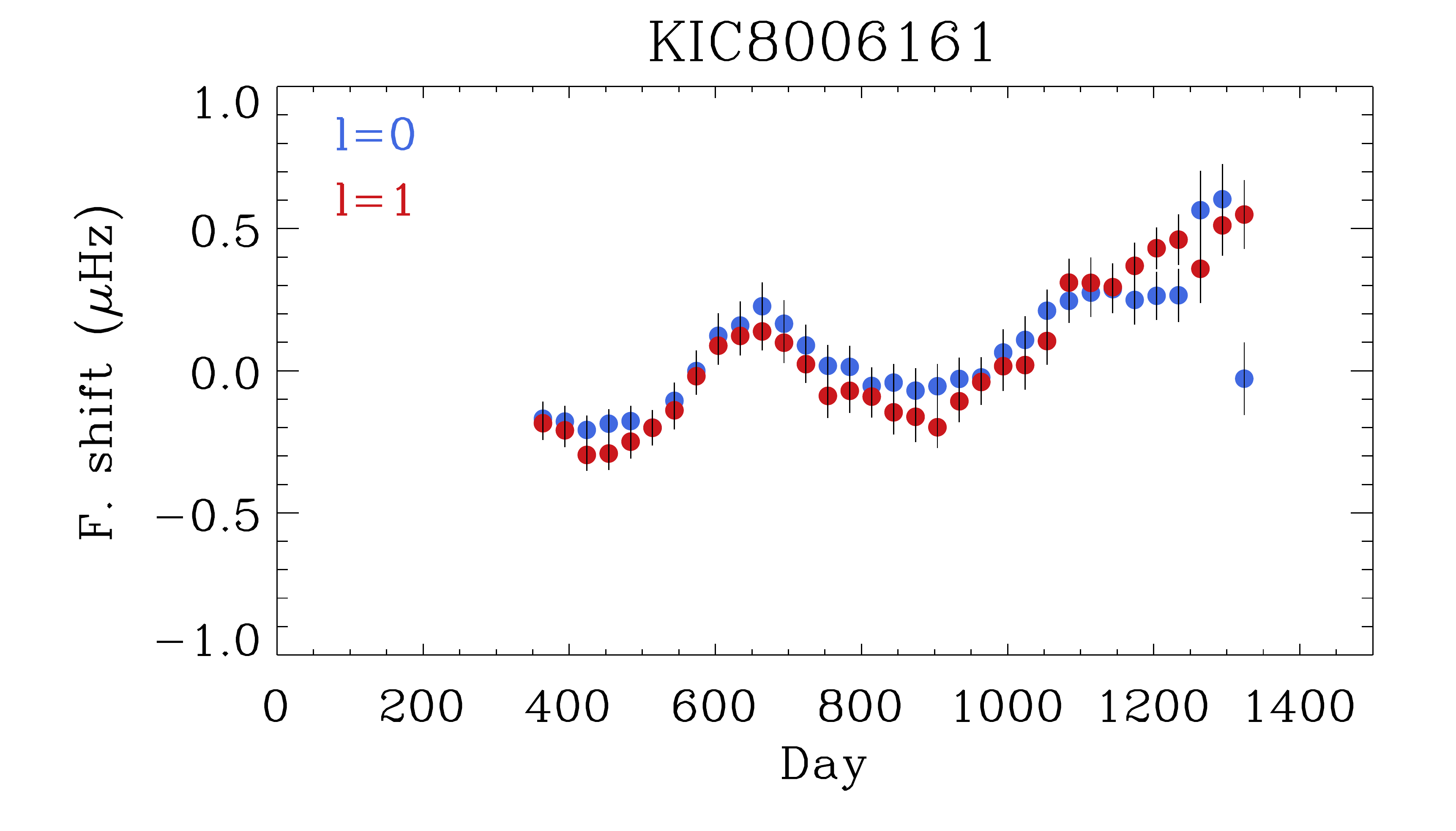}
\includegraphics[width=0.49\textwidth,angle=0]{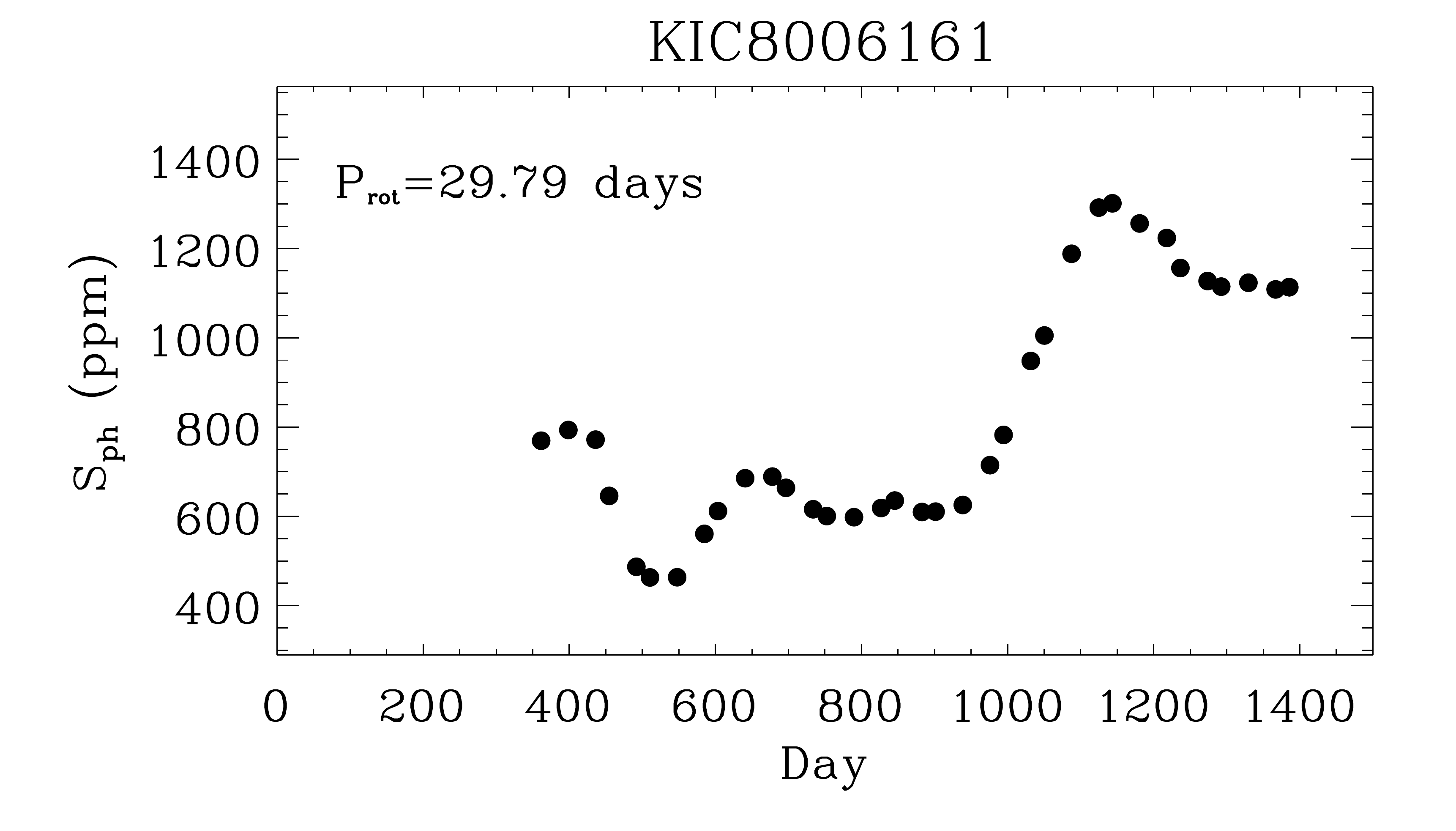}
\includegraphics[width=0.49\textwidth,angle=0]{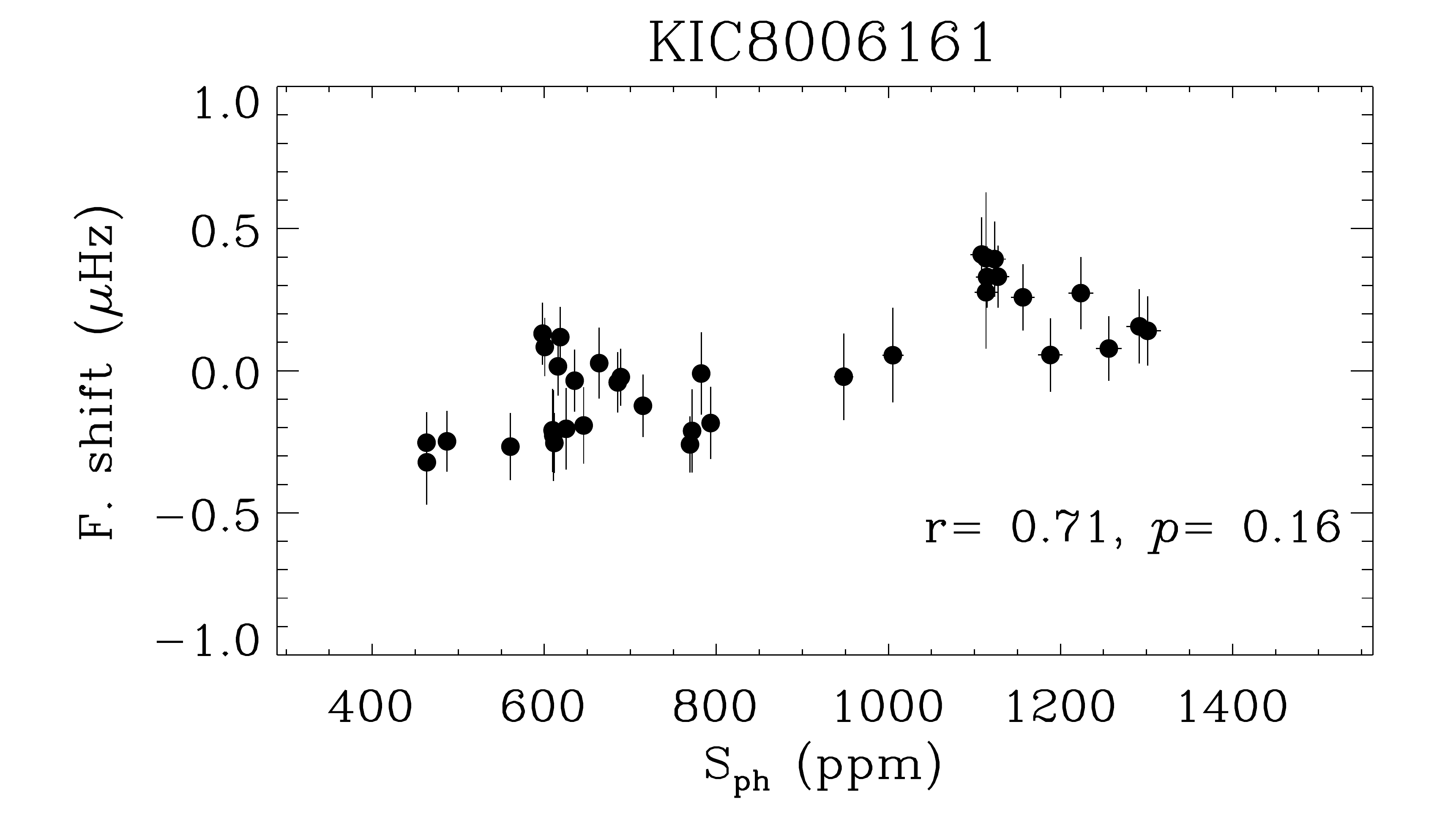}
\end{center}
\caption{\label{fig:kic80} {\it Upper-left panel}: Comparison of the temporal variability of the frequency shifts (in $\mu$Hz) of KIC\,8006161 extracted from the cross-correlation (Method\,\#1, in green) and peak-fitting (Method\,\#2, in black) analyses. {\it Upper-right panel}: Frequency shifts of the individual $l=0$ (blue) and $l=1$ (red) modes from Method\,\#2. {\it Bottom-left panel}: Temporal variability of the photospheric activity proxy $S_\textrm{ph}$ (in ppm). {\it Bottom-right panel}: Frequency shifts extracted from Method\,\#1 as a function of the photospheric magnetic proxy $S_\textrm{ph}$. The corresponding correlation coefficient $r$ and two-sided significance of the deviation from zero $p$ are also indicated.}
\end{figure*} 

\begin{figure*}[tbp]
\begin{center} 
\includegraphics[width=0.49\textwidth,angle=0]{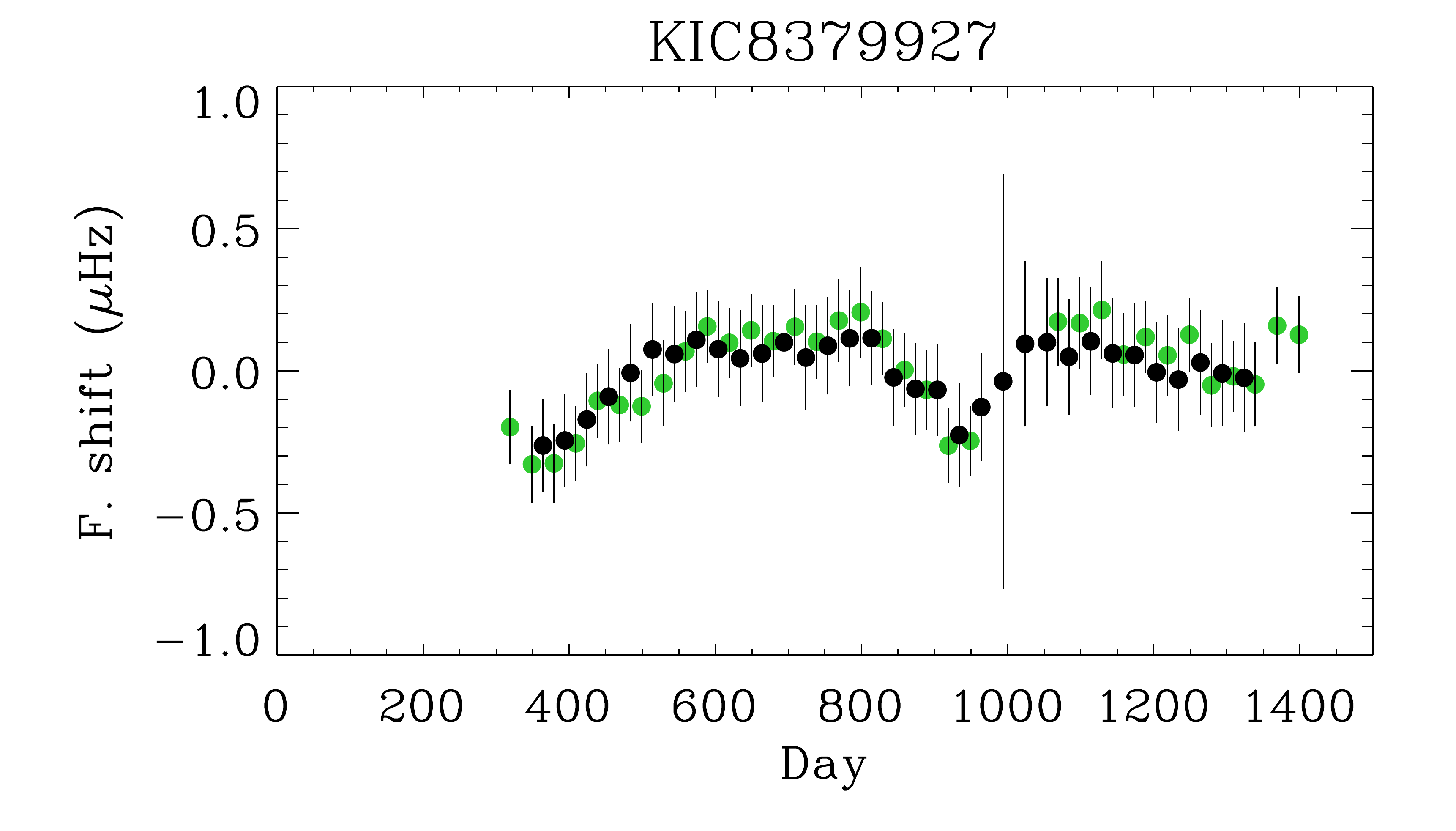}
\includegraphics[width=0.49\textwidth,angle=0]{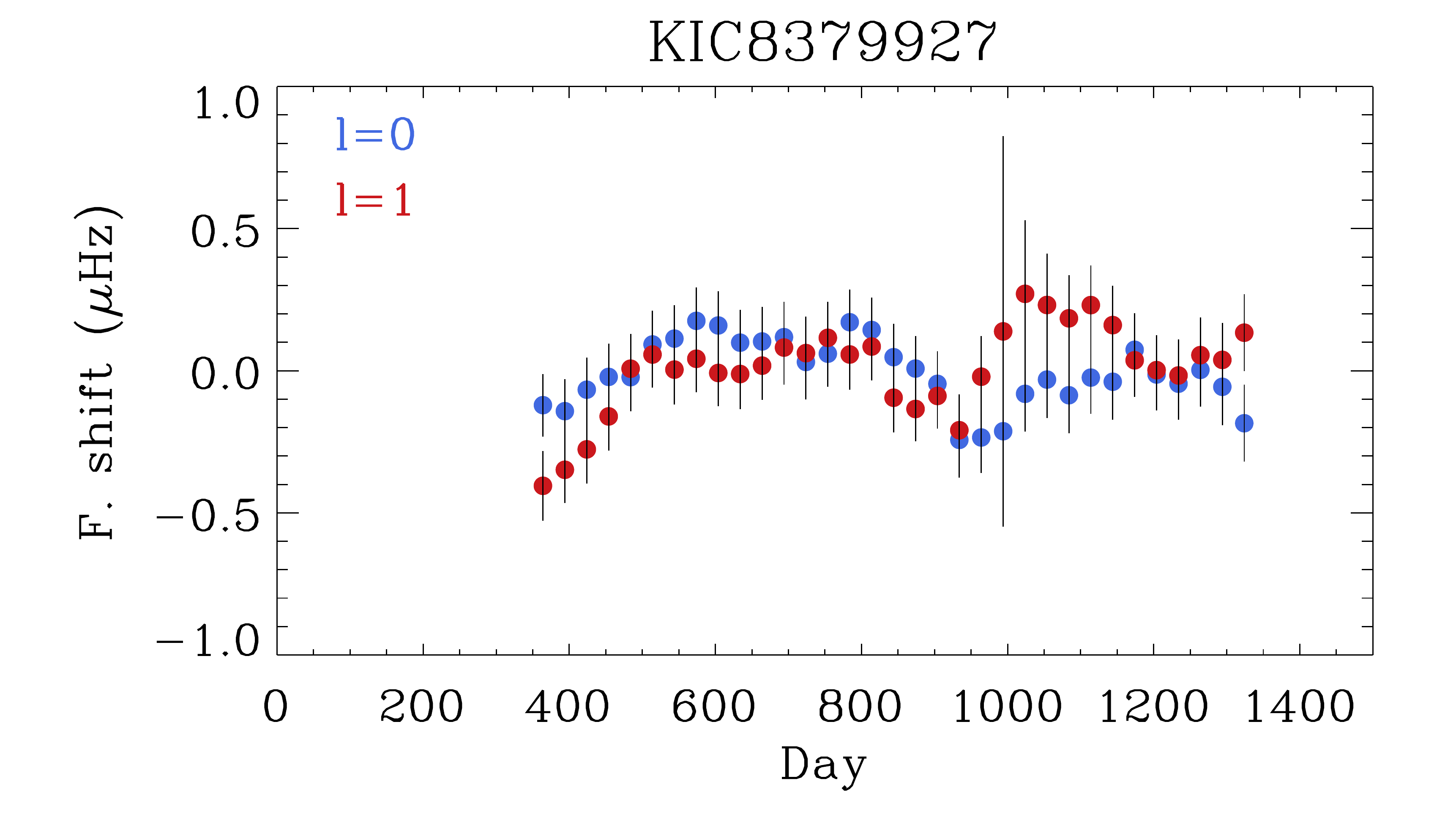}
\end{center}
\caption{\label{fig:kic83} {\it Left panel}: Comparison of the temporal variability of the frequency shifts (in $\mu$Hz) of KIC\,8379927 extracted from the cross-correlation (Method\,\#1, in green) and peak-fitting (Method\,\#2, in black) analyses. {\it Right panel}: Frequency shifts of the individual $l=0$ (blue) and $l=1$ (red) modes from Method\,\#2. }\end{figure*} 

\begin{figure*}[tbp]
\begin{center} 
\includegraphics[width=0.49\textwidth,angle=0]{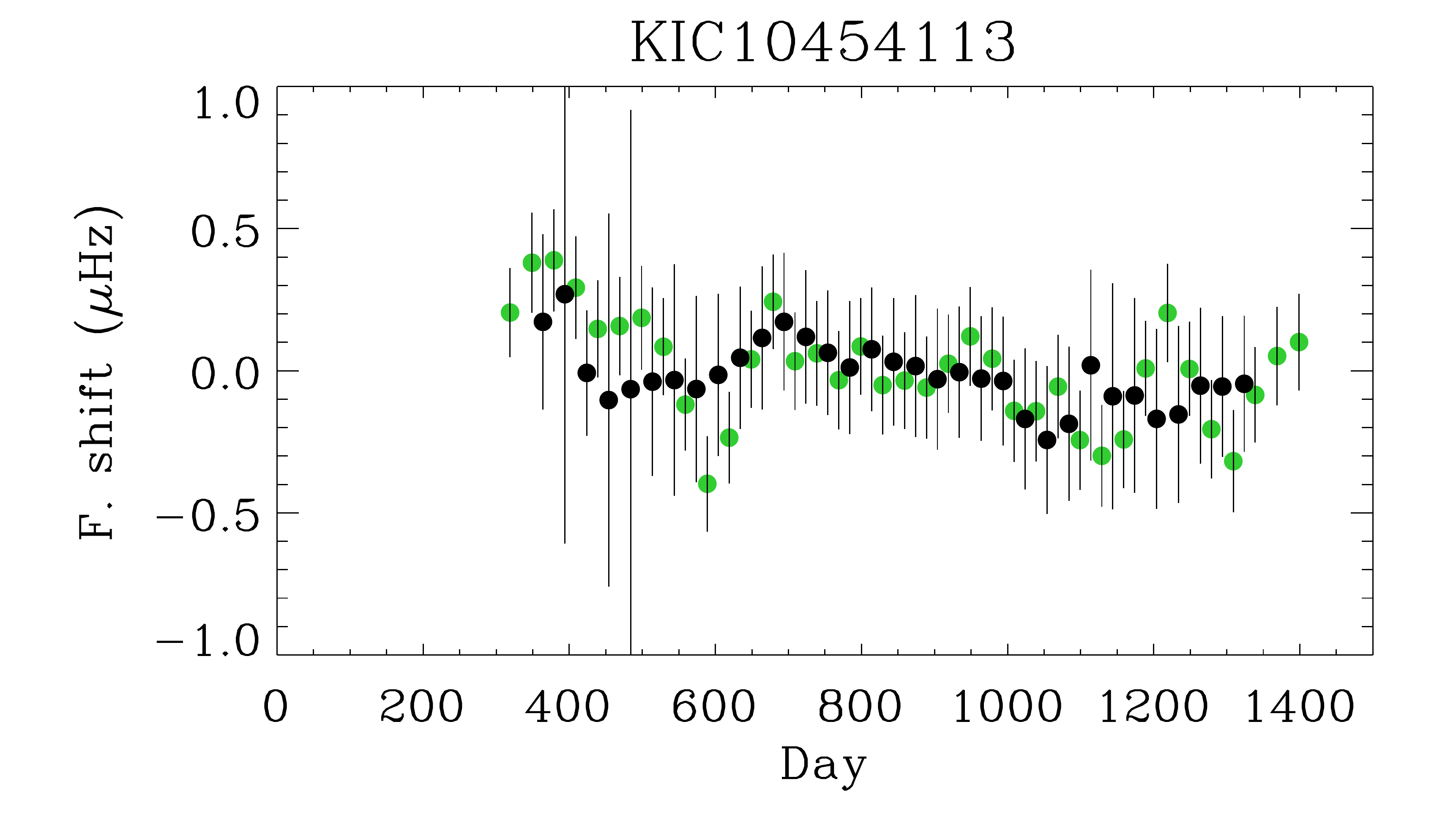}
\includegraphics[width=0.49\textwidth,angle=0]{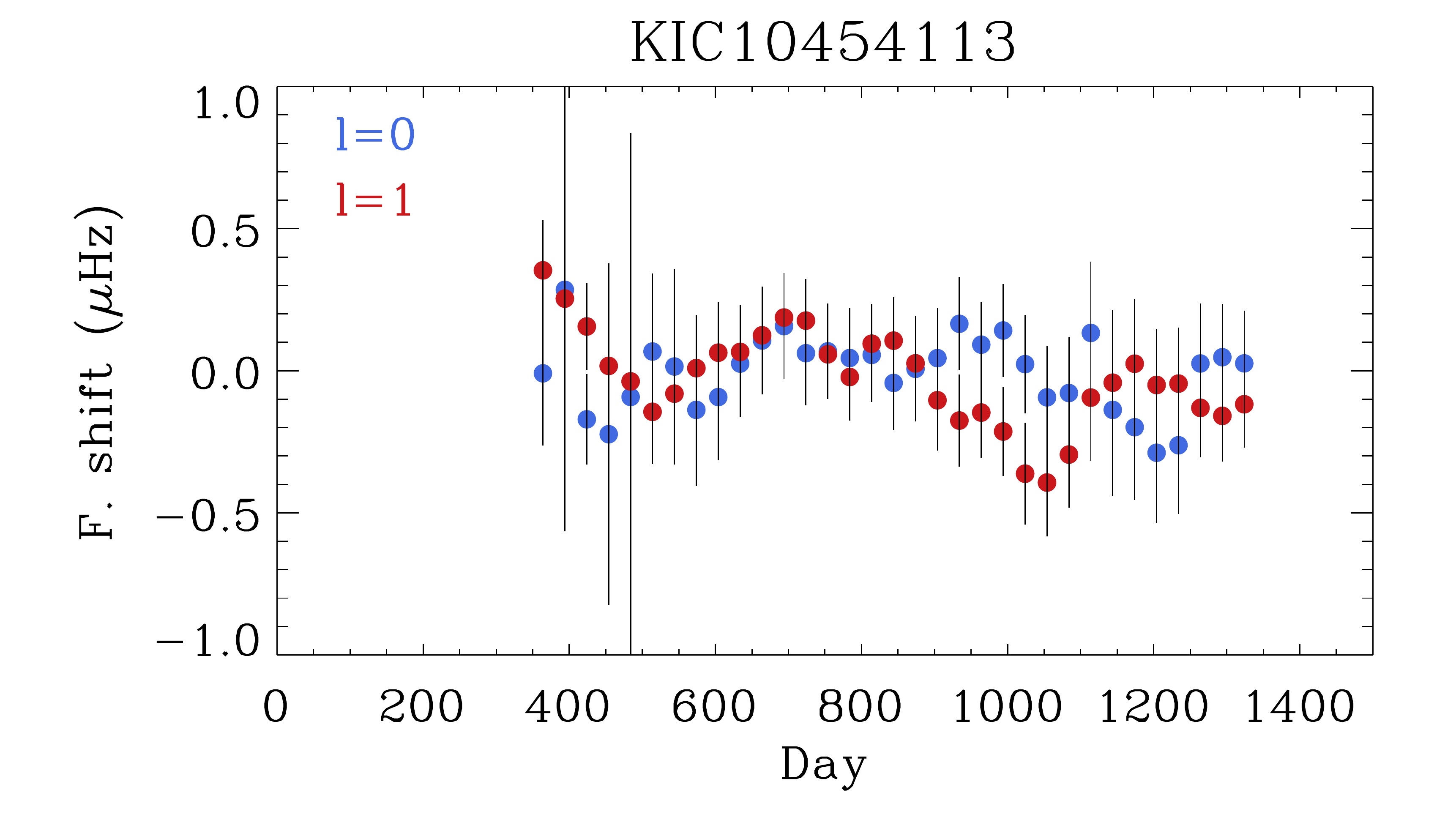}
\end{center}
\caption{\label{fig:kic104} {\it Left panel}: Comparison of the temporal variability of the frequency shifts (in $\mu$Hz) of KIC\,10454113 extracted from the cross-correlation (Method\,\#1, in green) and peak-fitting (Method\,\#2, in black) analyses. {\it Right panel}: Frequency shifts of the individual $l=0$ (blue) and $l=1$ (red) modes from Method\,\#2. }\end{figure*} 

\begin{figure*}[tbp]
\begin{center} 
\includegraphics[width=0.49\textwidth,angle=0]{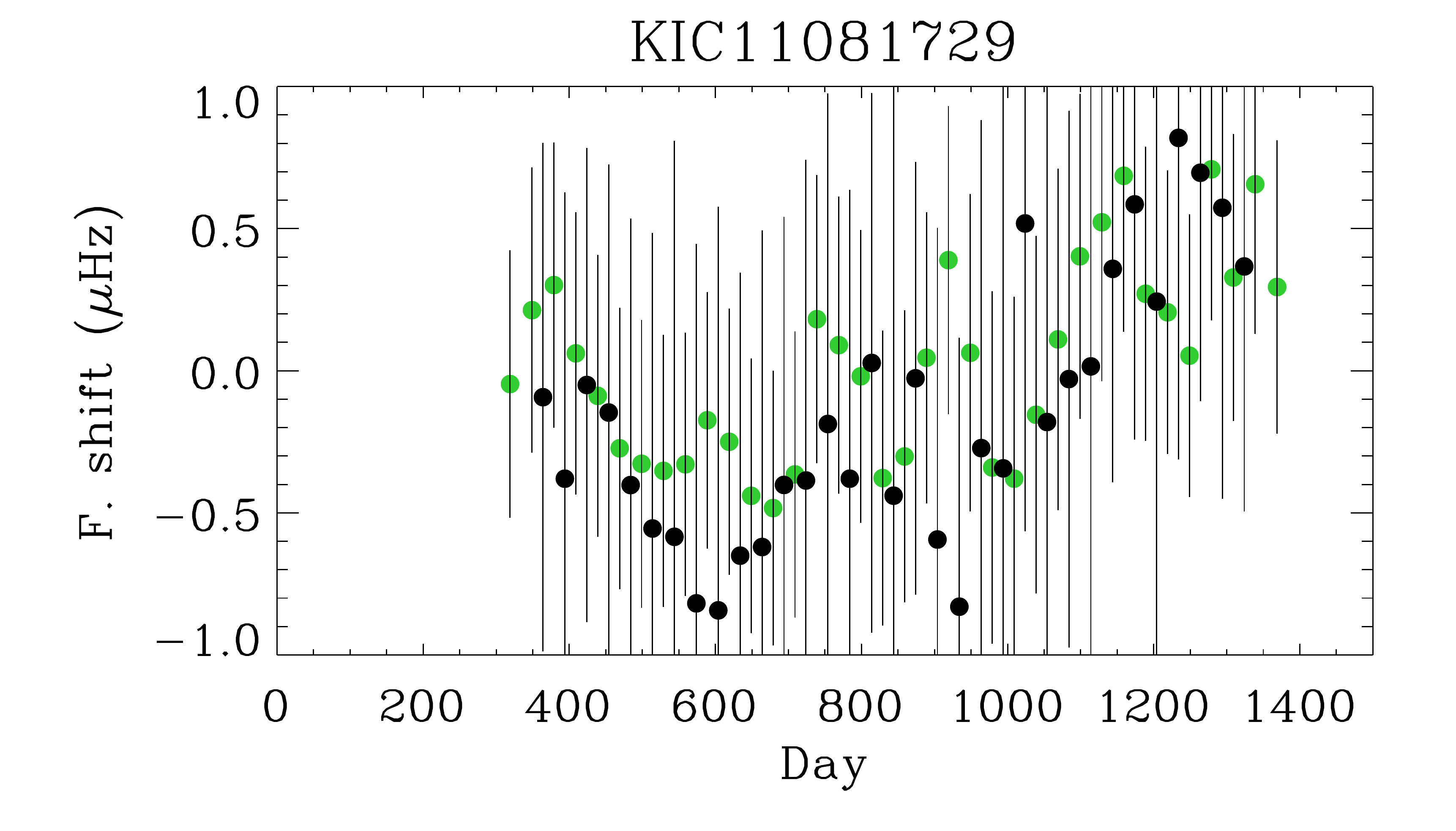}
\includegraphics[width=0.49\textwidth,angle=0]{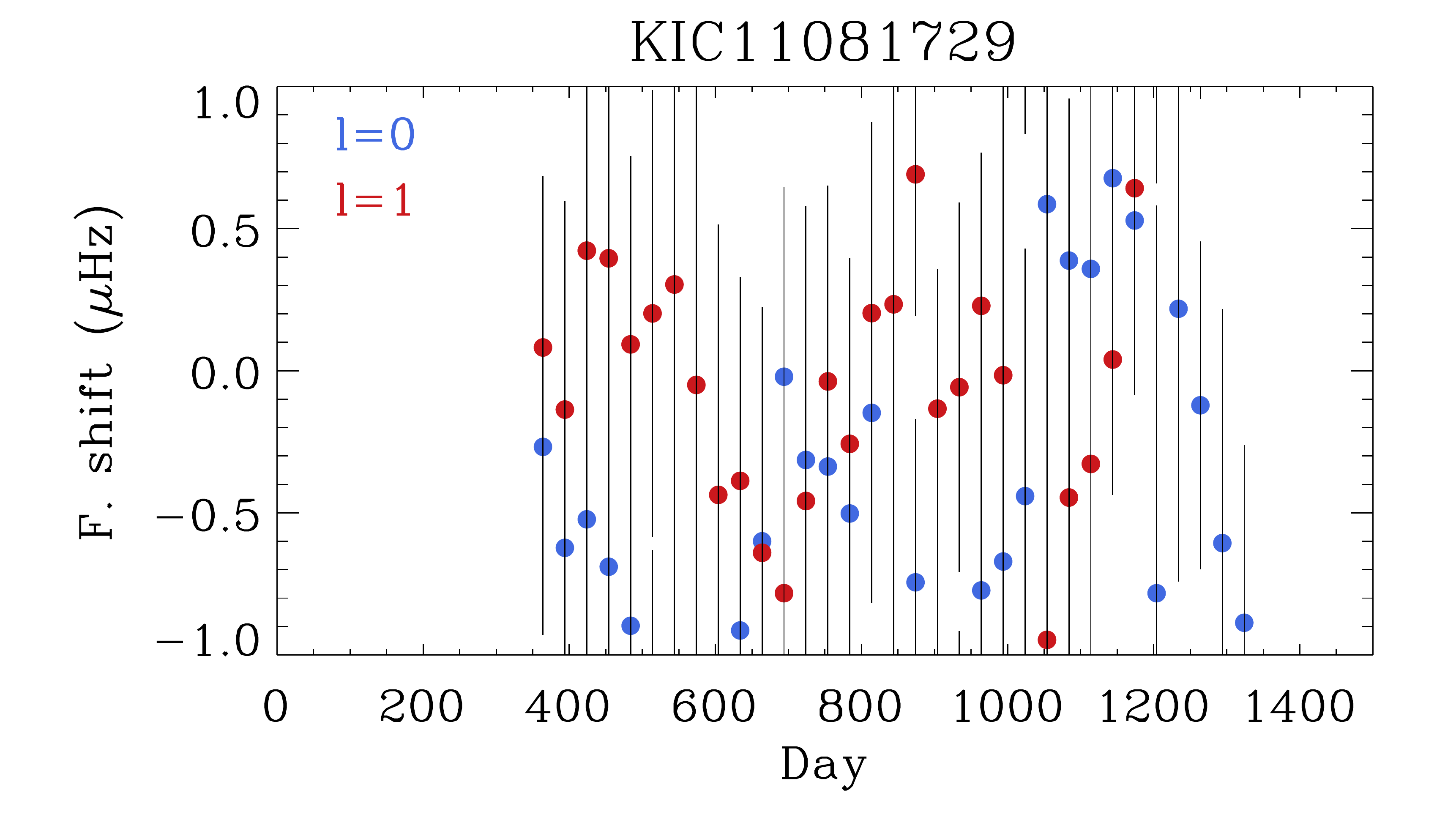}
\includegraphics[width=0.49\textwidth,angle=0]{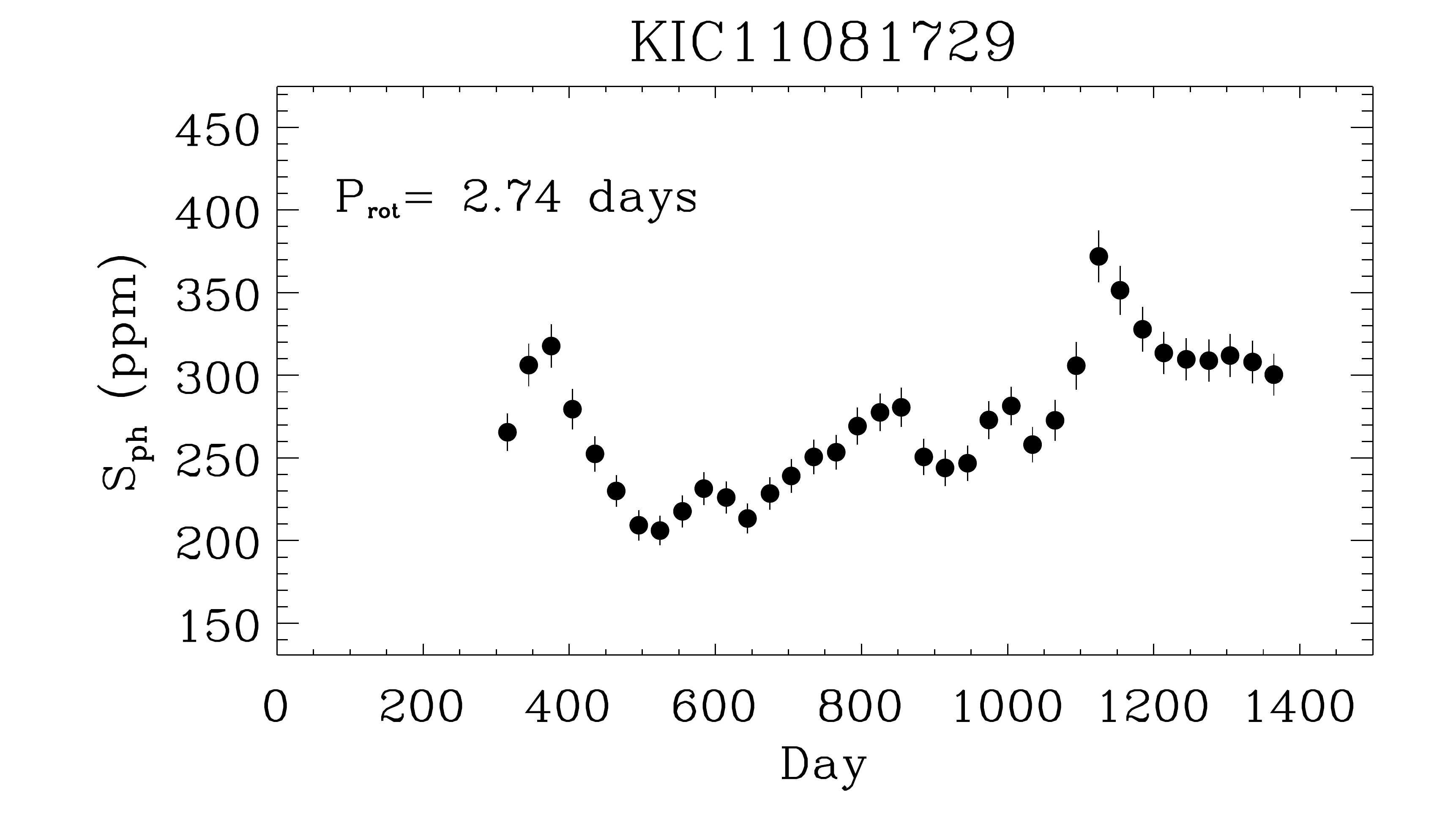}
\includegraphics[width=0.49\textwidth,angle=0]{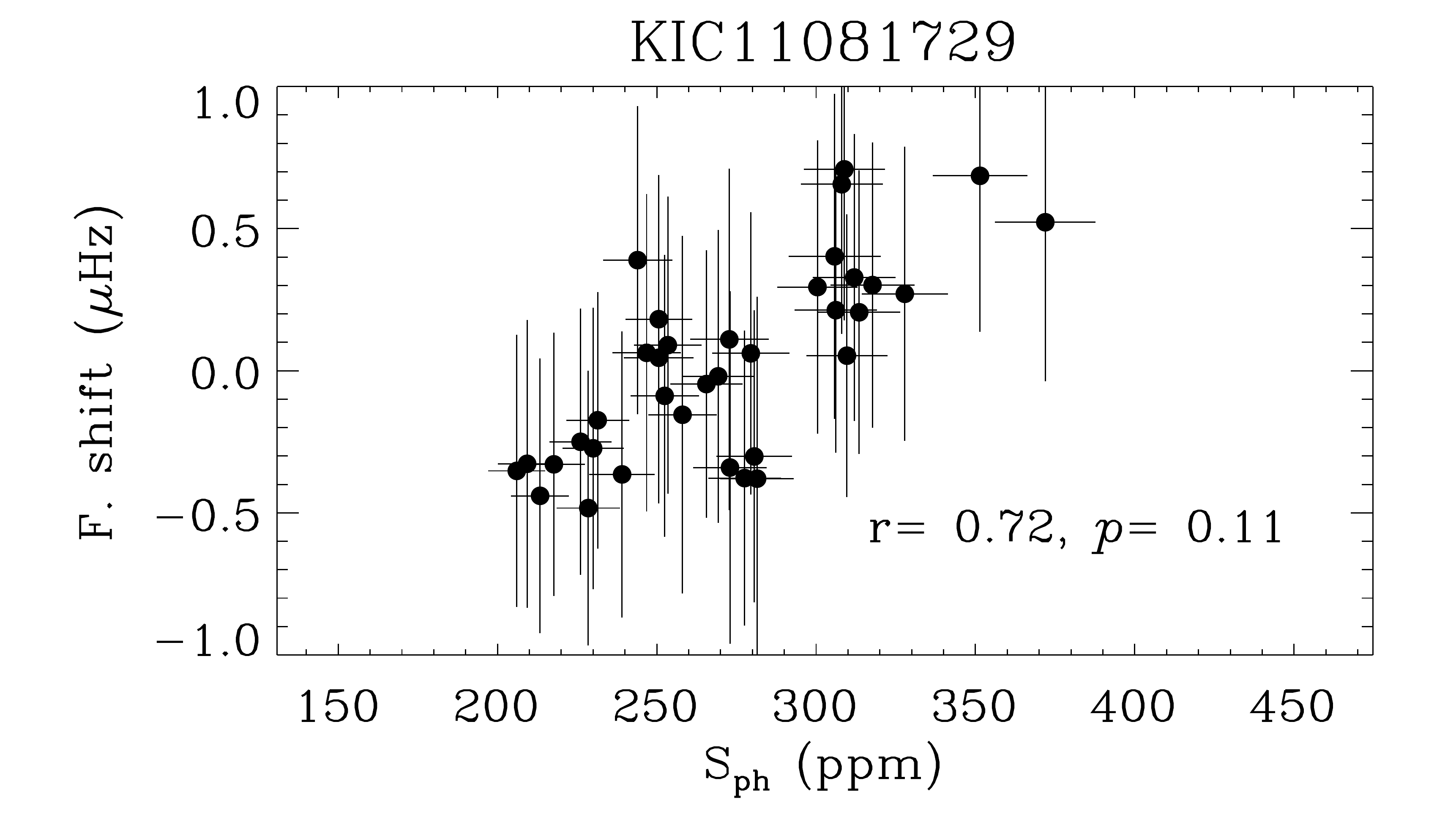}
\end{center}
\caption{\label{fig:kic110} {\it Upper-left panel}: Comparison of the temporal variability of the frequency shifts (in $\mu$Hz) of KIC\,11081729 extracted from the cross-correlation (Method\,\#1, in green) and peak-fitting (Method\,\#2, in black) analyses. {\it Upper-right panel}: Frequency shifts of the individual $l=0$ (blue) and $l=1$ (red) modes from Method\,\#2. {\it Bottom-left panel}: Temporal variability of the photospheric activity proxy $S_\textrm{ph}$ (in ppm). {\it Bottom-right panel}: Frequency shifts extracted from Method\,\#1 as a function of the photospheric magnetic proxy $S_\textrm{ph}$. The corresponding correlation coefficient $r$ and two-sided significance of the deviation from zero $p$ are also indicated.}
\end{figure*} 

\subsection{Case of the more evolved star KIC\,7747078}
\label{sec:mixed}
The star KIC\,7747078 is the only star in our sample with mixed modes that fulfilled the selection criterion $\lambda_1$ with a detection percentage $p(\lambda_1)$ of 84\% indicating that frequency shifts are measured. In evolved stars with mixed modes, the $l=1$ modes do not follow the asymptotic relation and have usually low amplitudes. However, the $l=1$ mixed modes are well visible in KIC\,7747078 \citep[see e.g.,][]{app12} which allows us to extract precise frequency shifts. 
The individual frequency shifts of the $l=0$ and $l=1$ modes  extracted from the peak-fitting analysis (Method\,\#2) are shown in Fig.~\ref{fig:mixed}. The temporal variations of the mixed mode $l=1$ are interestingly comparable to the variations of the $l=0$ mode. Such unobserved behavior yet could bring new inferences for the study of magnetic field in more evolved stars than the Sun and will deserve a close look in the entire {\it Kepler} sample of evolved solar-like stars.

\begin{table}
\caption{Stars selected for a detailed analysis of the frequency dependence of their frequency shifts.}
\label{table:4}
\centering 
\begin{tabular}{c c c c c} 
\hline\hline
KIC & $\lambda_1$  & $\lambda_2$ & $\lambda_3$  & $r$   \\
\hline
5184732   &  *  &  *  &  *  &  0.52  \\
8006161   &  *  &  *  &  *  &  0.71  \\
8379927   &  *  &  *  &   &  \\
10454113  &  *  &  *  &   &  \\
11081729  &  *  &   &  *   &  0.72  \\
\hline                                                 
\end{tabular}
\tablefoot{For each star, the selection criteria, $\lambda_1$, $\lambda_2$, and $\lambda_3$ are marked with an asterisk when fulfilled. The Spearman's correlation coefficients, $r$, between the frequency shifts and the $S_\text{ph}(t)$ activity proxy are given for stars with a measured surface rotation period, $P_\textrm{rot}$.}
\end{table}

\subsection{Detailed analysis of the best stars}
\label{sec:pkf_freq}
Based on Table~\ref{table:shifts}, the subset of the most promising stars was selected for a detailed analysis of the frequency dependence of the frequency shifts using the peak-fitting results from Method\,\#2. Indeed, in the case of the Sun, the frequency dependence \citep[see e.g.,][]{anguera92,chaplin98,salabert04} provides inferences to differentiate among the possible physical mechanisms responsible for the frequency variations with the activity cycle \citep[e.g.,][]{gough90}.

This subset, given in Table~\ref{table:4}, corresponds to the five stars which fulfill at least two of the selection criteria, $\lambda_1$, $\lambda_2$, and $\lambda_3$.
Four of them are classified as simple stars (KIC\,5184732, KIC\,8006161, KIC\,8379927, and KIC\,10454113), and one as a F-like star (KIC\,11081729). Although KIC\,11081729 has a low detection percentage of 32\%, which is typical for a F star, it has one of the largest frequency shifts among the selected stars (see Table~\ref{table:shifts}), which is moreover well correlated with the $S_\textrm{ph}(t)$ proxy.

Figures~\ref{fig:kic51}, \ref{fig:kic80},  \ref{fig:kic83}, \ref{fig:kic104}, and \ref{fig:kic110} show the comparison of the extracted frequency shifts between the cross-correlation analysis (Method\,\#1) and the peak-fitting analysis (Method\,\#2) for this subset of five stars. In the case of  Method\,\#2, the averaged values between the individual angular degrees $l=0$ and $l=1$ modes, as well as the individual $l=0$ and $l=1$ frequency shifts, are represented. When a surface rotation period $P_\textrm{rot}$ was measured, the temporal variation of the $S_\textrm{ph}(t)$ activity proxy interpolated over the same periods of time is also represented, along the relation between both observables. The Spearman's correlation coefficients $r$ and the associated two-sided significances of the deviation from zero $p$ are also indicated on Figs.~\ref{fig:kic51}, \ref{fig:kic80}, and \ref{fig:kic110}, and in Table~\ref{table:4} too.  We recall that these values were computed using independent points only.

Both methods \#1 and \#2, for which different lengths of subseries were used -- 90 and 180~days respectively -- are in excellent agreement with differences well within the uncertainties. For the simple stars, small differences between the frequency shifts of the individual $l=0$ and $l=1$ modes can be observed for KIC\,5184732, KIC\,8006161, and KIC\,8379927, while larger disparity is present in KIC\,10454113 along larger errors. Indeed, as well observed for the Sun, oscillation modes with different angular degrees respond differently to the activity cycle \citep[see e.g.,][and references therein]{palle89,anguera92,howe99,chano01,salabert15} in relation to the spatial distribution of the surface magnetic field \citep{howe02,chano04}. This is what could be observed here for these stars. Again, the differences between the $l=0$ and $l=1$ modes for the F-like star KIC\,11081729 are difficult to interpret given the large uncertainties.

\begin{figure*}[tbp]
\begin{center} 
\includegraphics[width=0.49\textwidth,angle=0]{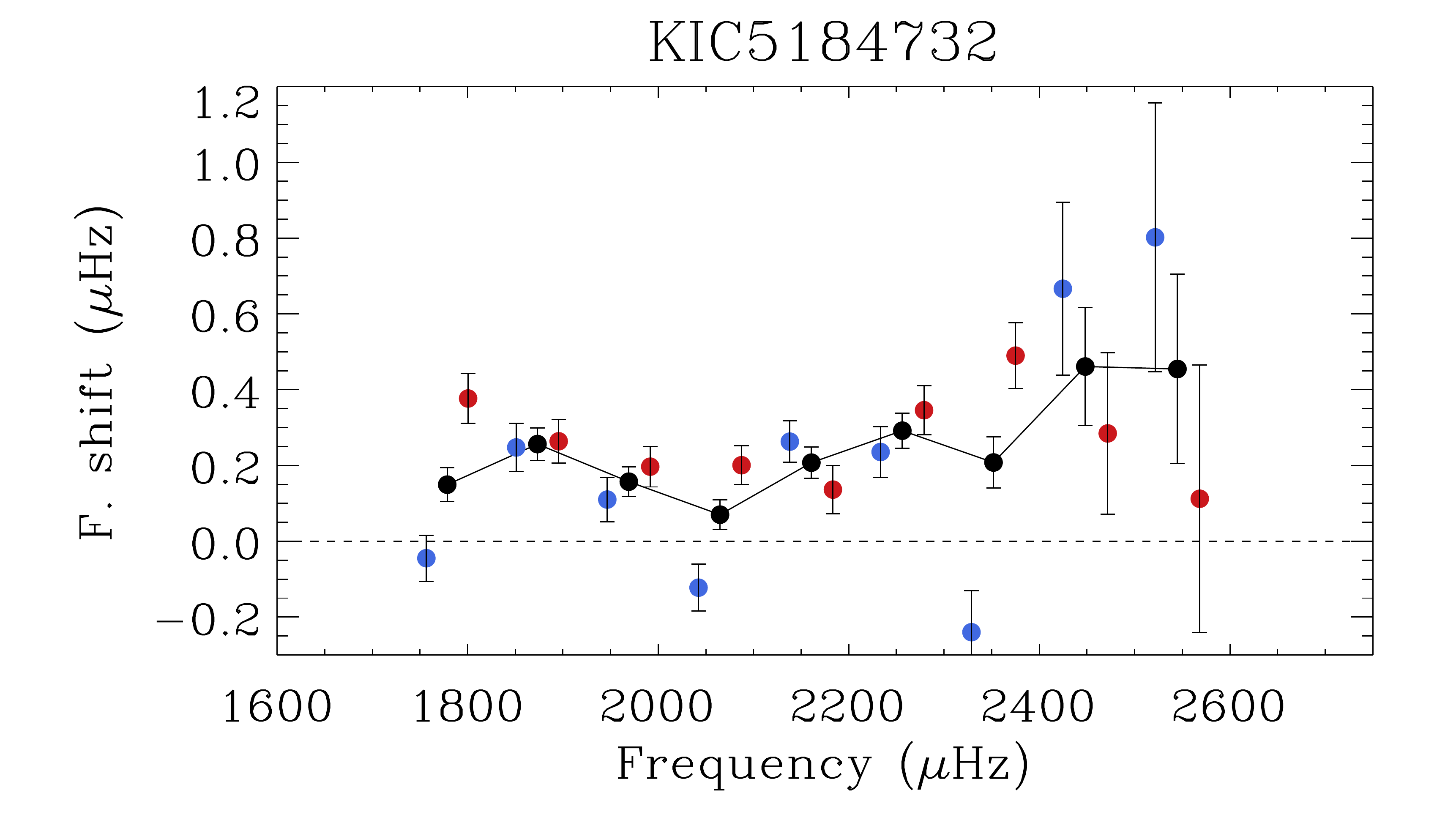}
\includegraphics[width=0.49\textwidth,angle=0]{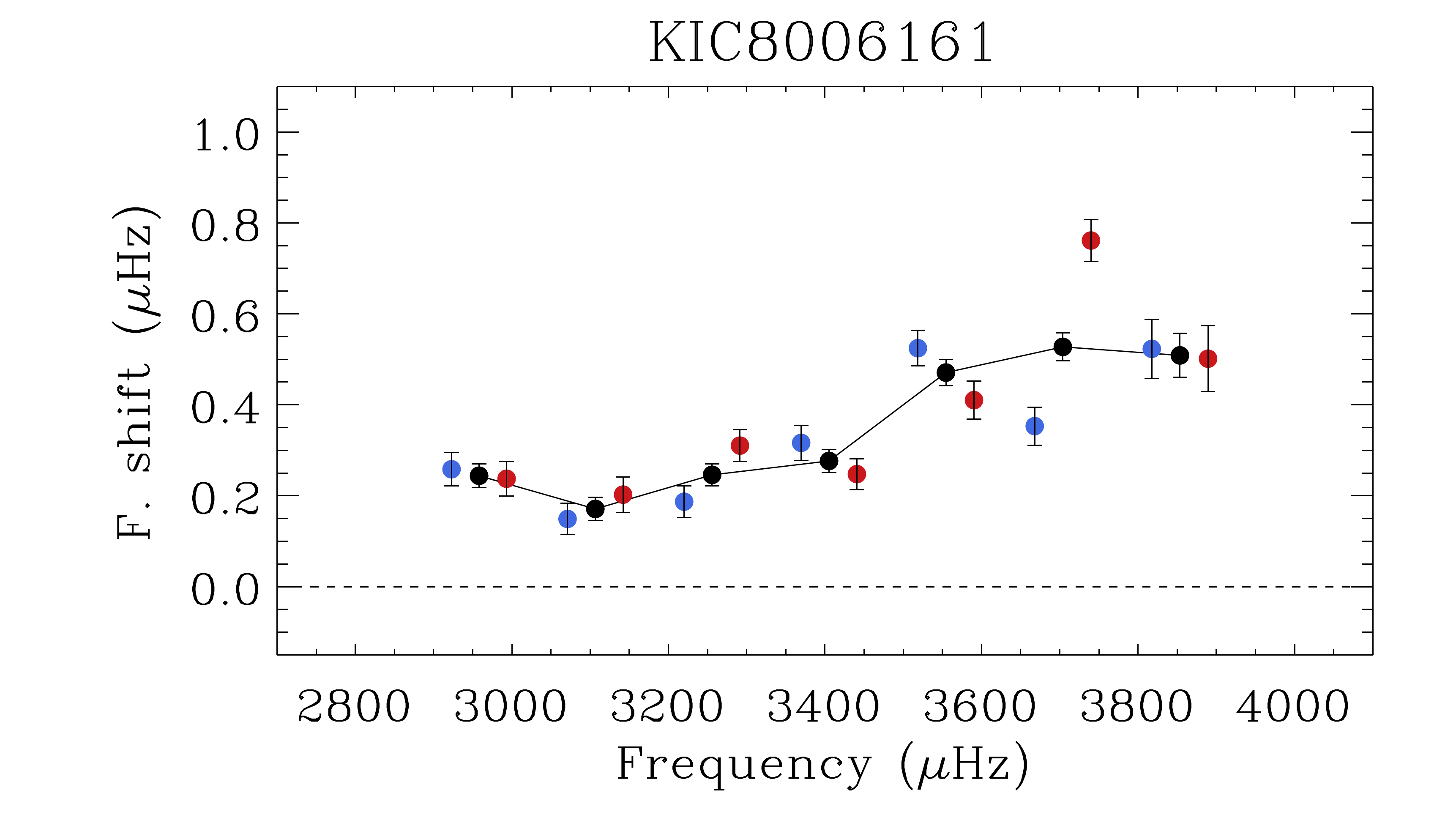}
\includegraphics[width=0.49\textwidth,angle=0]{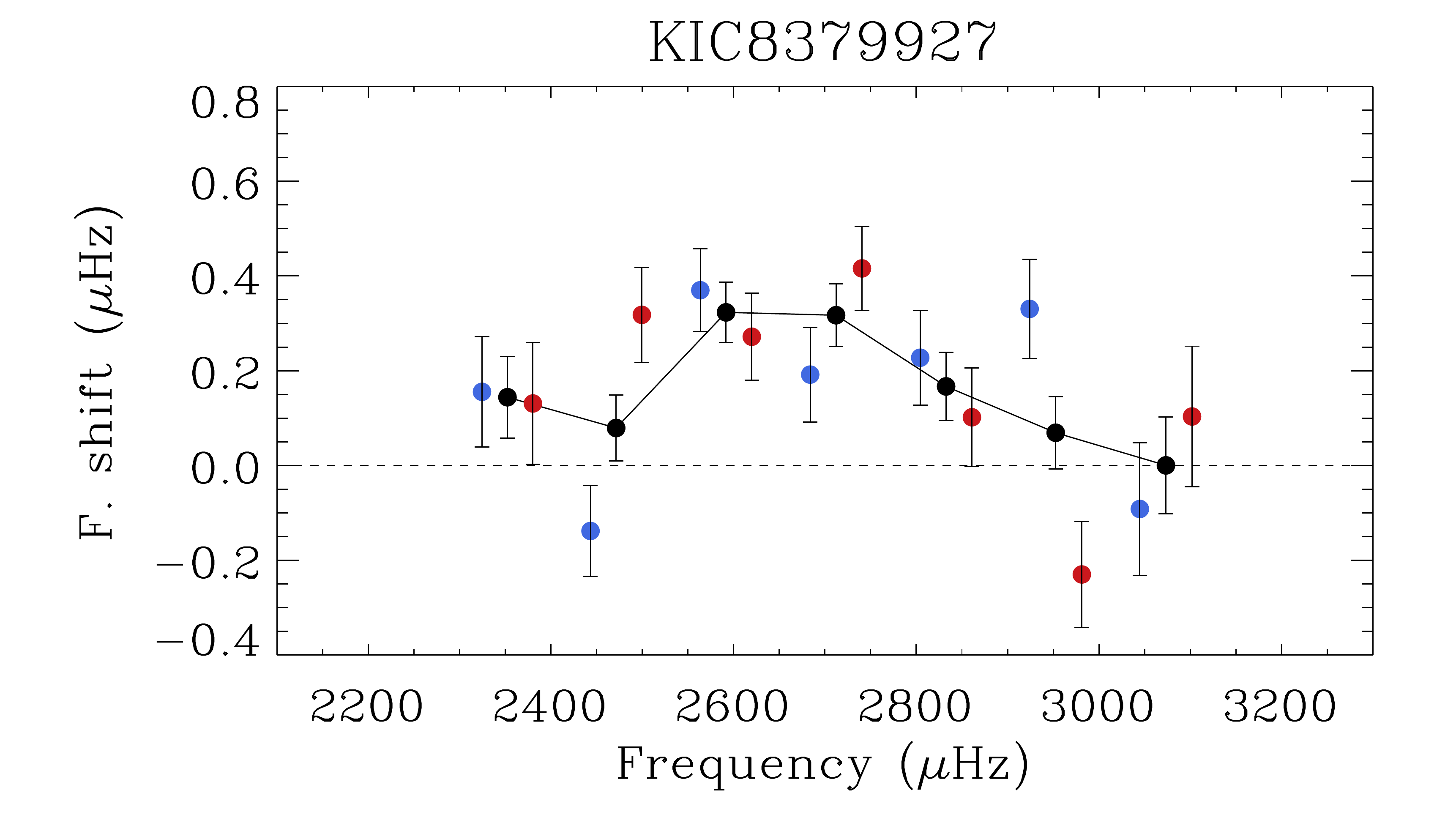}
\includegraphics[width=0.49\textwidth,angle=0]{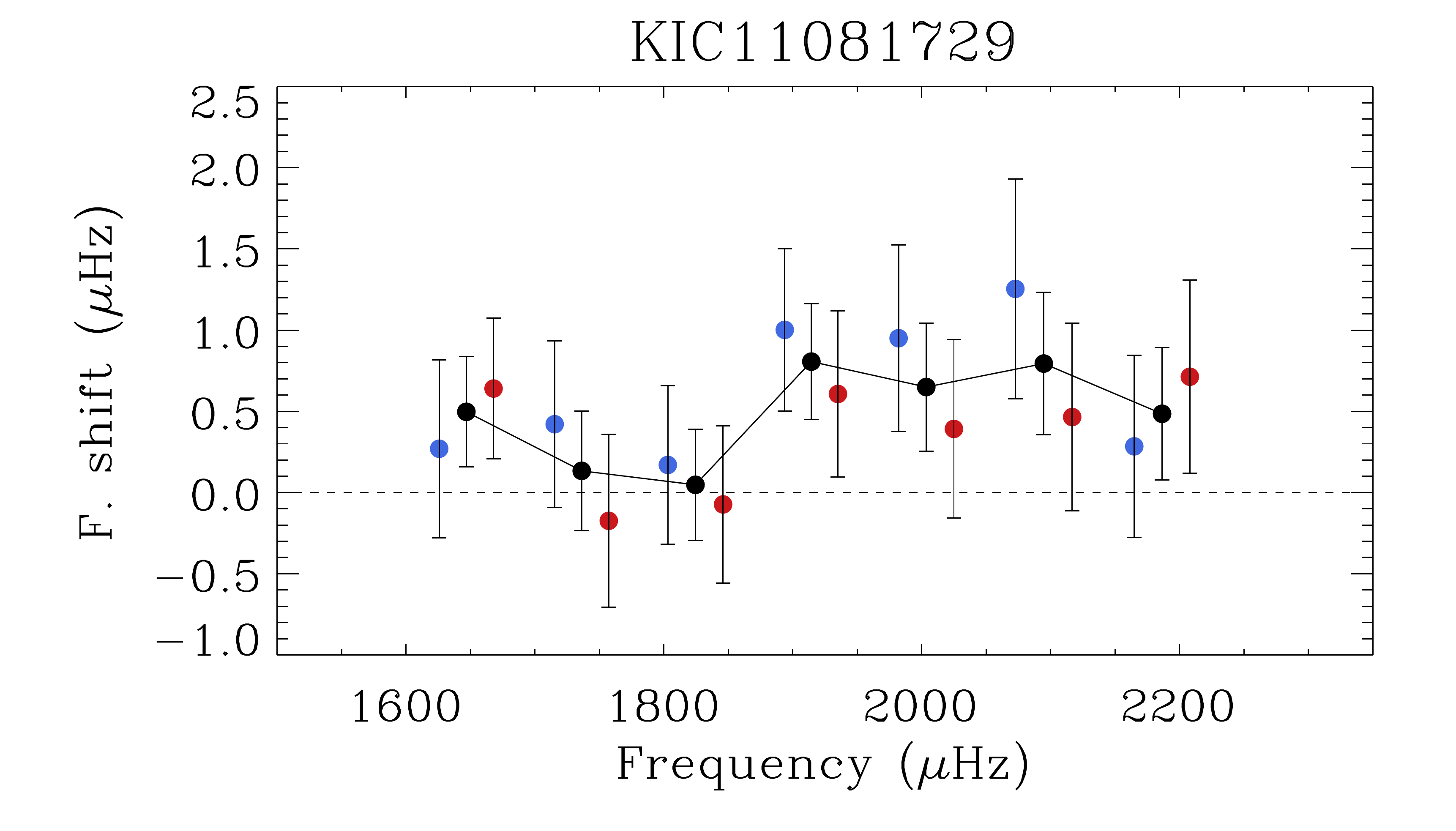}
\end{center}
\caption{\label{fig:freqvsfreq} Frequency shifts of the individual $l = 0$ (blue) and $l=1$ (red) modes as a function of frequency for KIC\,5184732, KIC\,8006161, KIC\,8379927, and KIC\,11081729 from Table~\ref{table:4}. The black dots correspond to the averaged shifts between consecutive $l = 0$ and $l=1$ modes. The horizontal dashed lines represent a zero shift.}
\end{figure*} 

In order to measure the frequency dependence of the frequency shifts, two periods of low and high activity around the median value of the frequency shifts were defined for each star. The weighted means of each fitted frequency from each 180-day subseries were estimated for both periods. The frequency shifts at each radial order were thus obtained by simply subtracting the frequencies measured during high activity by the corresponding frequencies at low activity. We considered both the $l=0$ and the $l=1$ modes independently. It is important to keep in mind that the two periods of activity thus defined are dependent on the unknown observed phase of magnetic variability for each given star. Despite this, the frequency dependence of the frequency shifts thus measured is found to be significant at $3\sigma$, $12\sigma$, $3\sigma$, and $2\sigma$ from the mean shift for KIC\,5184732, KIC\,8006161, KIC\,8379927, and KIC\,11081729 respectively, where $\sigma$ is the mean error on the shifts. This corresponds to confidence levels higher than 95\%. However, although KIC\,10454113 fulfills the criteria $\lambda_1$ and $\lambda_2$, no frequency dependence of the frequency shifts could be found at the level of precision of the data with a $1\sigma$ significance (i.e. a 68\% confidence). 

Figure~\ref{fig:freqvsfreq} shows the frequency shifts as a function of frequency for KIC\,5184732, KIC\,8006161, KIC\,8379927, and KIC\,11081729. The shifts of the individual $l=0$ and $l=1$ modes are represented as well as the corresponding mean values. Except for KIC\,8006161 which shows a frequency dependence comparable to what is observed in the case of the Sun \citep[see e.g.,][]{anguera92,chaplin98,salabert04}, as well as for KIC\,10644253 \citep{salabert16} and HD\,49933 \citep{salabert11}, the three other stars, KIC\,5184732, KIC\,8379927, and KIC\,11081729 follow quite different behaviors.

\begin{table*}[t]
\caption{Stellar parameters derived in Section~\ref{sec:model} of the modeled stars. } 
\label{table:table_stellarparam}
\centering
\begin{tabular}{l c c c c c c c}
\hline
KIC & $R$ $(R_{\odot})$ & $M$ $(M_{\odot})$ & Age (Gyr) & $T_{\rm{eff}}$ $(K)$ & log~$g$ & [Fe/H] & $\tau_{\rm{HeII}}$ $(s)$  \\
	  
\hline
5184732  & $1.37\pm 0.02$ & $1.29\pm 0.04$ & $4.00\pm 0.29$  & $5870\pm 120$ & $4.272\pm 0.005$ & $+0.29\pm 0.20$ & $916\pm10$  \\ 
8006161  & $0.94\pm 0.01$ & $1.01\pm 0.03$ & $4.98\pm 0.17$  & $5338\pm 66$ & $4.497\pm 0.005$ & $+0.64\pm 0.04$ & $621\pm10$   \\
8379927  & $1.14\pm 0.01$ & $1.17\pm 0.03$ & $1.55\pm 0.09$  & $6045\pm 110$ & $4.393\pm 0.003$ & $-0.10\pm 0.15$ & $683\pm10$   \\
11081729 & $1.38\pm 0.01$ & $1.27\pm 0.05$ & $2.21\pm 0.07$  & $6349\pm 28$ & $4.261\pm 0.011$ & $+0.11\pm 0.10$ & $798\pm10$   \\
\hline
\end{tabular}
\end{table*}
        
\section{Discussion}
\label{sec:model}
The degree dependence of the frequency shifts observed in the Sun from the low- to high-degree angular modes is removed once the 
shifts are corrected by the mode inertia \citep[see e.g.,][]{libbrecht90,chaplin01,howe02}. 
The shifts thus follow a smooth increase with frequency. This indicates that the structural changes 
occurring with magnetic activity that produce the p-mode frequency shifts are located in a thin layer close to the surface.
Different observational and theoretical works were carried out in order to better understand the observed dependences of the 
frequency shifts \citep[see e.g.,][for a review]{basu16}. 
For instance, the frequency dependence through the activity cycle was shown to be consistent with a decrease of less than 2\% in 
the radial component of the turbulent velocity in the outer layers of the Sun \citep{kuhn00,dzieb01,dzieb05}. 
This mechanism is only efficient in the uppermost layers of the convective zone where the turbulent velocities become comparable 
to the local sound speed and explains the smooth increase of the shifts with frequency. 

Nonetheless, detailed analyses of the frequency dependence of the p-mode frequency changes with magnetic activity in the Sun revealed further information. Indeed, \citet{goldreich91} first drew the attention of an apparent oscillatory component superimposed on the dominant smooth frequency trend of the frequency shifts. Based on the period of the signal, the authors surmised that it could be caused by changes in the magnetic field with the solar activity cycle confined to a thin layer just above the He~\textsc{i} ionization zone. \citet{gough94} estimated that the period of this oscillatory feature was roughly twice the inverse of the acoustic depth of the He~\textsc{ii} ionization zone. He thus proposed to relate this oscillatory feature with a temporal variation in the depression of the first adiabatic exponent located in a deeper layer than \citet{goldreich91} had suggested, and to associate it with changes in the acoustic glitch caused by the He~\textsc{ii} ionization. \citet{basu04} and \citet{verner06} did also detected variations with activity in the amplitude of the He~\textsc{ii} acoustic glitch. However, the physical mechanism responsible for such periodic changes is still not clear. See \citet{gough13} for a more detailed discussion.

Moreover, helioseismic inversions using the Michelson Doppler Imager \citep[MDI;][]{scherrer95} observations on board the SoHO satellite 
provided inferences on the radial dependence of the sound speed variations between the solar maximum and minimum 
\citep{baldner08,rs12}. \citet{mullan12} compared their results with the sound speed changes expected from a model of magnetic 
inhibition of convective onset 
and found an agreement for at least the outer half of the convection zone.
Since these structural inversions require intermediate- and high-degree modes, such detailed analysis is unfortunately not 
possible for other stars than the Sun. However, we can compare the frequency dependence between the stars discussed in 
Section~\ref{sec:pkf_freq} once their shifts are corrected by their mode inertia 
in relation to what it is observed in the Sun.

The mode inertia was obtained from structural models computed with MESA \citep{paxton11}. The OPAL opacities \citep{iglesias96} and the GS98 metallicity mixture \citep{grev98} were used, and the microscopic diffusion of elements was included, otherwise the standard input physics from MESA was applied. 
The theoretical low-degree, p-mode frequencies were computed in the adiabatic approximation using the ADIPLS code \citep{jdc08}. A $\chi^2$ minimization including simultaneously p-mode frequencies and spectroscopic data was applied. The procedure is comparable to the one described in \citet{perez16}, except that no gravity-like mixed modes were considered here. The input spectroscopic parameters such as, effective temperature ($T_{\rm{eff}}$), surface gravity ($\log g$), metallicity ([Fe/H]), and luminosity which was derived from the Hipparcos and Gaia parallaxes, were retrieved from the SIMBAD astronomical database. 
The stellar parameters thus obtained (see Table~\ref{table:table_stellarparam}) are consistent with the inputs parameters and are comparable to those found by \citet{metcalfe14}, \citet{creevey17}, and \citet{silva17}.

\begin{figure*}[tbp]
\begin{center} 
\includegraphics[width=0.49\textwidth,angle=0]{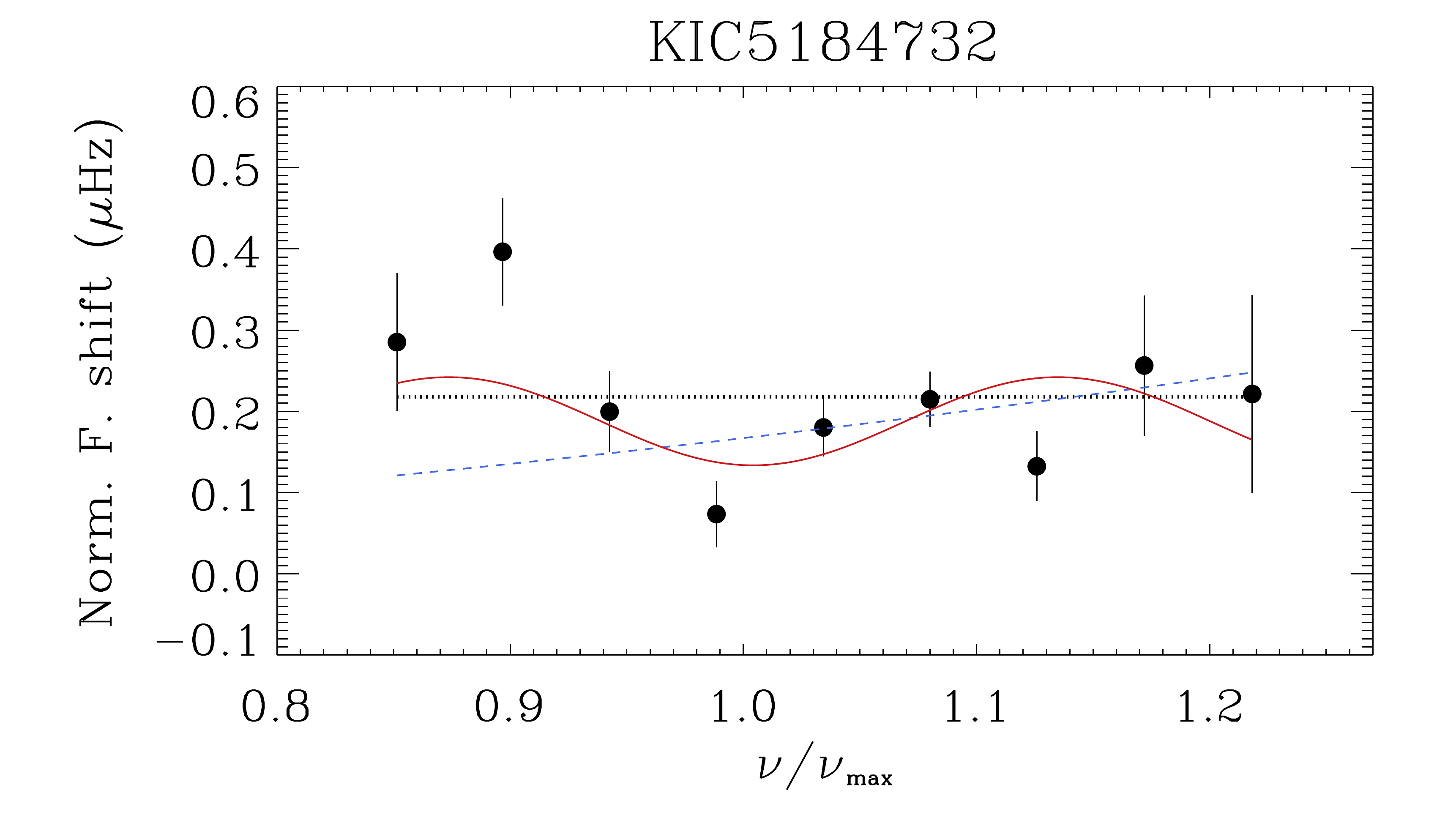}
\includegraphics[width=0.49\textwidth,angle=0]{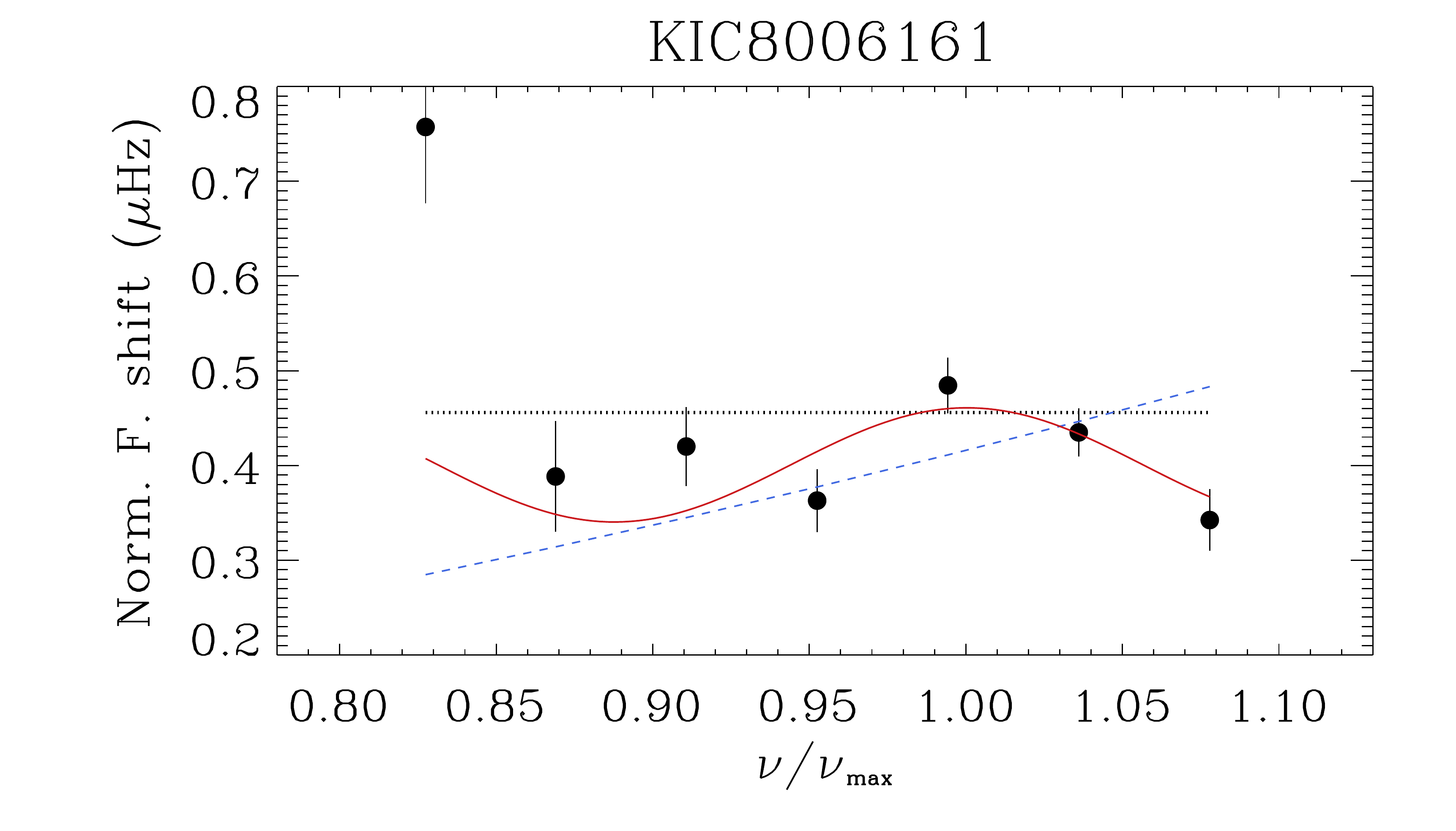}
\includegraphics[width=0.49\textwidth,angle=0]{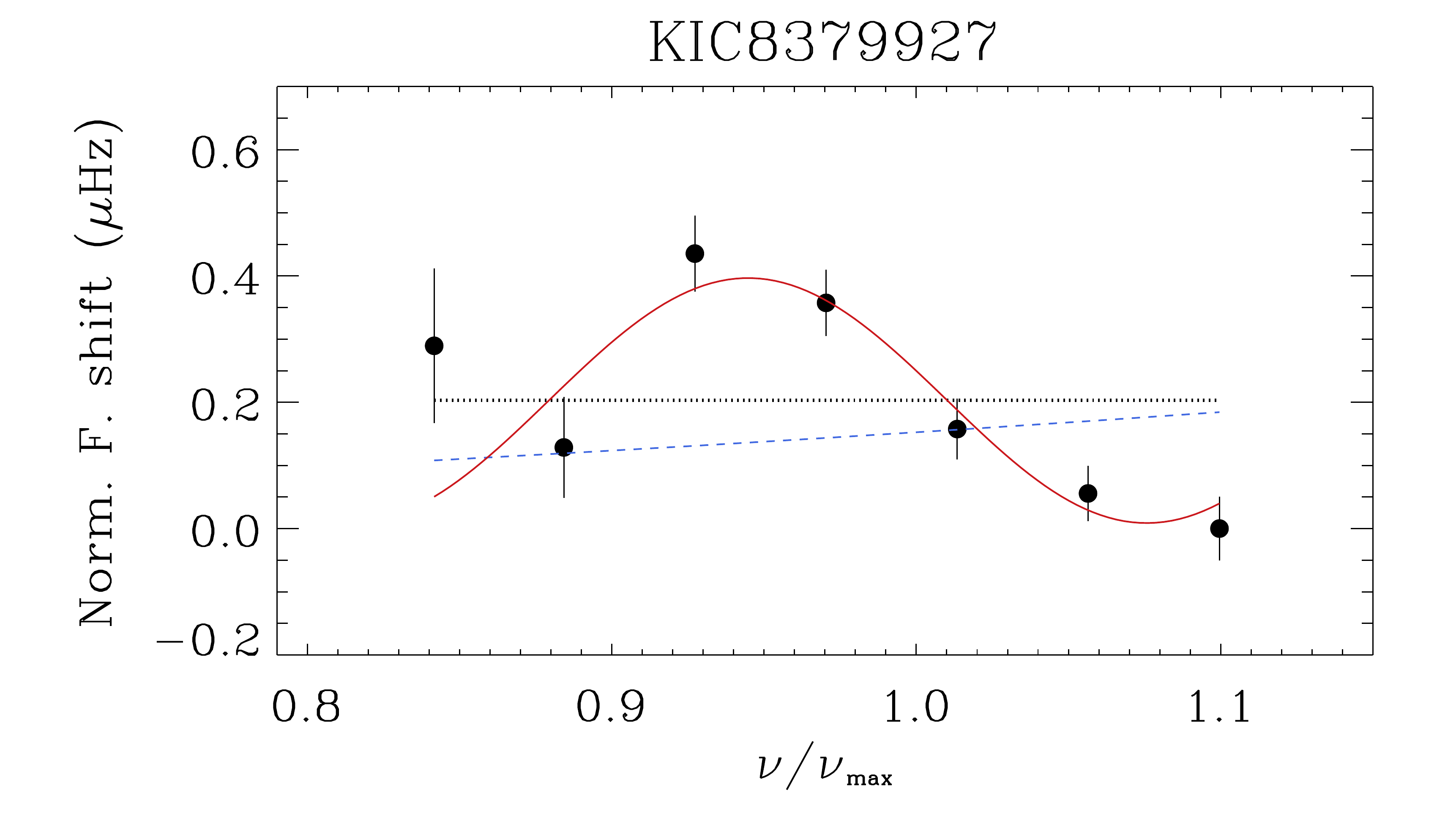}
\includegraphics[width=0.49\textwidth,angle=0]{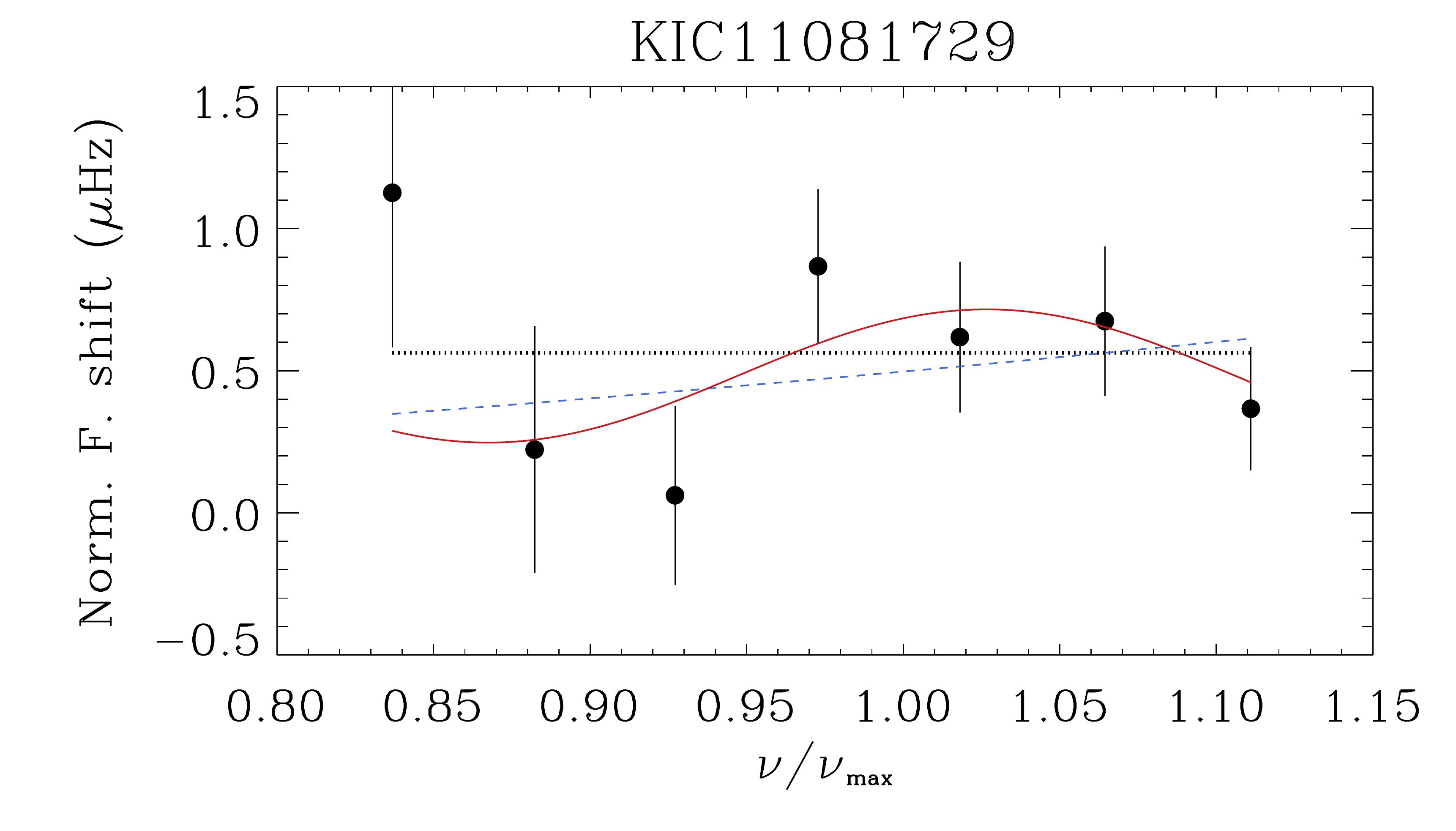}
\end{center}
\caption{\label{fig:models}
Normalized frequency shifts corrected by the mode inertia as a function of $\nu/\nu_\mathrm{max}$ for KIC\,5184732, KIC\,8006161, KIC\,8379927, and KIC\,11081729 as discussed in Section~\ref{sec:model}. The polynomial descriptions of the frequency shifts are represented by the horizontal dotted lines for $\alpha=1$ and by the blue dashed lines for $\alpha=2$. The solid red lines correspond to the results from the sine-wave fitting.}
\end{figure*}

We then followed \citet{salabert16} and used the normalized frequency shifts defined as $(I_{n,l}/I_\textrm{max}) \times \delta \nu_{n,l}/\nu_{n,l} \times \nu_\mathrm{max}$, where $I_{n,l}$ is the mode inertia and $I_\textrm{max}$ the mode inertia interpolated at the frequency $\nu_\textrm{max}$ of the maximum oscillation power excess. The scaling with $\nu_{\mathrm{max}}$ was introduced as it is linearly related to the acoustic cut-off frequency \citep{belkacem11}.
To minimize the errors, the mean values of the frequency shifts between adjacent $l=0$ and $l=1$ modes were considered here. The normalized frequency shifts thus corrected by the mode inertia are shown in Fig.~\ref{fig:models} as a function of $\nu/\nu_\mathrm{max}$ for each star 
in Table~\ref{table:4}. For the Sun, the frequency dependence can be described by a polynomial of the 
form $\delta \nu \propto  \nu^\alpha$, with $1\leq \alpha \leq 2$ \citep{salabert16}. The horizontal dotted lines in Fig.~\ref{fig:models} correspond to $\alpha=1$ and the blue dashed lines to $\alpha=2$. The mean normalized frequency shifts ($\alpha=1$) are estimated to be 0.22, 0.46, 0.20, and $0.56\,\mu$Hz for KIC\,5184732, KIC\,8006161, KIC\,8379927, and KIC\,11081729 respectively. For comparison, the corresponding value for the Sun is $0.24\,\mu$Hz.

Nonetheless, Fig.~\ref{fig:models} suggests that a linear formulation of the frequency dependence of the normalized frequency shifts is not sufficient to properly describe the variations within their uncertainties. Such discrepancies could indicate the presence of an additional or a different source of frequency perturbation. If the frequency perturbation caused by structural changes arises from a thin layer inside the resonant cavity of the acoustic modes, the frequency dependence of the frequency shifts will then follow a sinusoidal function of the form $A \cos (\omega 2 \tau + \varphi)$, where $\omega=2 \pi \nu$ is the p-mode angular frequency and $\tau$ the acoustic depth of the perturbation. The variables $A$ and $\varphi$ correspond to the amplitude and the phase of the sine wave respectively. 
The minimization of this sinusoidal function was performed by fitting the parameters $A$ and $\varphi$ as well as an additional offset to take into account the non-zero mean of the shifts. We tried also to fit the parameter $\tau$, but given the small number of measurements for each star and their associated errors, the determination of the acoustic depth of an hypothetic oscillatory signal from a non-linear fit depends strongly on its initial value. Therefore, $\tau$ was kept fixed and only the other parameters were varied. In order to have a reference to represent the sine-wave dependence of the normalized frequency shifts for each star, the parameter $\tau$ was fixed to the acoustic depth of the He~\textsc{ii} ionization zone $\tau_{\textrm{He\textsc{ii}}}$ given in Table~\ref{table:table_stellarparam}. 
This choice appears natural because, as mentioned earlier on, there are observational evidences of such oscillatory feature in the frequency shifts of the Sun \citep[see e.g.,][]{goldreich91,gough94,gough13}, even though its amplitude is much smaller than the main trend.
The corresponding fits thus obtained are represented in Fig.~\ref{fig:models} by the red solid lines. It is interesting to notice that as a first approximation the use of the He~\textsc{ii} acoustic depth can reproduce the observed dependence of the normalized frequency shifts. 

Such oscillatory behavior could suggest that the perturbations of the turbulent velocity in the outer layers, which is the main cause of the frequency shifts observed in the case of the Sun, is not the dominant factor in the four stars analyzed here and that it could be located within the resonant cavity instead. Given the quality of the data and the small number of stars which successfully passed our selection criteria, this is an intriguing hypothesis which would need to be investigated with a larger number of solar-like stars observed for instance in the future by the PLATO space mission \citep{rauer14}.

%
%
\section{Conclusions}
\label{sec:conclusion}
The photometric  {\it Kepler} observations of a sample of 87 main-sequence, solar-like pulsating stars were analyzed in order to study their magnetism through the measurements of their p-mode frequency shifts, obtained using two independent methods: a cross-correlation analysis of all the visible modes, and a peak-fitting analysis of the individual modes. The results from both methods were compared to each other and to the associated photometric activity proxy $S_\text{ph}$ as well. To determine if the measured variations are likely to be genuine signatures of magnetic variability or to be noise, we developed restrictive selection criteria based on Monte Carlo simulations. A subset of 20 stars was thus selected as potential candidates, for which the frequency dependence of the individual $l=0$ and $l=1$ frequency shifts was measured for the four best stars fulfilling the selection criteria. We used the normalized frequency shifts corrected by the mode inertia to study the frequency dependence of the shifts in relation to what is known for the Sun. Although the sample is small and the uncertainties large to derive firm conclusions, the results obtained here could suggest that the dominant source of the perturbation responsible for the frequency shifts with magnetic activity in these four stars is different than in the Sun.

Furthermore, within this subset of 20 potential candidates, the frequency shifts of the $l=1$ modes of a more evolved star with mixed modes were observed to be comparable to the $l=0$ frequency shifts.

%
%
\begin{acknowledgements}
The authors wish to thank the entire {\it Kepler} team, without whom these results would not be possible. Funding for this Discovery mission is provided by NASA's Science Mission Directorate. D.S. and R.A.G. acknowledge the financial support from the CNES GOLF and PLATO grants. D.S. acknowledges the Observatoire de la C\^ote d'Azur for support during his stays. C.R. and F.P.H. acknowledge the financial support from MINECO under grant ESP2015-65712-C5-4-R. This paper has made use of the IAC Supercomputing facility HTCondor (\url{http://research.cs.wisc.edu/htcondor/}), partly financed by the Ministry of Economy and Competitiveness with FEDER funds, code IACA13-3E-2493. This research has made use of the SIMBAD database, operated at CDS, Strasbourg, France.
\end{acknowledgements}

%
%

%
%
\begin{appendix}
\section{Temporal variability of the mode power and damping in KIC\,5184732 and KIC\,8006161}
\label{sec:height}
In this Appendix, we present the temporal variability of the p-mode heights and linewidths extracted from the peak-fitting analysis (Method\,\#2, Section~\ref{sec:pkf}) for two stars, KIC\,5184732 and KIC\,8006161. The mode power and energy supply rate were also estimated as described in \citet{salabert06}. These two stars are the most promising stars in our sample for showing variations in other p-mode parameters than frequencies.

\begin{figure*}[tbp]
\begin{center} 
\includegraphics[width=0.49\textwidth,angle=0]{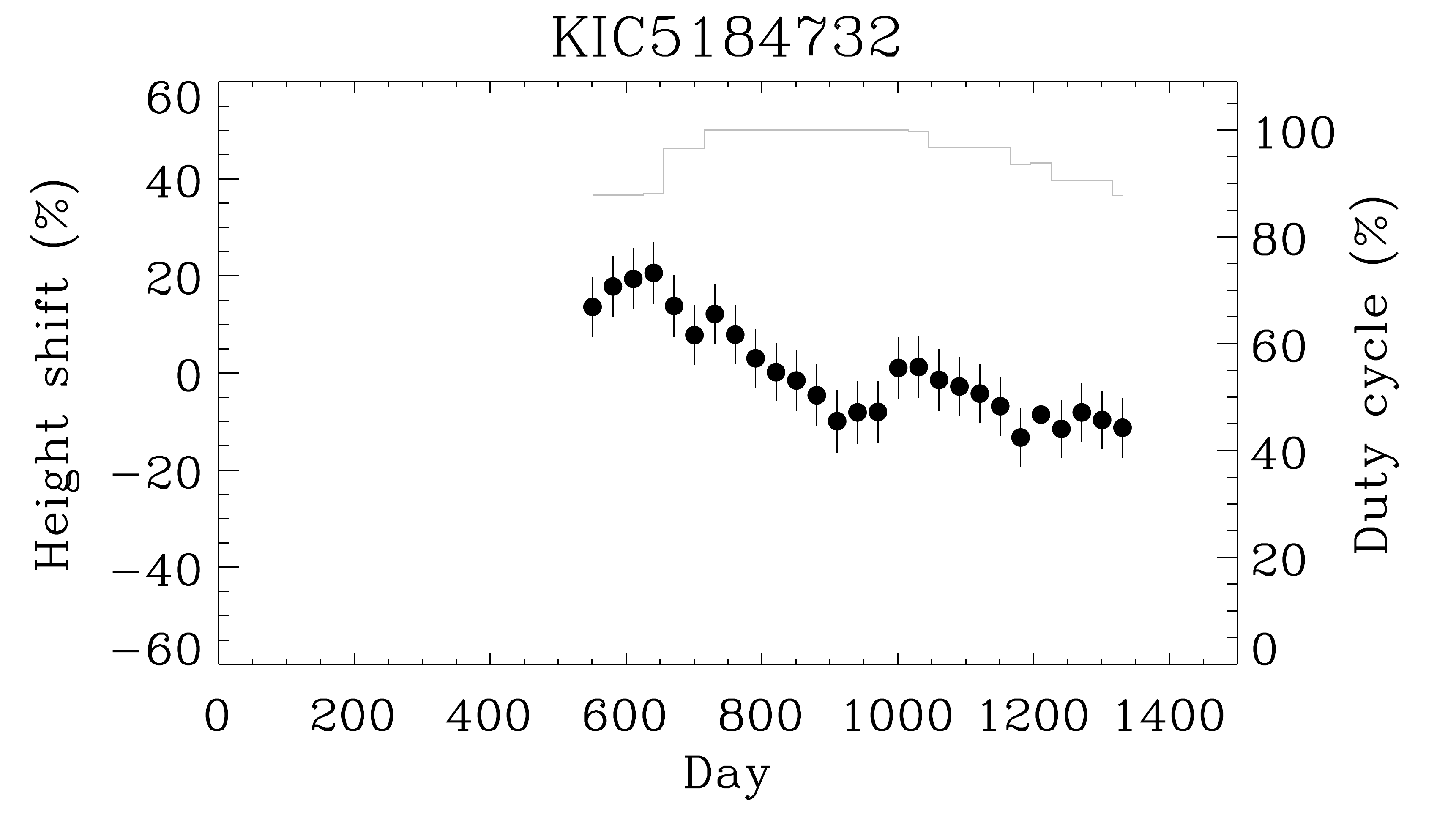}
\includegraphics[width=0.49\textwidth,angle=0]{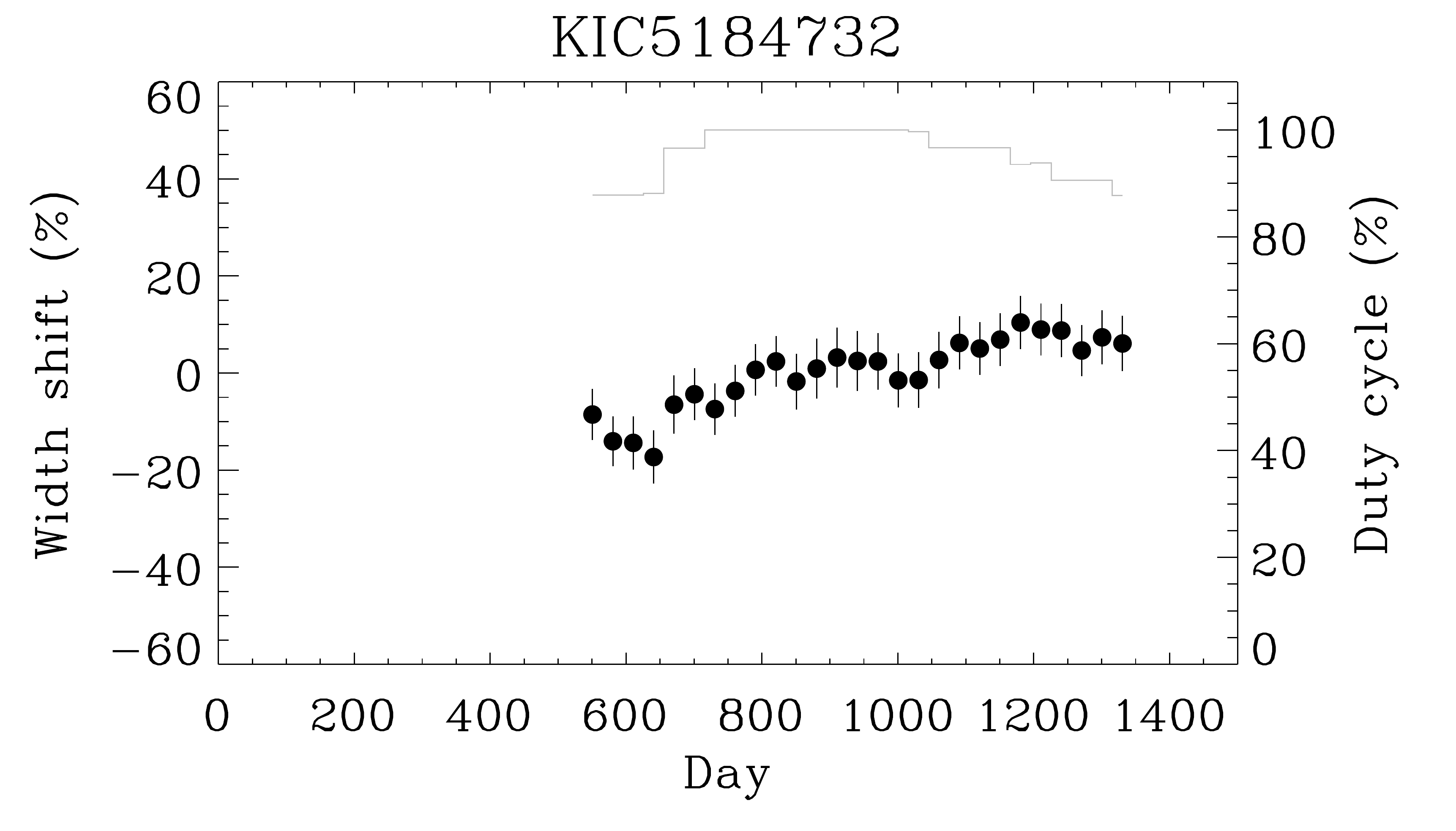}
\includegraphics[width=0.49\textwidth,angle=0]{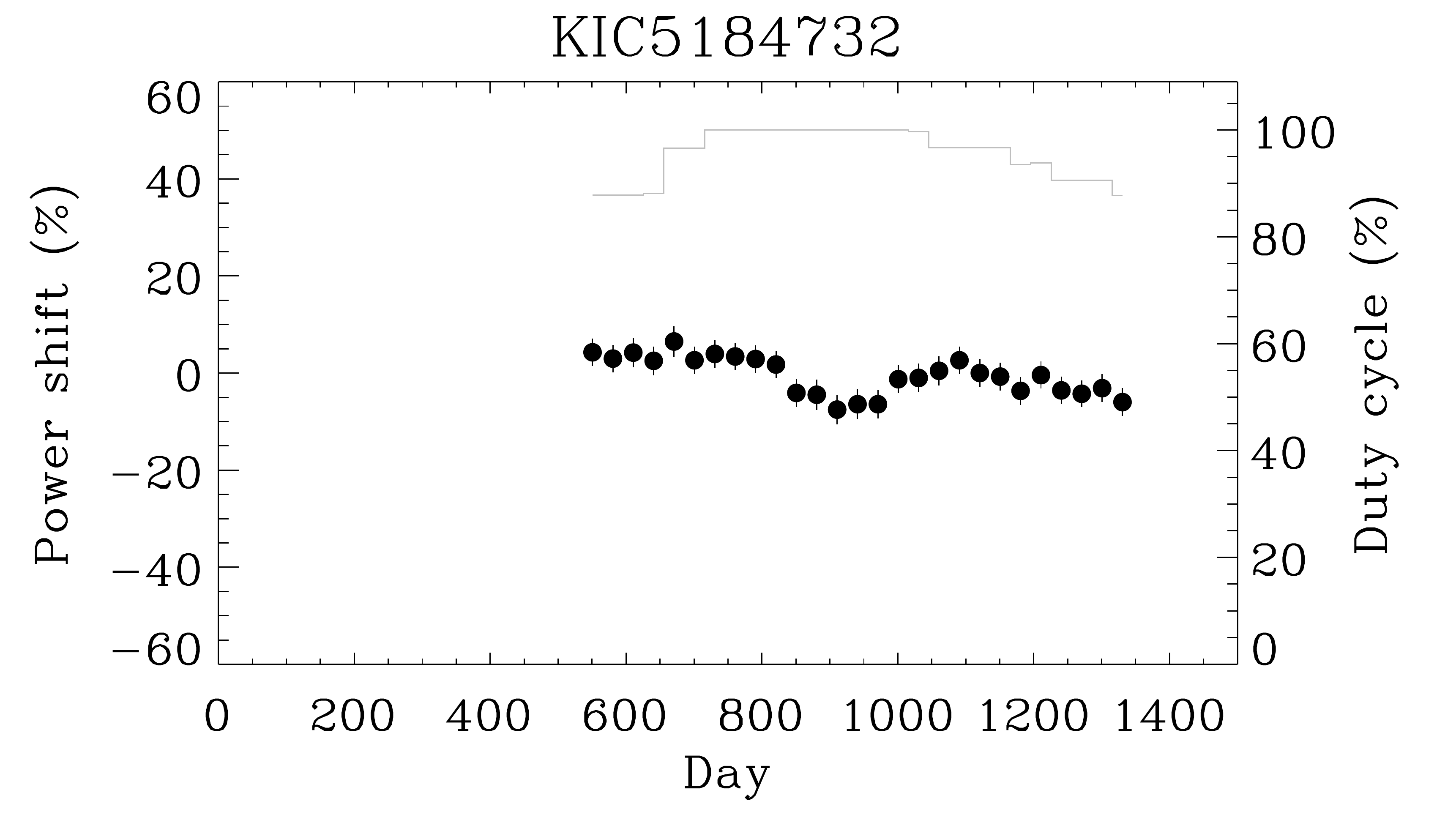}
\includegraphics[width=0.49\textwidth,angle=0]{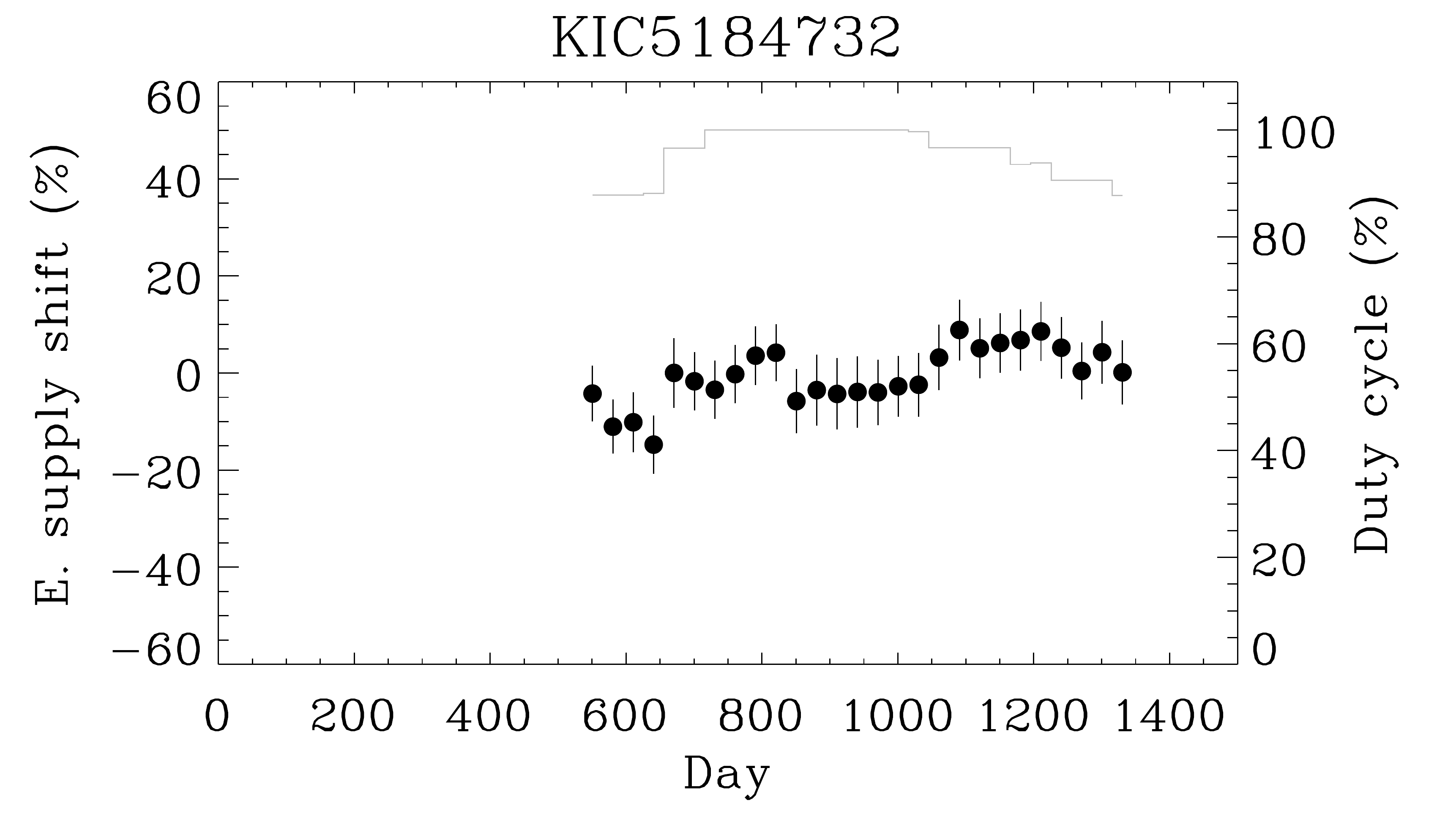}
\includegraphics[width=0.49\textwidth,angle=0]{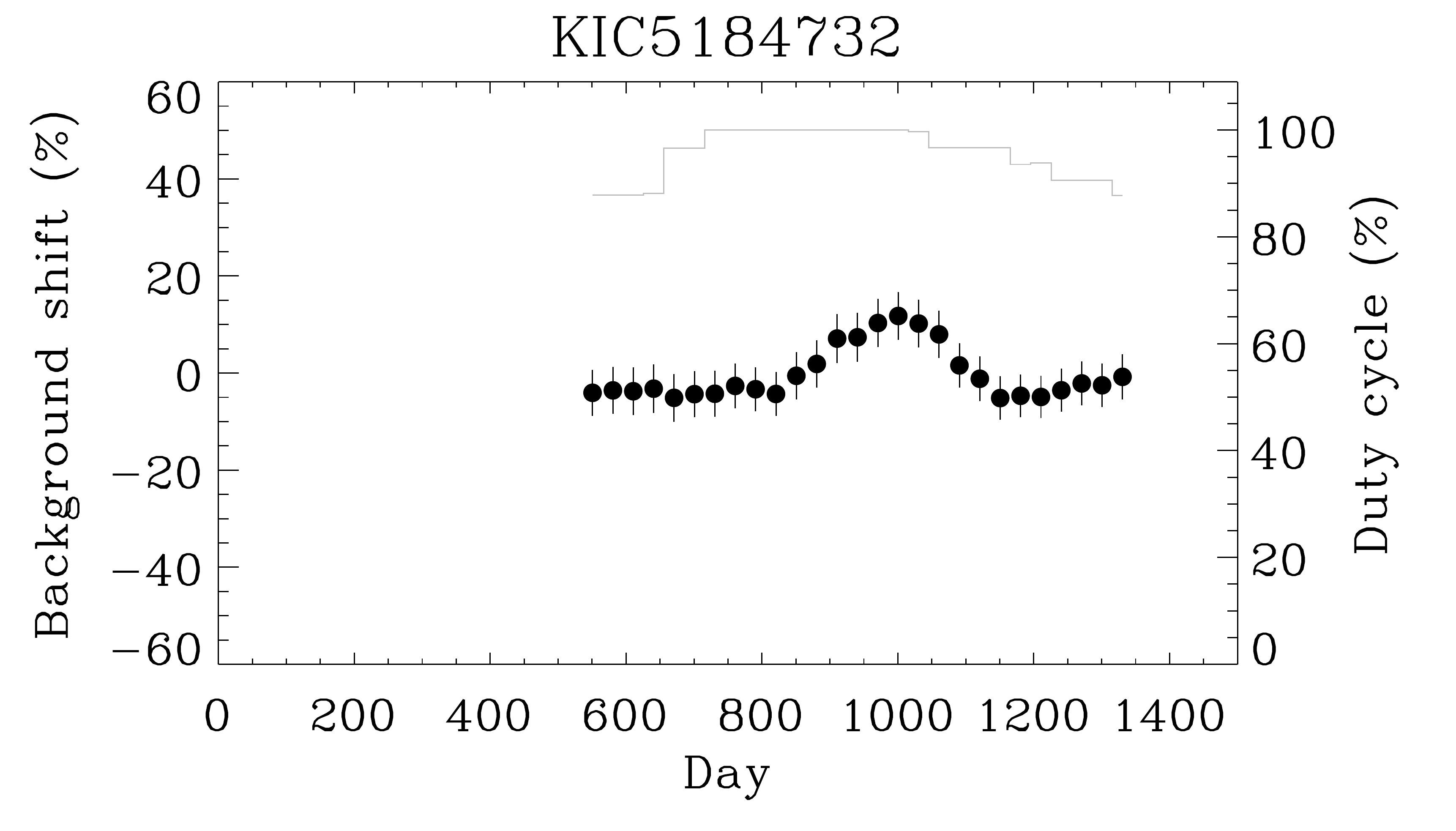}
\includegraphics[width=0.49\textwidth,angle=0]{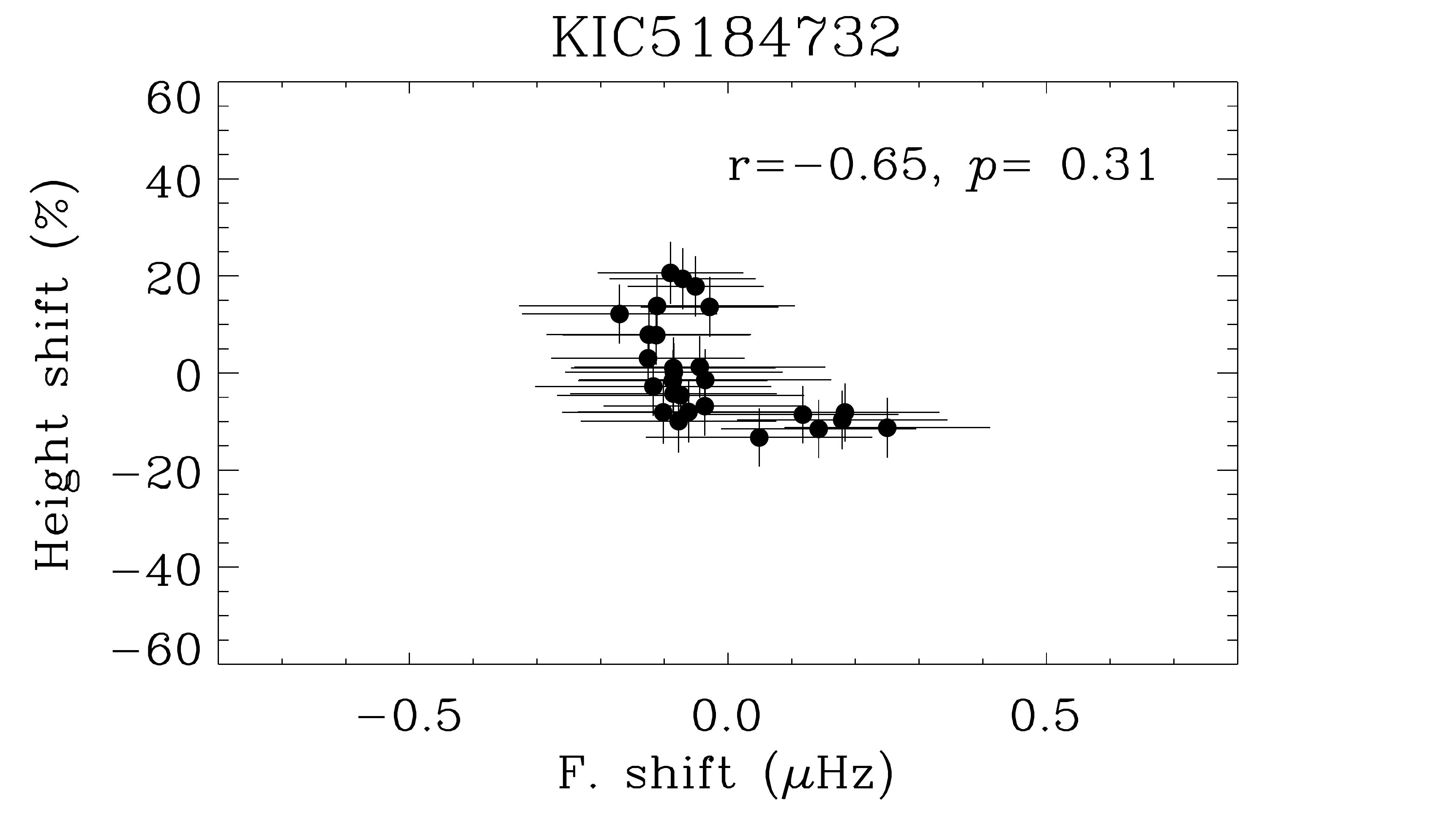}
\includegraphics[width=0.49\textwidth,angle=0]{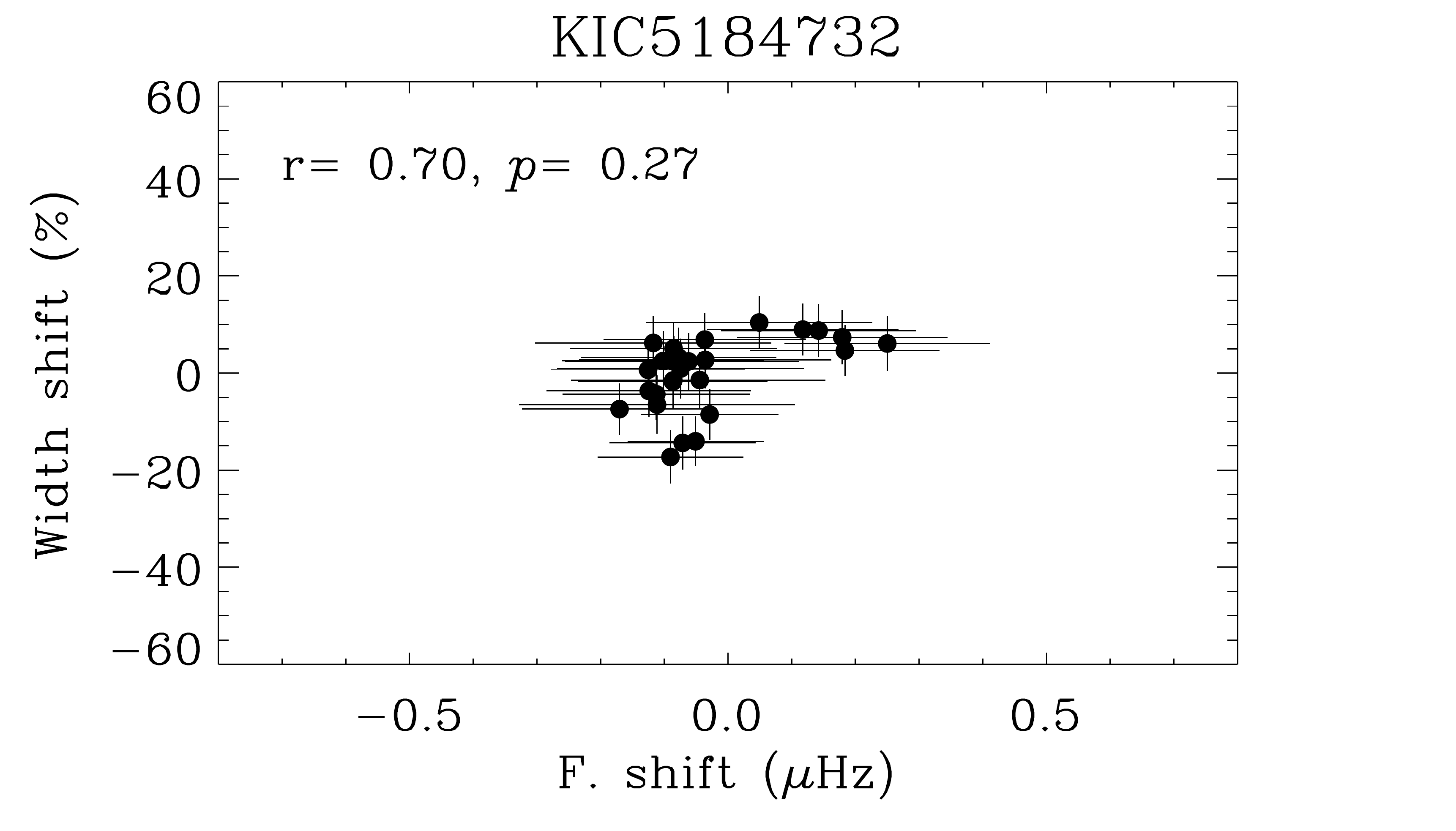}
\includegraphics[width=0.49\textwidth,angle=0]{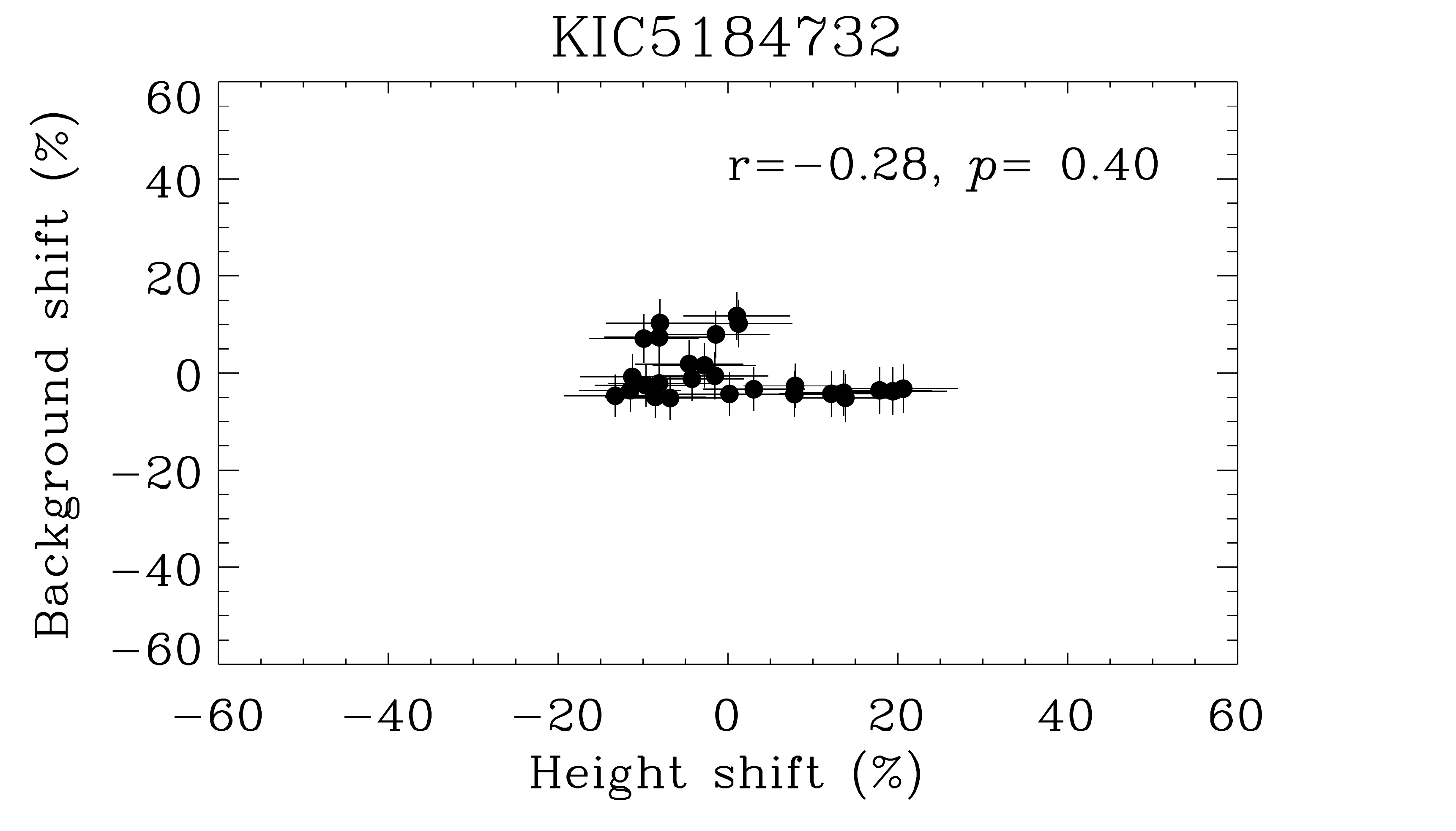}
\end{center}
\caption{\label{fig:height51}  
Fractional temporal variability (in \%) of the p-mode power and damping parameters extracted from the peak-fitting analysis (Method\,\#2) of KIC\,5184732. The duty cycle is shown in light grey. The corresponding correlation coefficients $r$ and two-sided significances of the deviation from zero $p$ are also indicated.}
\end{figure*}

\begin{figure*}[tbp]
\begin{center} 
\includegraphics[width=0.49\textwidth,angle=0]{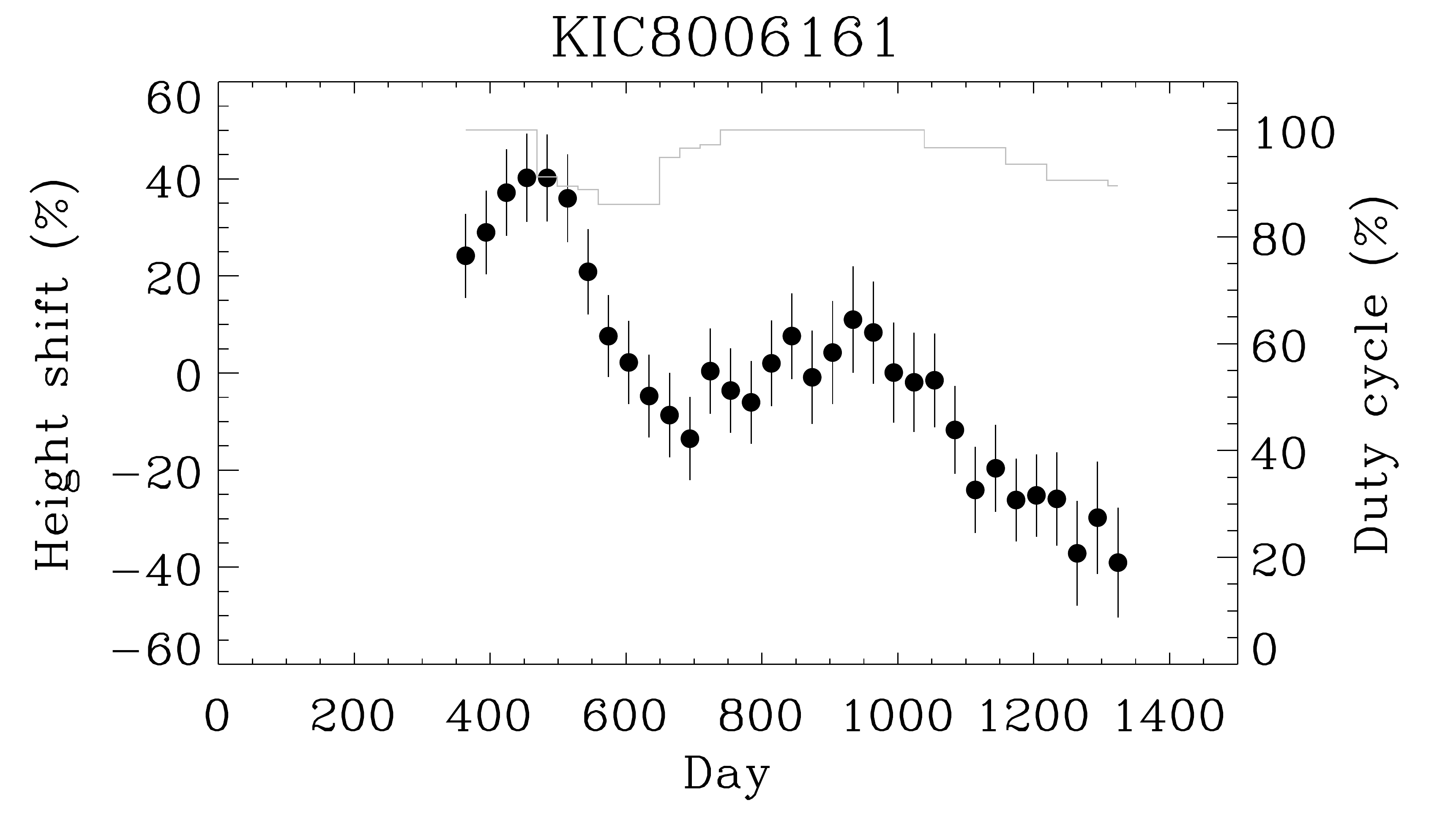}
\includegraphics[width=0.49\textwidth,angle=0]{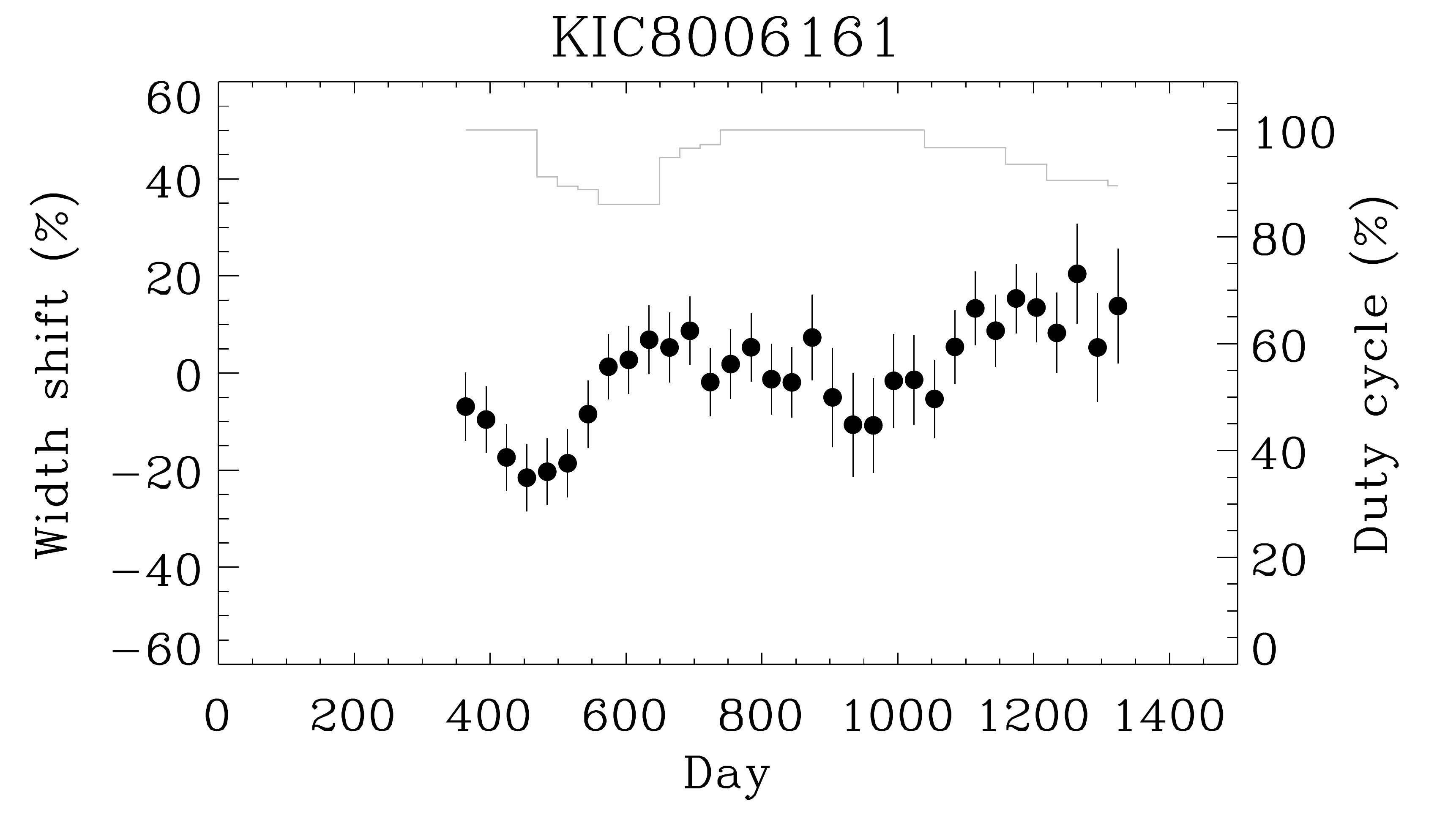}
\includegraphics[width=0.49\textwidth,angle=0]{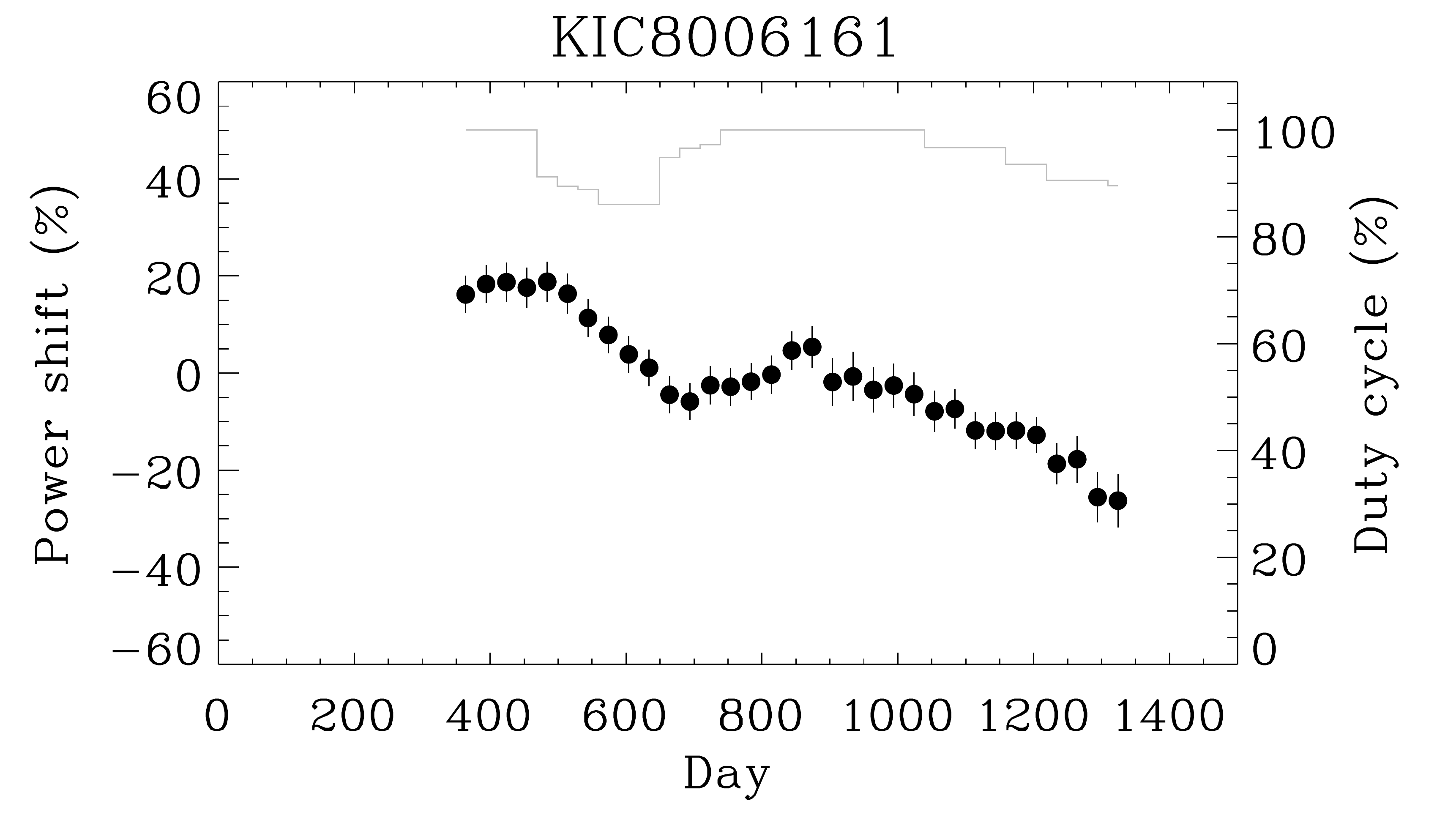}
\includegraphics[width=0.49\textwidth,angle=0]{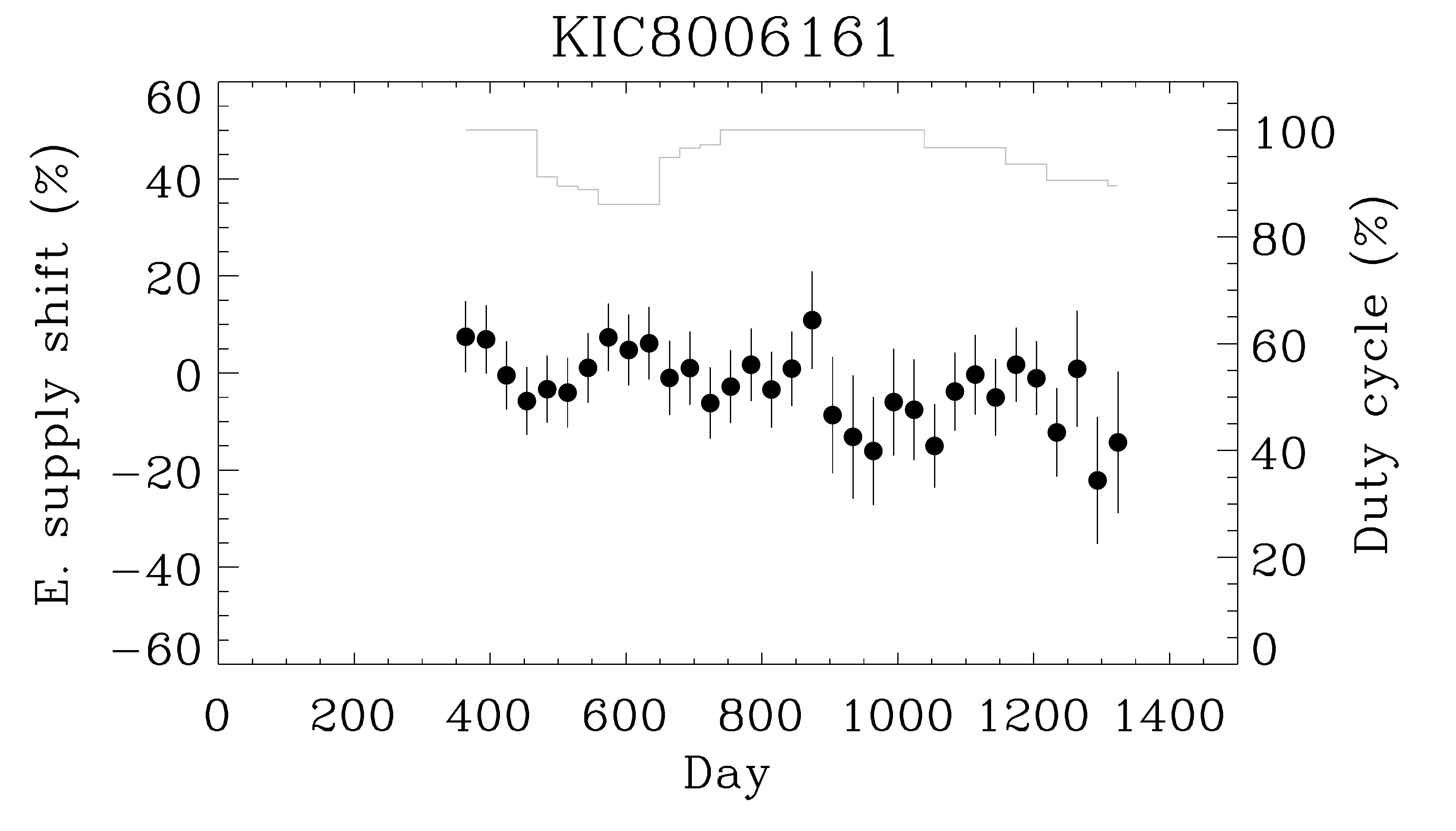}
\includegraphics[width=0.49\textwidth,angle=0]{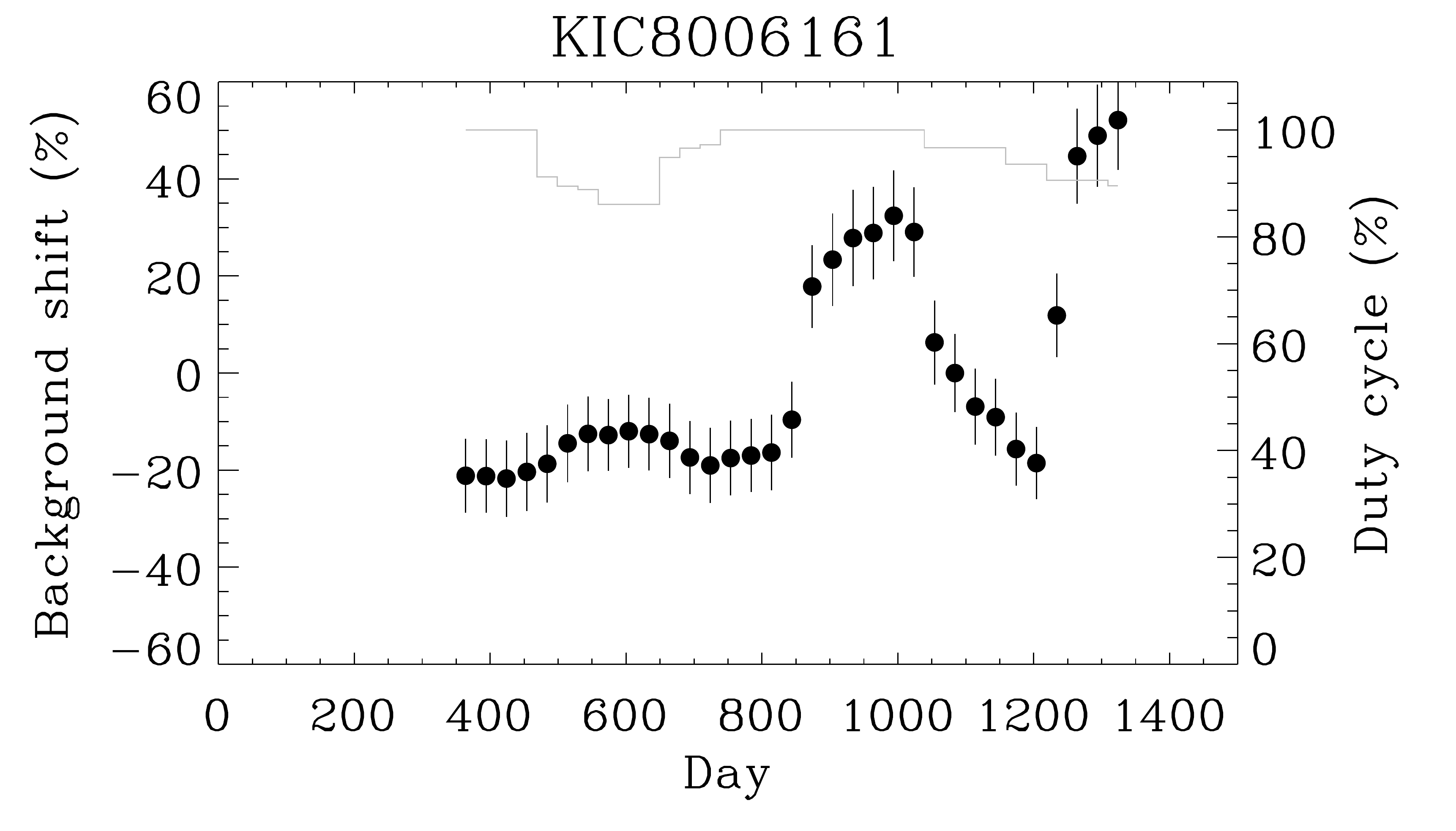}
\includegraphics[width=0.49\textwidth,angle=0]{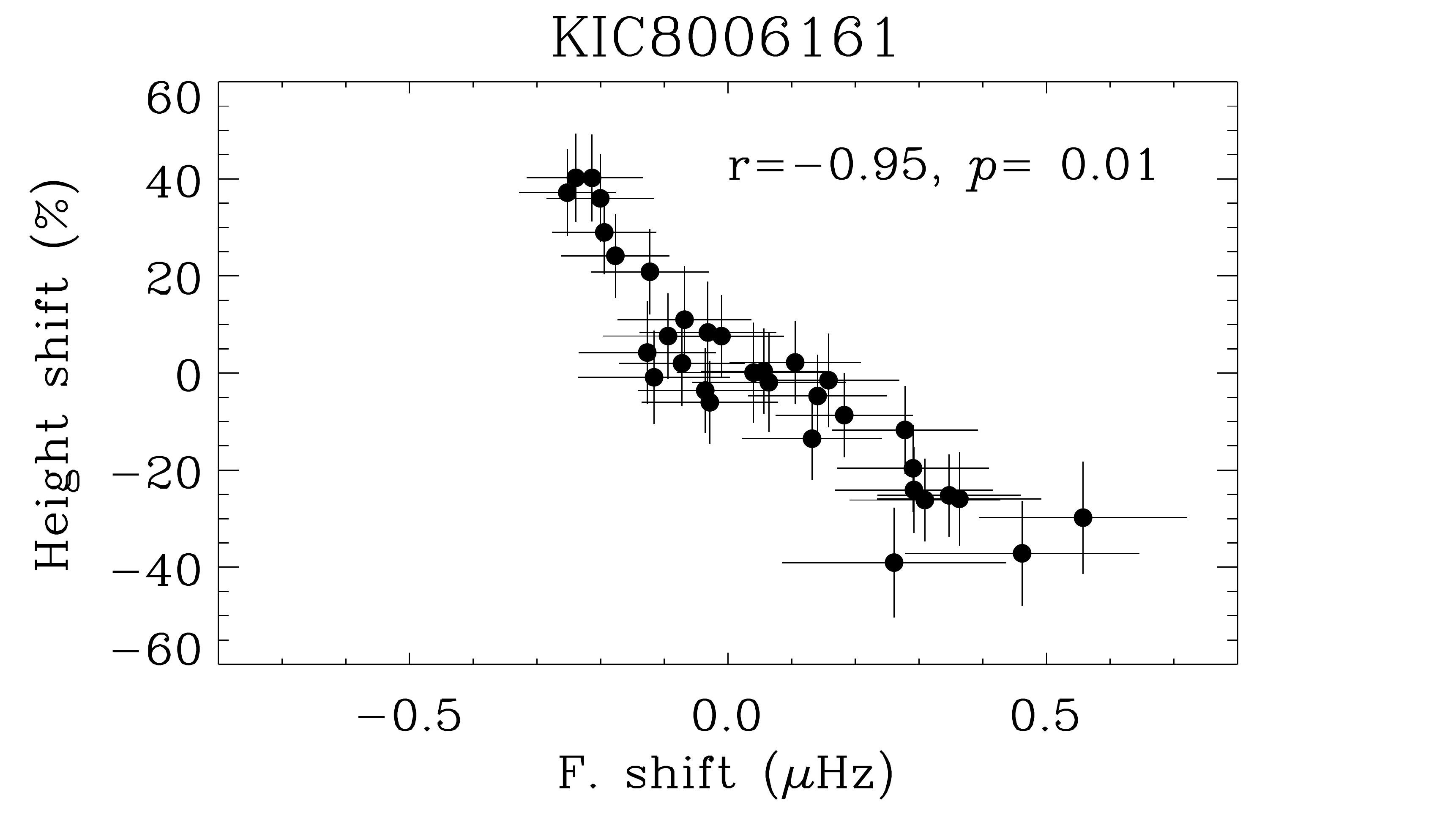}
\includegraphics[width=0.49\textwidth,angle=0]{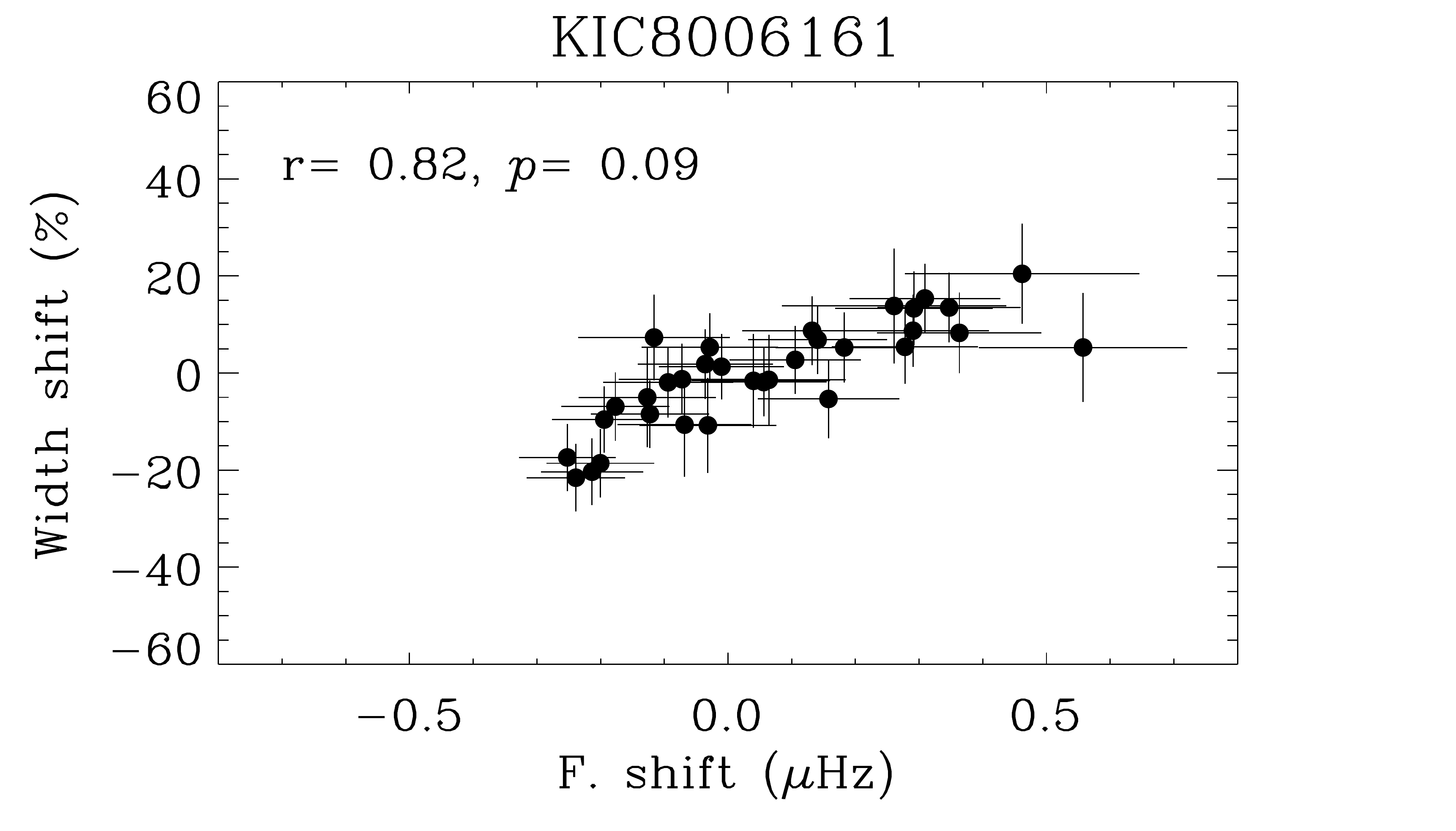}
\includegraphics[width=0.49\textwidth,angle=0]{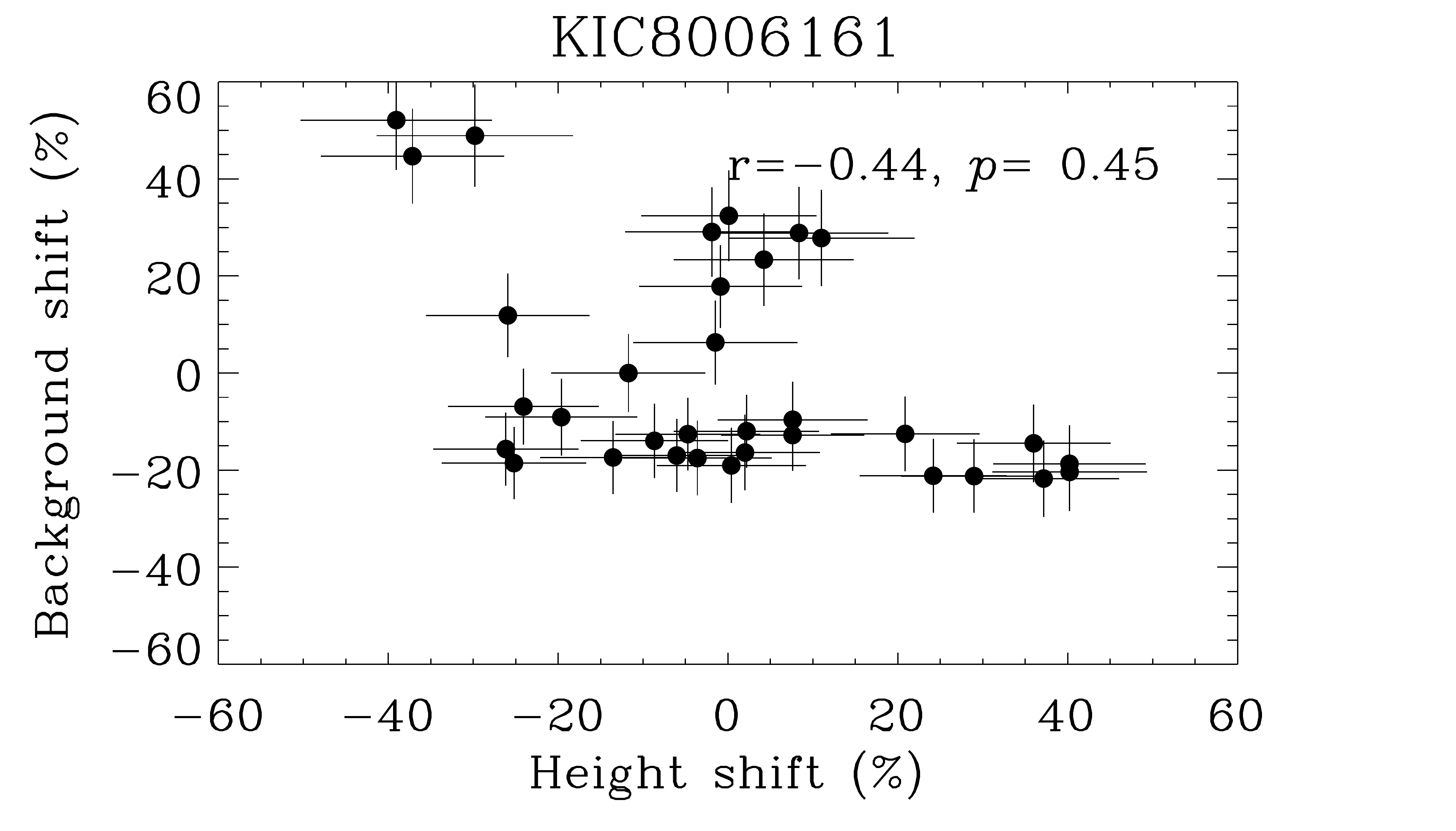}
\end{center}
\caption{\label{fig:height80}  
Fractional temporal variability (\%) of the p-mode power and damping parameters extracted from the peak-fitting analysis (Method\,\#2) of KIC\,8006161. The duty cycle is shown in light grey. The corresponding correlation coefficients $r$ and two-sided significance of the deviation from zero $p$ are also indicated.}
\end{figure*}

Figures~\ref{fig:height51} and \ref{fig:height80} show the fractional temporal evolutions likely associated to magnetic activity of the mode heights, linewidths, powers, and energy supply rates of KIC\,5184732 and KIC\,8006161 respectively. They are compared to the corresponding variations of the background noise and of the duty cycle of the analyzed {\it Kepler} subseries. The relations between mode heights and linewidths with the frequency shifts are also shown on Fig.~\ref{fig:height51} and \ref{fig:height80}. The associated Spearman's correlation coefficients $r$ and two-sided significances of the deviation from zero $p$ are given as well.  We note that they were computed using independent points only. Finally, the relations between the background noise and the mode heights are presented. The results between correlation and anti-correlation of the variations with magnetic activity are similar to what is observed for the Sun. In the case of KIC\,8006161, we confirm the results from \citet{kiefer17} on the anti-correlation between the frequency shifts and the mode height variability.

\begin{figure*}[tbp]
\begin{center} 
\includegraphics[width=0.49\textwidth,angle=0]{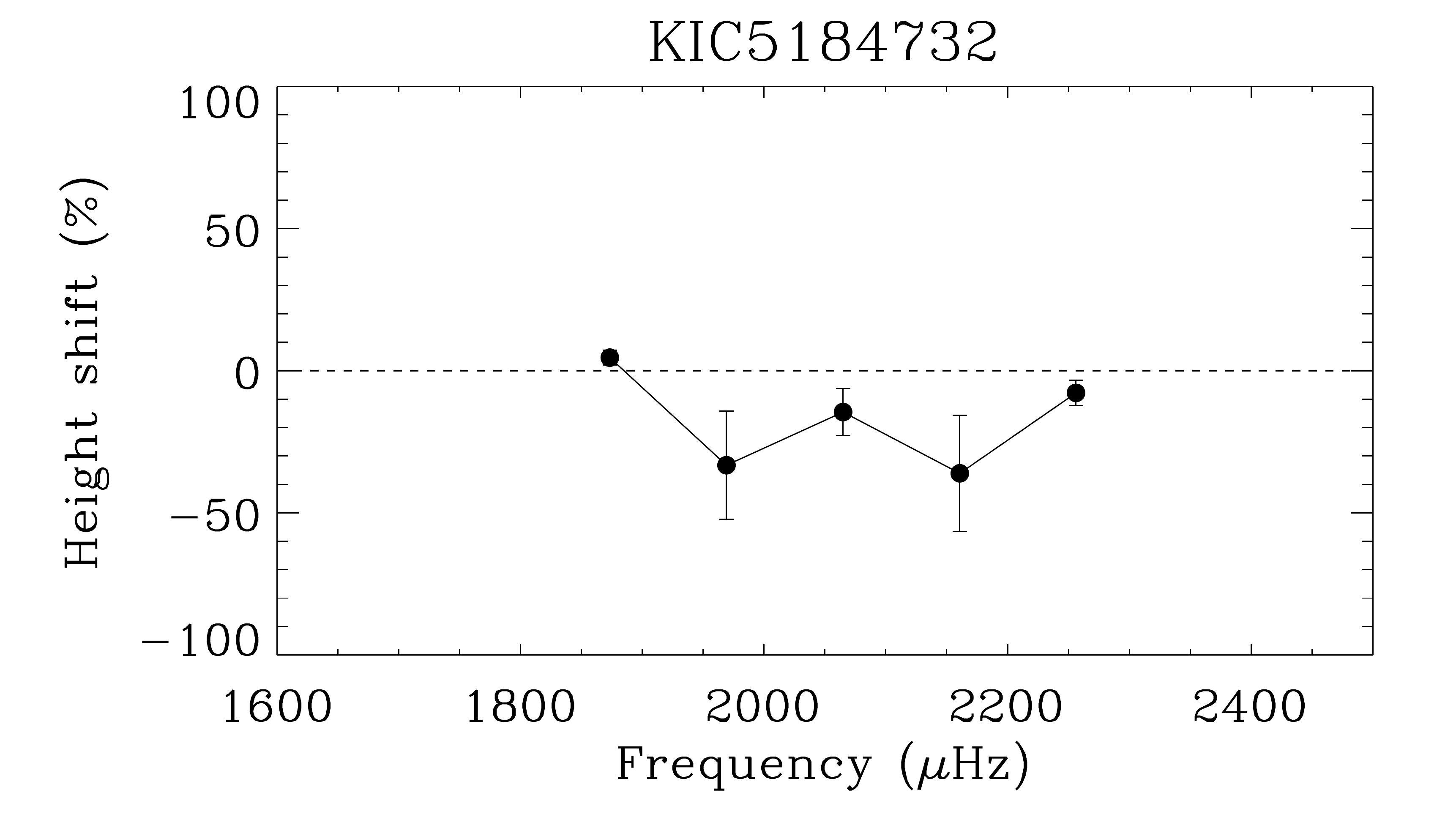}
\includegraphics[width=0.49\textwidth,angle=0]{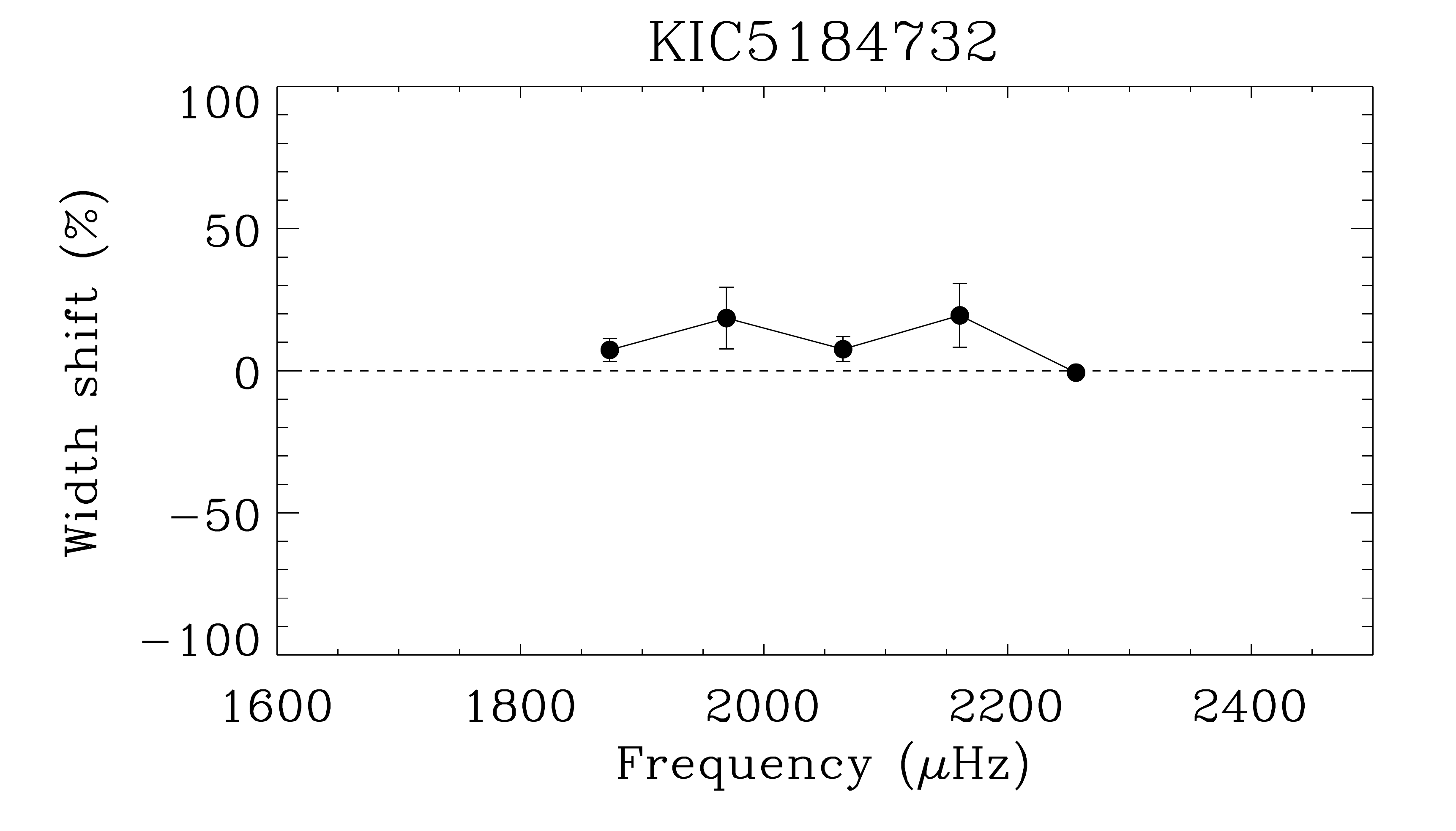}
\includegraphics[width=0.49\textwidth,angle=0]{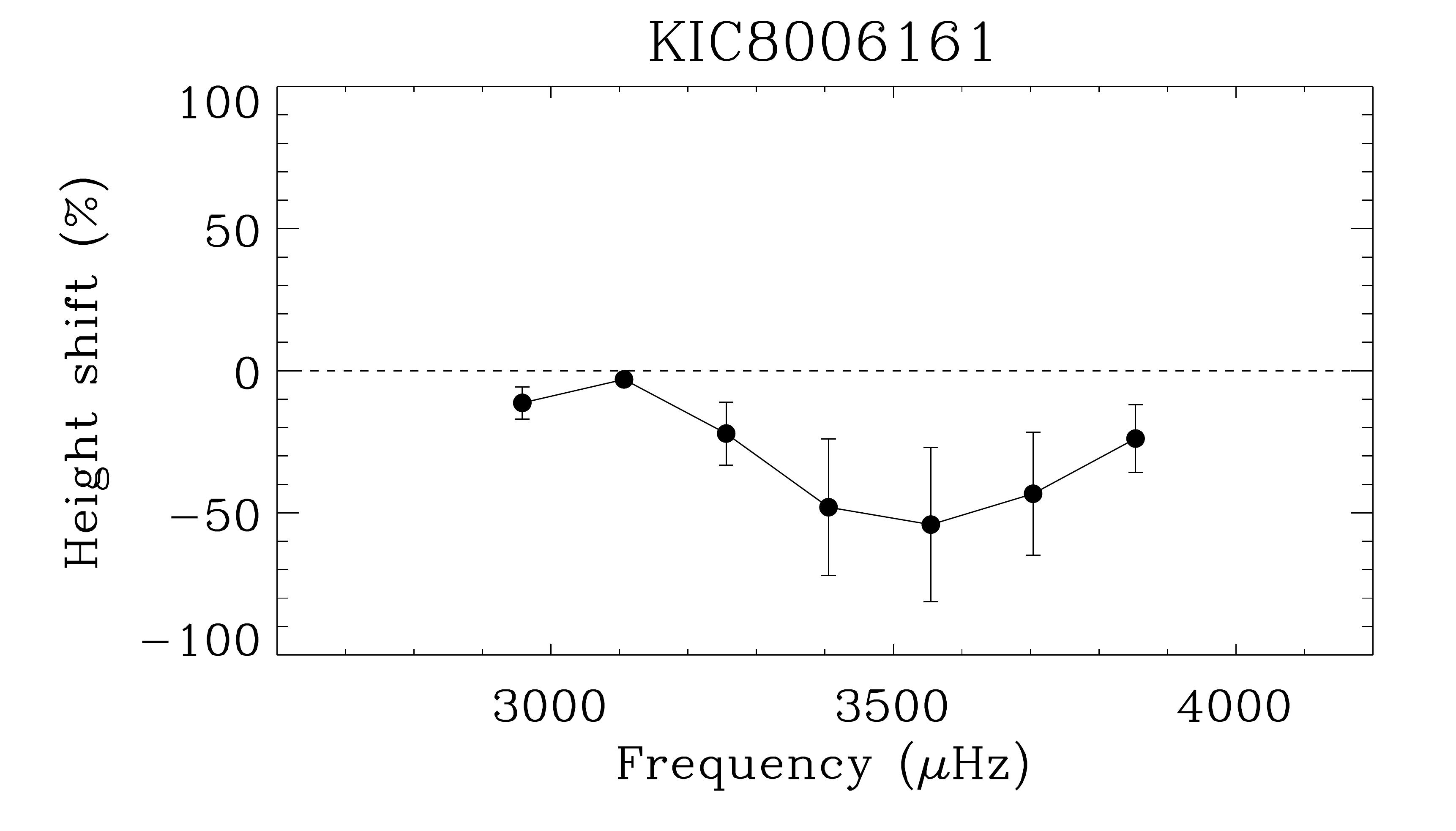}
\includegraphics[width=0.49\textwidth,angle=0]{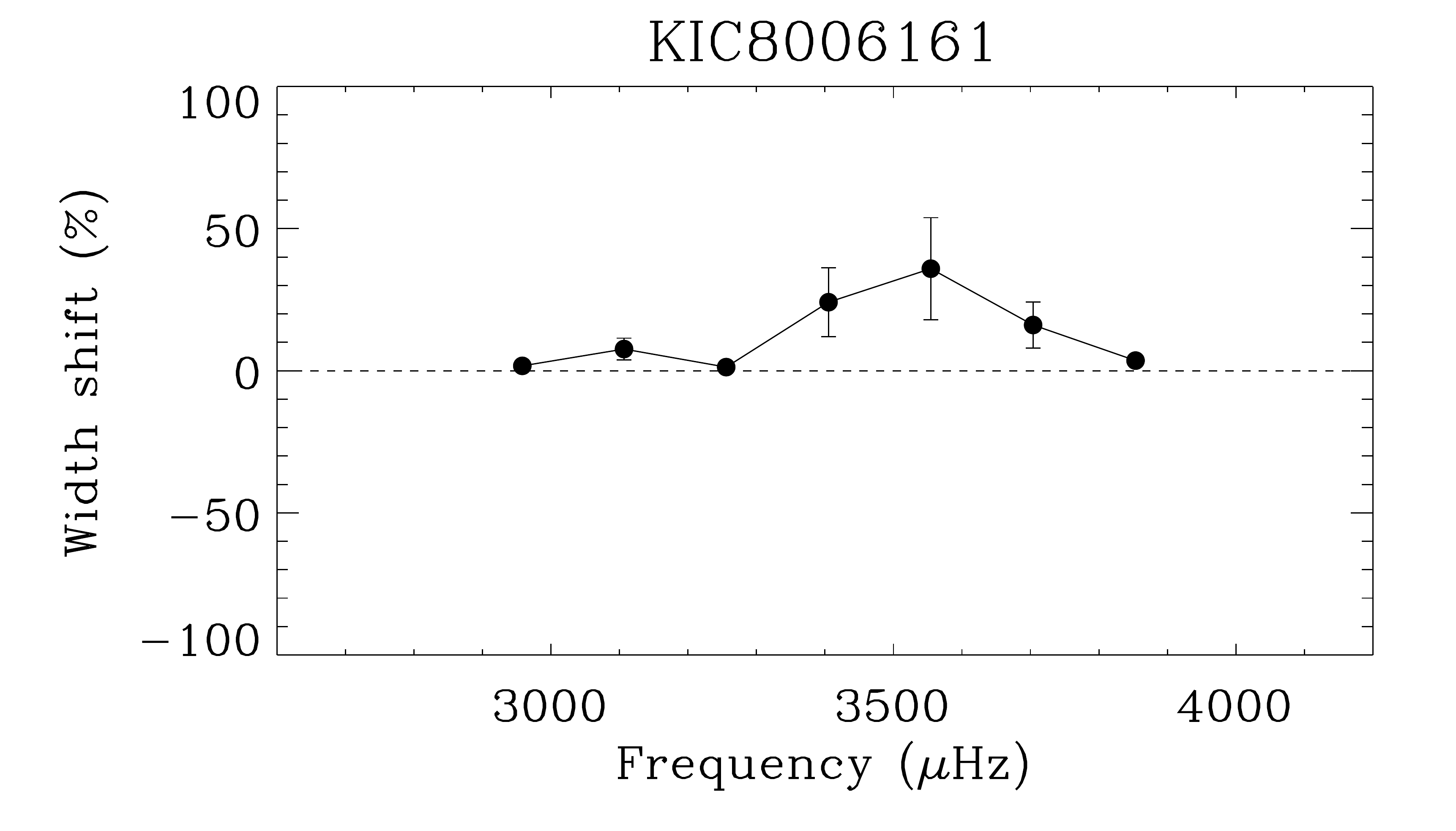}
\end{center}
\caption{\label{fig:heightvsfreq}  
Frequency dependence of the fractional variations measured in the p-mode heights and linewidths of KIC\,5184732 ({\it top}) and KIC\,8006161 ({\it bottom}). The horizontal dashed lines represent a null variation.}
\end{figure*}

Figure \ref{fig:heightvsfreq} shows the frequency dependence of the mode height and linewidth variabilities. Although the results are clearer for KIC\,8006161, the observed dependences are comparable to what observed in the Sun \citep[see e.g.,][]{salabert06}. It is the first time that such a frequency dependence of the mode heights and linewidths with magnetic activity is observed in distant stars. 

\end{appendix}

\end{document}